\documentclass[11pt]{article}
\usepackage[T1]{fontenc}
\usepackage[utf8]{inputenc}
\usepackage{amsmath,amssymb,amsthm,mathtools}
\usepackage[margin=1in]{geometry}
\usepackage{enumitem,booktabs,longtable,array}
\usepackage[colorlinks=true,linkcolor=blue,citecolor=blue,urlcolor=blue]{hyperref}
\usepackage{bookmark}
\usepackage{authblk}

\hypersetup{pdftitle={Product expansion of three algebraic-geometry codes over a fixed field},
            pdfauthor={Cassie Ding, Min-Hsiu Hsieh}}
\numberwithin{equation}{section}

\newtheorem{theorem}{Theorem}[section]
\newtheorem{lemma}[theorem]{Lemma}
\newtheorem{proposition}[theorem]{Proposition}
\newtheorem{corollary}[theorem]{Corollary}
\newtheorem{externalinput}[theorem]{External input}
\newtheorem{question}[theorem]{Question}

\theoremstyle{definition}
\newtheorem{definition}[theorem]{Definition}
\newtheorem{remark}[theorem]{Remark}

\newtheorem{example}[theorem]{Example}

\theoremstyle{plain}
\newtheorem{mainthm}{Theorem}

\newtheorem{maincor}[mainthm]{Corollary}

\newcommand{\F}{\mathbb F}
\newcommand{\PP}{\mathbb P}
\newcommand{\OO}{\mathcal O}
\newcommand{\II}{\mathcal I}
\newcommand{\LL}{\mathcal L}
\renewcommand{\AA}{\mathcal A}

\newcommand{\NN}{\mathcal N}
\newcommand{\BB}{\mathcal B}
\newcommand{\MM}{\mathcal M}
\newcommand{\QQ}{\mathcal Q}
\newcommand{\cF}{\mathcal F}
\newcommand{\cG}{\mathcal G}
\newcommand{\Hom}{\operatorname{Hom}}
\newcommand{\cHom}{\mathcal{H}\!om}
\newcommand{\Pic}{\operatorname{Pic}}
\newcommand{\Supp}{\operatorname{Supp}}
\newcommand{\supp}{\operatorname{supp}}
\newcommand{\im}{\operatorname{im}}
\newcommand{\length}{\operatorname{length}}
\newcommand{\Ann}{\operatorname{Ann}}
\newcommand{\codim}{\operatorname{codim}}
\newcommand{\wt}{\operatorname{wt}}

\newcommand{\ev}{\operatorname{ev}}
\newcommand{\Res}{\operatorname{Res}}
\newcommand{\res}{\operatorname{res}}

\newcommand{\Xj}[1]{X^{(#1)}}
\newcommand{\Cj}[1]{C^{(#1)}}

\DeclareMathOperator{\gon}{gon}
\newcommand{\lm}{\operatorname{lm}}
\newcommand{\LME}{\LL_{M,E'}}

\newcommand{\rhoGS}{\rho_{\mathrm{GS}}}
\newcommand{\boxprod}{\mathop{\boxtimes}\limits_{j=1}^{3}}

\title{Uniform Three-Fold Product Expansion of Algebraic-Geometry Codes over a Fixed Field}
\author[1]{Cassie Ding\thanks{\texttt{dkexin@engineering.upenn.edu}}}
\author[2]{Min-Hsiu Hsieh\thanks{\texttt{min-hsiu.hsieh@foxconn.com}}}
\affil[1]{Department of Electrical and Systems Engineering, University of Pennsylvania}
\affil[2]{Hon Hai (Foxconn) Research Institute, Taipei, Taiwan}
\date{\today}

\begin{document}
\maketitle

\begin{abstract}
A collection of classical codes $C_1,\dots,C_m$ has \emph{product expansion} if every low-weight word in the sum of their extensions to the product space can be written as a sum of words supported on few lines, each line being a codeword of the corresponding $C_i$.
For $m=2$ product expansion is equivalent to robust testability of the tensor code, while for $m\ge 3$ it is strictly stronger; it is the hypothesis on the local codes behind recent constructions of good quantum LDPC codes, good quantum locally testable codes, and quantum codes with transversal non-Clifford gates.
No explicit family over a fixed field was previously known to achieve constant rate, linear distance for both the code and its dual, and 3-fold product expansion at growing length.

We prove that for the one-point evaluation codes $C_i$ of length $N_i=728\cdot27^{\,i-1}$ and pole order $\lfloor N_i/5\rfloor$ on the Garcia--Stichtenoth tower over $\F_{729}$, the triple $(C_i,C_i,C_i)$ has product expansion with a constant independent of $i$. The dual codes $C_i^\perp$ then form an explicit fixed-field family of rate tending to $109/130$ and relative distance at least $8/65$.
The proof is geometric: we view the words of the sum code as arrays on the product of three copies of the $i$-th curve of the tower, where the parity checks are sections of a line bundle.
An exact generation theorem describes the sum-code words supported on the grid footprint of a suitable divisor. A support-localisation argument then shows that every low-weight array lies on such a divisor, after deleting a small number of coordinates. 
We also give a sufficient criterion for three-fold product expansion in terms of degree and genus ratios, yielding explicit families over every non-prime finite field of order at least $512$.
The principal conclusions have been formally verified in Lean~4.
\end{abstract}


\section{Introduction} \label{sec:intro}


A quantum low-density parity-check (qLDPC) code is a stabilizer code in which every check acts on a bounded number of qubits and every qubit enters a bounded number of checks. Such codes are a leading candidate for fault-tolerant quantum computation.
Whether asymptotically good qLDPC codes exist, with rate and relative distance bounded away from zero, was open for two decades. It was settled by Panteleev and Kalachev~\cite{PK22}, Leverrier and Z\'emor~\cite{LZ22} and Dinur, Hsieh, Lin and Vidick~\cite{DHLV23}. In each of these constructions the quantum code is assembled from a bounded-degree expanding complex and a small collection of \emph{local} classical codes. The proof that the quantum distance is linear reduces to a property of the local codes alone.

That property is \emph{product expansion}~\cite{PK22,KP22}. For codes $C_1,\dots,C_D\subseteq\F_q^n$, consider the $D$-dimensional arrays that can be written as a sum $u_1+\dots+u_D$, where all lines of $u_j$ in direction $j$ are codewords of $C_j$. The collection is $\rho$-product-expanding if every such array of Hamming weight $w$ admits such a decomposition using at most $w/(\rho n)$ nonzero coded lines. In words, cancellation between directions cannot produce a light array out of many nonzero lines. 
For two codes of linear distance, the properties of product expansion, robustness and agreement testability are equivalent up to constant factors.
This equivalence does not extend to three or more codes.

Suitable forms of these properties are a key ingredient in proving the distance and soundness analyses of the almost-good quantum locally testable codes~\cite{DLV24}, of the good quantum locally testable codes~\cite{GJ26,LLL26good}, and of the good classical locally testable codes~\cite{DEL+22,PK22}. 
Bafna, Li and Nguyen~\cite[Theorem~1.3]{BLN26} give a further construction of good quantum locally testable codes, conditional on a variant of the Reed--Solomon product-expansion conjecture, and the private PCPs of Bafna and Vyas~\cite{BV26} rest on product expansion as well.
A second line of work concerns logical gates. By the Eastin--Knill theorem~\cite{EK09} no code implements a universal gate set transversally. The non-Clifford gates needed for universality, such as $T$ or $\mathrm{CCZ}$, are the expensive part of every fault-tolerant architecture, and are traditionally supplied by magic-state distillation~\cite{BK05,BH12}. 
Wills, Hsieh and Yamasaki~\cite{WHY25} obtained the first constant-overhead magic-state distillation, and asymptotically good quantum codes with a transversal $\mathrm{CCZ}$ gate were constructed by Golowich and Guruswami~\cite{GG24} and Nguyen~\cite{Ngu25}.
All three constructions exploit the multiplicative structure of algebraic codes, namely Reed--Solomon or algebraic-geometry codes, and the resulting quantum codes are not LDPC: their checks have high weight.

The subsequent constructions of qLDPC codes, or codes with sublinear check weight, carrying transversal non-Clifford gates~\cite{GL24,Lin24,Zhu25a,Zhu25b,GG25b,LLL26} are products of three classical codes, or codes on three-dimensional complexes. The reason is that a $\mathrm{CCZ}$ gate is carried by a trilinear form, and a trilinear form lives on a three-fold product. In these constructions the product expansion of three codes, or of a mixed collection of codes and dual codes, is exactly the hypothesis that controls the distance of the quantum code. Golowich and Guruswami observe that stronger product-expansion results would permit extensions to products of higher order~\cite[Section~1]{GG25b}, and no-go theorems for logical gates on product codes~\cite{FZLL25} delineate what such products can achieve.

The constructions use product expansion in two structurally different regimes. In the first, the local codes have constant length $n$, fixed independently of the length $N$ of the quantum code. A random choice over a fixed field then supplies the required expansion with positive probability, and the finite search for suitable local codes does not depend on $N$. This is the regime of~\cite{PK22,LZ22,DHLV23,DLV24,LLL26}.
The second regime, introduced by the homological product codes of Golowich and Guruswami~\cite{GG24b}, requires product expansion for component codes of growing length. 
Their subsystem product codes~\cite{GG25b} exemplify this setting: the component length and check weight scale as \(\Theta(N^{1/t})\), while the gate arity is independent of \(t\). 
Here the component codes must be explicit, since their length grows and no finite search is possible. Reed--Solomon codes qualify, but they need an alphabet size $q\ge n$ that grows with $N$. Whether explicit codes of growing length over a \emph{fixed} field can have the required product expansion is therefore the binding constraint of this regime, and it is a question about classical codes.

\subsection{What is known, and the challenge} \label{subsec:intro_challenge}
 
For two codes, product expansion is equivalent up to constants to the robust testability of the tensor code of Ben-Sasson and Sudan~\cite{BSS06}, provided both codes have linear distance~\cite{KP22,KP25}.
In this form it is known for Reed--Solomon codes of rates $R_1+R_2<1$, by the bivariate testing theorem of Polishchuk and Spielman~\cite{PS94}; for pairs of random codes~\cite{PK22,KP22,LZ22,DHLV23}; for tensor products with a suitable LDPC factor~\cite{DSW06}; and for algebraic-geometry codes over a fixed field, with rate $O(n^{-1/2})$ by Garg, Sudan and Wu~\cite{GSW25} and with constant rate by Garg and Sengupta~\cite{GargSengupta25}. The last of these results gives the first explicit family of growing length over a fixed field with constant rate, linear dual distance and robust testability of the tensor square. 
This result supplies the two-code case on which the present work builds. Robust testability is not automatic: there are pairs of codes of large distance whose tensor product is not robustly testable~\cite{Val05,CR05,GM12}.
For three or more codes, product expansion is strictly stronger than robust testability of the tensor product under the axis-parallel line tester. Kalachev~\cite{Kal23} showed that for the primitive Reed--Solomon code of rate $\tfrac13$ over $\F_{2^{2t}}$, three copies have product expansion at most $1/n$, although the tensor cube is robustly testable. The witness is an array that meets every line at most once. By monotonicity, the same collapse holds for every Reed--Solomon code containing the one above. 

The positive results at $D\ge3$ are of three kinds. Random collections of codes are product-expanding, with constant $\rho$, over fields of characteristic two whose size grows with $n$~\cite{KP25} and over prime fields whose size grows with $n$~\cite{CRS26}. Punctured tensor Reed--Solomon codes with one dual factor~\cite[Theorem~2.4]{GG25b}, and punctured Reed--Solomon codes on random evaluation sets together with their duals~\cite{LLL26}, are product-expanding over fields of size exponential in $n$. Collections of constant-soundness locally testable codes with bounded column weight are product-expanding, explicitly and over a fixed field, for any $D$~\cite[Lemma~14]{KP25}, \cite[Lemma~8.7]{GG25b}. The last family is explicit and fixed-field, but its parity checks have bounded weight, so the dual of each code has  distance $O(1)$. Using such a code as the dual of the component code destroys the distance of the component code itself.

The challenge is the following conjunction. One wants an explicit family $\{C_n\}$ over a fixed field with constant rate and linear distance, such that the three-fold collection of \emph{duals} $(C_n^\perp,C_n^\perp,C_n^\perp)$ is product-expanding with a constant independent of $n$. Each known positive result fails one requirement. Random codes are not explicit and need growing fields. Reed--Solomon codes need $q\ge n$ and fail for a range of rates by Kalachev's separation. Locally testable codes give bounded distance for $C_n$. The algebraic-geometry result of Garg and Sengupta is a two-direction statement. Its proof, an intersection-theoretic divisibility argument on the product of two curves, does not extend to three directions, precisely because product expansion and robust testability part ways there.
\subsection{The central question and the main result} \label{subsec:intro_main}

\begin{question}[explicit fixed-field three-fold expansion] \label{q:main}
Is there a prime power $q$ and an explicit family of linear codes $\{C_n\subseteq\F_q^{\,n}\}_{n\to\infty}$ over the fixed field $\F_q$ such that, simultaneously and with all constants independent of $n$,
\begin{enumerate}[label=\textup{(\roman*)},leftmargin=2.5em,itemsep=1pt,topsep=2pt]
\item \textbf{constant rate:} $\dim C_n\ge Rn$ for some $R>0$;
\item \textbf{linear distance:} $d(C_n)\ge\delta n$ for some $\delta>0$;
\item \textbf{product expansion of the duals at $D=3$:} the collection $(C_n^\perp,C_n^\perp,C_n^\perp)$ is $\rho_0$-product-expanding for some $\rho_0>0$?
\end{enumerate}
\end{question}

The present paper answers Question~\ref{q:main} affirmatively, with an explicit family of algebraic-geometry codes over a fixed field. Let $k_0=\F_{729}$ and let $\cdots\to X_3\to X_2\to X_1$ be the Garcia--Stichtenoth tower of curves over $k_0$ defined by the Artin--Schreier equations $z_{i+1}^{27}+z_{i+1}=x_i^{28}$. Let $E_i$ be the $N_i=728\cdot27^{\,i-1}$ rational points of $X_i$ above $x_1\in k_0^\times$, let $P_{\infty,i}$ be the point at infinity, and let\begin{equation} \label{eq:intro_family}
C_i:=C_L\bigl(E_i,\,\lfloor N_i/5\rfloor\,P_{\infty,i}\bigr)\subseteq k_0^{N_i}
\end{equation}
be the one-point evaluation code: its codewords are the vectors $(f(P))_{P\in E_i}$ obtained from rational functions $f$ on $X_i$ defined over $k_0$, with no poles away from $P_{\infty,i}$ and with pole order at most $\lfloor N_i/5\rfloor$ there.

We regard arrays as functions $x:E_i^3\to k_0$. A coordinate line in direction $j\in\{1,2,3\}$ is obtained by fixing the two coordinates other than the $j$th. A \emph{coded line} is an array that vanishes outside one coordinate line and whose restriction to that line is a nonzero codeword of $C_i$. The \emph{three-direction sum code} $S_3(C_i)$ consists of all sums of coded lines in the three directions. For $x\in S_3(C_i)$, let $L(x)$ be the smallest number of coded lines whose sum is $x$, with $L(0)=0$, and let
 \[
\wt(x):=\bigl|\{(P,Q,R)\in E_i^3:x(P,Q,R)\ne0\}\bigr|
\] be its Hamming weight.
Thus $(C_i,C_i,C_i)$ is $\rho$-product-expanding precisely when every $x\in S_3(C_i)$ admits a representation using at most $\wt(x)/(\rho N_i)$ coded lines.

\begin{mainthm}[uniform $D=3$ certificate] \label{thm:A}
There is a constant $\rho>0$, independent of $i$, such that for every floor $i\ge1$ and every array $x\in S_3(C_i)$,
\[
N_i\,L(x)\le\rho^{-1}\wt(x).
\]
That is, the collections $(C_i,C_i,C_i)$ are $\rho$-product-expanding uniformly in $i$.
\end{mainthm}

\begin{maincor}[answer to Question~\ref{q:main} at $q=729$] \label{cor:B}
The family $C_n:=C_i^\perp\subseteq\F_{729}^{\,n}$, $n=N_i$, satisfies \textup{(i)}--\textup{(iii)} of Question~\ref{q:main}: on every floor its relative distance exceeds $\tfrac{8}{65}$ and its rate lies in $[\tfrac45-\tfrac1n,\tfrac{109}{130})$, its rate tends to $\tfrac{109}{130}$, its dual $C_n^\perp=C_i$ has relative distance at least $\tfrac45$, and $(C_n^\perp,C_n^\perp,C_n^\perp)$ is $\rho$-product-expanding. Moreover $C_n^{*3}=\F_{729}^{\,n}$ on every floor.
\end{maincor}

The proof of Theorem~\ref{thm:A} goes through a quantitatively sharper statement, formulated in terms of the \emph{pair filling mass} $P_i(x)$, the least total weight of a decomposition $x=y_{12}+y_{13}+y_{23}$ in which every slice of $y_{jk}$ perpendicular to the third direction lies in the two-direction sum code (Definition~\ref{def:pair_mass}). One always has $\wt(x)\le P_i(x)$, and the two-direction input gives $P_i(x)\ge\eta N_iL(x)$ for a constant $\eta>0$. The following theorem says that light arrays can be filled, two directions at a time, at essentially no cost beyond their own weight.

\begin{mainthm}[asymptotically sharp filling] \label{thm:B}
For every $\mu>0$ there are $\lambda_0>0$ and $i_0$ such that for every $i\ge i_0$ and every nonzero $x\in S_3(C_i)$ with $\wt(x)\le\lambda_0N_i^3$,
\[
\wt(x)\;\le\;P_i(x)\;\le\;(1+\mu)\wt(x).
\]
The constants $\lambda_0$ and $i_0$ are explicit in $\mu$ and in the two-direction expansion constant $\eta$ \textup{(}Theorem~\textup{\ref{thm:sharp_filling}}\textup{)}.
\end{mainthm}

Theorem~\ref{thm:A} follows from Theorem~\ref{thm:B}. For words of weight above $\lambda_0N_i^3$ the trivial bound $L(x)\le3N_i^2$ suffices, and on the finitely many floors below $i_0$ the ratio $N_iL(x)/\wt(x)$ is bounded because $S_3(C_i)$ is finite (Remark~\ref{rem:light_words}). With $\mu=1$ one obtains an explicit finite initial floor and an explicit certificate; the sufficient condition for $i_0=4$ is given in Corollary~\ref{cor:explicit_certificate}.
\paragraph{Scope.}
Theorems~\ref{thm:A} and~\ref{thm:B} are statements  about classical codes. 
To our knowledge, they provide the first explicit family over a fixed field satisfying the conjunction of Question~\ref{q:main}, which shows that growing-length algebraic codes over a fixed field can have three-fold product expansion. This is evidence that the regime required by growing-component constructions such as~\cite{GG24b,GG25b} can be met by structured codes.

Two elementary observations delimit what Theorem~\ref{thm:A} can say (Section~\ref{subsec:consequences}). If the Schur cube $C^{*3}$ of a code is a proper subspace of $\F_q^n$, then the dual collection $(C^\perp,C^\perp,C^\perp)$ has product expansion at most $1/n$: a vector orthogonal to $C^{*3}$, placed along the main diagonal of the cube, is a light array meeting every line at most once. Consequently, simultaneous three-fold product expansion of a code and of its dual, with constants independent of $n$, requires both Schur cubes to be full. For the low-pole-order AG family considered here the Schur cube of the evaluation code is proper, so the expansion is a \emph{one-sided} property: it holds for the evaluation codes and forces Schur fullness of their duals. Uniform expansion also forces every nonzero array in the corresponding sum code to meet some coordinate line in $\Omega(n)$ cells (Lemma~\ref{lem:line_weight}). This is a statement about sum-code arrays; a logical-gate interpretation requires separate stabilizer and logical multiplication conditions.

We stress the limits of the result. It does not by itself construct a quantum code. The symmetric triple $(C_i,C_i,C_i)$ is not the mixed collection of codes and duals that Theorem~\ref{thm:subsystem_product} below requires,  and the codes $C_i$ are not LDPC: every nonzero parity check is a nonzero word of $C_i^\perp$ and therefore has weight at least $d(C_i^\perp)$, which grows linearly with $N_i$. Whether the mixed collections required by~\cite{GG25b}, or the two-way collections required by~\cite{LLL26}, can be realised over a fixed field by algebraic-geometry codes is left open, as is the extension to four or more directions (Remark~\ref{rem:four_directions}). 
In particular, the present theorem does not verify the hypotheses of the quantum locally testable codes of Dinur, Lin and Vidick~\cite{DLV24}, which involve four codes and two-way robustness. Moreover, the theorem concerns exactly the codes $C_i$ in~\eqref{eq:intro_family}. Product expansion does not automatically transfer to another algebraic-geometry code, even one contained in a related evaluation space, or one that contains $C_i$.

The family is, however, not special. Section~\ref{sec:general} states the sufficient criterion that the proof establishes for arbitrary triples of evaluation codes on possibly different curves over a finite field (Theorem~\ref{thm:general}). 
In its symmetric form, the criterion reads $2\beta<\theta<\min\{(1-8\beta)/3,\,\tfrac12-7\beta\}$, where $\theta$ is the degree ratio and $\beta$ is the genus ratio. It is satisfied by an explicit tower over every non-prime finite field of order at least $512$ (Corollary~\ref{cor:field_coverage}).
Question~\ref{q:main} therefore has a positive answer over all these fields.
\subsection{Technical overview} \label{subsec:intro_method}

Product expansion is a statement of linear algebra over $\F_{729}$. The proof is geometric. It reads an array of the sum code as an object on the threefold $Y=X_i\times X_i\times X_i$ and on its three pair surfaces, and it turns the combinatorial claim, that a light array is a sum of few coded lines, into statements about divisors on $Y$ and about sections of line bundles on curves inside $Y$. This subsection describes the four ideas on which the proof rests and the tools built for them. We believe that three of these tools are of use beyond this paper, and we say why.

\emph{Parity checks are sections.} Codewords of $C_i$ are functions in a Riemann--Roch space on $X_i$. By residue duality, parity checks of $C_i$ are differentials on $X_i$ with prescribed poles. Taking box products, the parity checks of the three-direction sum code are the global sections of a line bundle $\LL$ on $Y$, evaluated by residues at the grid $E_i^3$ of rational points. An array lies in the sum code exactly when it is orthogonal to every global section of $\LL$.\par
The decisive consequence is this. If $R\subset Y$ is a union of coordinate lines, an array supported on the grid points of $R$ is a sum of coded lines on $R$ precisely when it is orthogonal to the sections of $\LL|_R$. So the gap between ``in the sum code'' and ``a sum of coded lines on $R$'' is the cokernel of the restriction map $H^0(Y,\LL)\to H^0(R,\LL|_R)$. Decomposing an array becomes extending sections, which is a cohomological problem.
\\\

\emph{Carriers.} The restriction map is not surjective for an arbitrary union of lines, and the lines carrying a light array must be chosen with care. A \emph{carrier} is a divisor $J=\sum_{j<k}p_{jk}^*D_{jk}$ on $Y$ made of three cylinders over effective divisors $D_{jk}$ on the pair surfaces $X_i\times X_i$. The $D_{jk}$ are chosen to pass through a prescribed sparse set of grid points of the surface, to have no components along the coordinate axes, to have small degree on each coordinate, and to meet every row and column of the grid in few points.\par 
Their existence is a dimension count in the sections of a box product of two line bundles on $X_i$ of degree about $g+s-1$, where $s^2$ exceeds the number of points to be enclosed. Bundles with the required properties do not exist over $\F_{729}$ once $s$ is large, for the same reason that long MDS codes do not exist over a small field. The construction therefore takes place over the function field of a product of Picard varieties of $X_i$, and a scalar-extension lemma brings the conclusion back to $\F_{729}$ at the end.\par
The coordinate lines contained in $J$ are its \emph{fibres}, and the grid points of $J$ are its \emph{footprint}. A carrier is the three-dimensional, positive-genus counterpart of the auxiliary bivariate polynomial in the bivariate testing theorem of Polishchuk and Spielman: a divisor takes the place of the polynomial, and the fibres of the cylinders take the place of its lines.
\\\

\emph{Exact generation on a carrier.} The central geometric theorem of the paper says that for a carrier $J$ whose factor degrees are small compared to the code length, the restriction map $H^0(Y,\LL)\to H^0(R,\LL|_R)$ is surjective, $R$ being the union of the fibres of $J$. Equivalently, an array supported on the footprint of $J$ lies in the sum code if and only if it is a sum of codewords placed on the fibres of $J$. Nothing is assumed about the reducedness of the $D_{jk}$ or the transversality of their intersections.\par
The proof extends sections of $\LL$ in two stages, from $R$ to $J$ and from $J$ to $Y$. The second stage is a K\"unneth computation. The first is the heart of the matter, because the three cylinders cross. Their pairwise intersections, with a canonical scheme structure, form a Cohen--Macaulay curve $K\subset Y$, and the extension is routed through $K$: a section on $R$ is first extended from $R\cap K$ to $K$, then glued and extended to $J$. The sections of $\LL$ on $K$ vanishing on $R\cap K$ are filtered in three layers, each a line bundle on one of the intersection curves twisted by a finite subscheme, and the vanishing of $H^1$ is proved layer by layer.\par
The vanishing on these possibly non-reduced curves rests on an interpolation criterion that we prove for projective Gorenstein curves: if a line bundle $\cF$ on such a curve $Z$ satisfies $\deg(\cF|_V)>2p_a(V)-2+\length(T\cap V)$ for every Cohen--Macaulay subcurve $V\subseteq Z$, then sections of $\cF$ take arbitrary prescribed values on the finite subscheme $T$. The arithmetic genera $p_a(V)$ are bounded by adjunction and the Hodge index theorem on the pair surfaces and by a Koszul-type genus bound for Cohen--Macaulay subcurves of the threefold. What results is a numerical criterion, in the degree of the codes, the genus of the curve and the factor degrees of the carrier, under which every array on a carrier is generated by its fibres.

We regard this theorem, together with the interpolation criterion, as the main technical contribution. In the bivariate setting of Polishchuk and Spielman, and in its algebraic-geometry form due to Garg and Sengupta, the corresponding statement is a divisibility lemma, proved by counting zeros with B\'ezout's theorem on a surface. On a threefold there is nothing to divide by, and a carrier is a reducible, in general non-reduced, surface. The extension through $K$ and the Gorenstein interpolation criterion replace the zero count by cohomology.\par
The criterion involves only arithmetic genera, which adjunction bounds from above; components of negative self-intersection, which abound on products of curves of positive genus, only lower them. This is what makes the genus harmless at this stage, and it is the reason we expect the theorem to apply to other products of curves and other evaluation codes.
\\\

\emph{Support localisation.} Exact generation applies to arrays supported on the footprint of a carrier with small factor degrees, and a carrier absorbs only few grid points per line. A light array need not be of this form: some of its lines may carry many nonzero cells. The localisation argument shows that, after the deletion of a vanishing fraction of the coordinates, every light array is of this form.\par
There are two steps. Heavy lines, those carrying a constant fraction of their cells in the support, are few by counting, and a first carrier $J_1$ absorbs them. Multiplying the array by the equation of $J_1$ produces an array of a shifted sum code, built from codes of slightly larger pole order, all of whose lines are light.\par
The support $T$ of the new array is a tiny fraction of the grid, so the sections of an auxiliary box-product line bundle on $Y$ that vanish on $T$ form a linear system of large dimension, whose fixed divisor $\Delta$ has small degree on every coordinate axis. An interpolation argument on a complete-intersection curve cut out by that system shows that $T$ lies on $\Delta$. The support of a light array with light lines therefore lies on a divisor of small axis degrees.\par
Finally, the grid points of such a divisor are covered by few coordinate lines. This is where the curve enters in an essential way. For the divisors $\Delta_{P,Q}$ cut by $\Delta$ on the lines $\{P\}\times\{Q\}\times X_i$, the theorem of the cube gives the linear equivalence $\Delta_{P,Q}+\Delta_{P_0,Q_0}\sim\Delta_{P,Q_0}+\Delta_{P_0,Q}$ between effective divisors on $X_i$ of degree twice the axis degree of $\Delta$. When that degree is below the gonality of $X_i$, a linear equivalence of effective divisors is an equality, so the fibre divisors are determined by those over one row and one column of the pair surface, and the grid points of $\Delta$ lie on few lines. The gonality bound is a point-counting statement over $\F_{729}$, which survives the passage to the working field by a specialisation argument. A second carrier $J_2$ absorbs the covering lines, and exact generation on $J_1+J_2$ applies. \par
The gonality of the projective line is one, so this step has no counterpart for Reed--Solomon codes, and the gap is real rather than an artefact of the method. We show this by an \emph{additive witness} (Proposition~\ref{prop:rs_witness}): the three-fold product expansion of the Reed--Solomon code on $\F_q$ or $\F_q^*$ collapses to $O(1/n)$ at every rate above $\tfrac14$ in characteristic $2$ and above $\tfrac13$ in characteristic $3$, extending Kalachev's separation to a range of rates, while these same families satisfy every two-direction hypothesis used in this paper (Corollary~\ref{cor:H_insufficient}). Algebraic-geometry codes are therefore not a substitute for Reed--Solomon codes here but the reason the method works: the curve supplies the gonality that the projective line lacks.
\\\

\emph{Two directions, transfer and filling.} The remaining ingredients are quantitative. The theorem of Garg and Sengupta gives two-direction expansion, with a single constant $\eta$, for every pair of codes obtained from $C_i$ by deleting at most a quarter of the coordinates. A transfer inequality bounds the pair filling mass of an array by that of its restriction to a smaller box plus the deleted weight, up to a factor close to one when few coordinates are deleted; this is what permits the deletions made by the two carriers.\par
A weighted filling lemma converts a decomposition into coded lines on fibres that meet one another in few points into a filling of cost $(1+o(1))$ times the weight. Together these give Theorem~\ref{thm:B}, and Theorem~\ref{thm:A} follows from Theorem~\ref{thm:B} and the two-direction bound $P_i(x)\ge\eta N_iL(x)$.

Every quantity in the argument enters through a ratio: the degree of the code divisor, the genus of the curve and the factor degrees of the carriers, each divided by the code length. The numerical conditions can therefore be stated directionally, for three codes on three possibly different curves of comparable length, and they give a sufficient criterion for uniform three-fold expansion in the form of five explicit inequalities between these ratios. The tower over $\F_{729}$ is one instance; the symmetric form of the criterion is met by explicit towers over every non-prime field of order at least $512$. Sections~\ref{sec:transfer}--\ref{sec:proof} carry out the argument in the order described here, and Section~\ref{sec:general} states the criterion.
\subsection{Status of the argument and external inputs} \label{subsec:intro_status}

Every result in this paper is proved in full from the standard background of Sections~\ref{sec:ag_background} and~\ref{sec:product_geometry} and the following external inputs, each of which is stated where it is first used:
\begin{enumerate}[label=(E\arabic*),leftmargin=2.8em,itemsep=1pt,topsep=2pt]
\item the two-code robustness theorem of Garg and Sengupta~\cite{GargSengupta25} (External Input~\ref{ext:two_code_robustness}); only the existence of a robustness constant $\rhoGS(\epsilon)$ depending on $\epsilon$ alone is used, and every constant below is expressed through $\alpha:=\rhoGS(\tfrac13)$;
\item the specialisation lemma of Lang and Nishimura, in the form of~\cite[Theorem~3.6.11]{Poonen17} (used in Lemma~\ref{lem:gonality}).
\end{enumerate}
Beyond these, the proof uses the K\"unneth formula, Serre duality, Riemann--Roch, Hodge index and adjunction on surfaces, the theorem of the cube, and the Hilbert--Burch theorem.

The field-coverage corollary of Section~\ref{sec:general} rests in addition on the towers of Garcia and Stichtenoth~\cite{GS95} and of Bassa, Beelen, Garcia and Stichtenoth~\cite{BBGS15}. The directional bookkeeping of that section was first drafted with the help of an AI system and then checked by hand, and its numerical examples were verified with exact rational arithmetic. The constants have been chosen for simplicity of verification rather than optimised. The paper has not been independently refereed.
\paragraph{Conventions.}
Throughout, $k_0=\F_{729}=\F_{3^6}$ and $\ell=27$. Weights $\wt(\cdot)$ and distances are unnormalised Hamming weights and distances; for a matrix or a three-dimensional array, the weight is the number of nonzero entries. From Section~\ref{sec:carriers} on, auxiliary objects are constructed over a working field $K=k_0(V)$, where $V$ is a finite product of Picard varieties of $X$ and affine spaces over $k_0$. We write $\Omega$ for an algebraic closure of the current $K$. Geometric cohomology arguments may be checked over $\Omega$, whereas the gonality bound and the grid-cover argument are used over $K$; Sections~\ref{subsec:carrier_conventions} and~\ref{subsec:fields} give the precise conventions.

Two unrelated objects are traditionally called $L$: $L(G)$, with a divisor as argument, is a Riemann--Roch space, whereas $L(x)$, with an array as argument, is a line count. Likewise $P$ with a point as subscript or argument is a point, while $P(x)$ is the pair filling mass. The letter $D$ with a pair of indices, $D_{jk}$, always denotes a divisor on a pair surface, and $D$ without indices an effective divisor on $Y$. The number of directions, three, is never denoted by a letter in the body of the paper. The remaining notation is collected in Table~\ref{tab:notation}.

\begin{table}[ht]
\centering
\small
\renewcommand{\arraystretch}{1.15}
\begin{tabular}{@{}lp{0.56\textwidth}l@{}}
\toprule
Symbol & Meaning & Fixed in \\
\midrule
$i$; $j,k,l$ & floor of the tower; directions of an array & \S\ref{sec:family}, \S\ref{sec:sum_codes}\\
$N_i$, $N$ & number of evaluation points on floor $i$; $N=N_i$ once a floor is fixed & Prop.~\ref{prop:tower_properties}\\
$n_j$, $n_{\min}$ & side lengths of a box; their minimum & \S\ref{sec:sum_codes}\\
$g$, $M$ & genus of the curve; pole order, $M=\lfloor N/5\rfloor$ on the tower & \S\ref{sec:family}\\
$\theta$, $\beta$ & pole fraction $\tfrac15$; limiting genus ratio $\tfrac1{26}$ & \S\ref{sec:family}\\
$\alpha$, $\eta$ & $\rhoGS(\tfrac13)$; $22\alpha/(59(1+\alpha))$, expansion constant for safe punctured pairs & \eqref{eq:eta}\\
$L(x)$, $P(x)$, $\lm(x)$ & line count; pair filling mass; largest number of nonzero entries on a line & Def.~\ref{def:lines_and_mass}, \ref{def:pair_mass}, Lemma~\ref{lem:line_weight}\\
$\varrho(x)$ & ratio $N_iL(x)/\wt(x)$ & Rem.~\ref{rem:light_words}\\
$t_j$, $t$, $\Gamma$ & indices deleted from axis $j$; $t=t_1+t_2+t_3$; transfer factor & Lemma~\ref{lem:transfer}\\
$Y$, $p_j$, $p_{jk}$ & the threefold $X^{\times3}$ and its projections & \S\ref{sec:product_geometry}\\
$\LME$, $\LL$ & line bundle on $Y$ whose sections are the parity checks of the sum code & \eqref{eq:LM}\\
$\Sigma$ & an integral curve in $Y$ (the letter $C$ is reserved for codes) & \S\ref{subsec:positivity}\\
$d_j(V)$ & projection degree of a subcurve $V$, with multiplicities & \S\ref{subsec:positivity}\\
$D_{jk}$, $\Pi_i$, $J$ & pair divisors; the cylinders $p_{jk}^*D_{jk}$; the carrier $J=\Pi_1+\Pi_2+\Pi_3$ & Prop.~\ref{prop:carrier}, \S\ref{sec:generation}\\
$s$, $a$, $m$ & carrier parameter with $s^2>m$; auxiliary degree $a=g+s-1$; number of prescribed lines & Prop.~\ref{prop:carrier}\\
$\nu$ & crossing bound $2(s-1)$ for the fibres of a carrier footprint & Prop.~\ref{prop:carrier}\\
$R$, $K$, $\Phi$ & reduced union of fibres in $J$; curve of pairwise intersections; footprint & \S\ref{sec:generation}\\
$\Delta$, $\delta$ & fixed divisor of the auxiliary system; its largest axis degree & \S\ref{subsec:fixed_divisor}\\
$\gamma_K(X)$ & gonality of $X$ over the field $K$ & Def.~\ref{def:gonality}\\
\bottomrule
\end{tabular}
\caption{Notation used across sections.}\label{tab:notation}
\end{table}

\paragraph{Organisation.}
Section~\ref{sec:ag_background} recalls evaluation codes and their duals. Section~\ref{sec:family} constructs the tower and the code family and records its parameters on every floor. Section~\ref{sec:sum_codes} defines sum codes, the line count and the pair filling mass, states the two-direction input, records the consequences of a certificate (Schur fullness, one-sidedness, dense gate tensors), states the targets, and shows by a Reed--Solomon witness that the two-direction input alone cannot reach them. Section~\ref{sec:transfer} proves the one-constant expansion for punctured pairs and the transfer inequality. Section~\ref{sec:product_geometry} translates the picture to $Y$ and proves the vanishing lemmas. Section~\ref{sec:carriers} constructs carriers. Section~\ref{sec:generation} proves exact generation on a carrier. Section~\ref{sec:localization} proves support localisation. Section~\ref{sec:proof} proves Theorems~\ref{thm:A} and~\ref{thm:B}. Section~\ref{sec:general} separates the numerical inputs of the proof by direction and states the resulting general criterion, with the symmetric region, the field-coverage corollary, and examples. Appendix~\ref{sec:appendix_ag_parameters} contains the proofs of the standard facts about evaluation codes, Appendix~\ref{sec:appendix_curves} the proof of the interpolation criterion on Gorenstein curves, and Appendix~\ref{sec:appendix_general} the proof of the general criterion.

\section{Curves, divisors and evaluation codes} \label{sec:ag_background}

This section fixes the language in which AG codes are described. Everything in it is standard. Our main reference is Stichtenoth's book~\cite[Chapters~1, 2 and~4]{Stichtenoth09}, where the proofs that are not given here can be found. AG codes go back to Goppa~\cite{Goppa81}. Other accounts are~\cite{TV91,HvLP98,TVN07,CR21}, and~\cite[Chapter~IV]{Hartshorne77} treats curves in the geometric language.

\subsection{Curves, places and divisors} \label{subsec:app_divisors}

Let $X$ be a smooth, projective, geometrically irreducible algebraic curve of genus $g$ over a finite field $\F_q$ (in the applications, $\F_q=k_0$). We denote the function field of a variety $V$ by $K(V)$, and we write $F:=K(X)$ for the curve. We use the geometric language of curves and the algebraic language of function fields interchangeably; the two are equivalent~\cite[Appendix~B]{Stichtenoth09}, \cite[Section~I.6]{Hartshorne77}. A closed point of $X$ is the same thing as a place $P$ of $F$. To each place belong a discrete valuation $v_P:F^\times\to\mathbb Z$, which records the order of vanishing of a function at $P$ (negative values are pole orders), and a residue field $k(P):=\OO_{X,P}/\mathfrak m_P$, where $\OO_{X,P}=\{f\in F^\times: v_P(f)\ge 0\}\cup\{0\}$ is the local ring of functions without pole at $P$ and $\mathfrak m_P$ its maximal ideal. The residue field is a finite extension of $\F_q$, and its degree is the \emph{degree} of the place, $\deg(P):=[k(P):\F_q]$. A function $f\in\OO_{X,P}$ has a value $f(P)\in k(P)$, namely its class modulo $\mathfrak m_P$. The places of degree one are the \emph{rational points}, and the set of rational points is denoted $X(\F_q)$. Every point at which we evaluate a function or take a residue in this paper is rational.

Geometric irreducibility is needed for the dimension bounds below: if $X$ split into $e>1$ components over $\overline{\F}_q$, the field of constants would be $\F_{q^e}$ and $\dim_{\F_q}L(0)=e$, so that the bound $\dim L(G)\ge\deg G+1-g$ would fail.

A \emph{divisor} on $X$ is a finite formal sum $D=\sum_P n_P\,P$ of places with integer coefficients. Its \emph{degree} is $\deg(D)=\sum_P n_P\deg(P)$, its \emph{support} $\Supp(D)$ is the set of places with $n_P\neq 0$, and $D$ is \emph{effective}, written $D\ge 0$, if every $n_P\ge 0$. A nonzero function $f\in F^\times$ has a \emph{principal divisor} $(f):=\sum_{P} v_P(f)\,P$, which lists its zeros and poles with multiplicity and has degree zero. Two divisors are \emph{linearly equivalent}, written $D_1\sim D_2$, if $D_1-D_2=(f)$ for some $f\in F^\times$.

\subsection{Riemann--Roch spaces and line bundles} \label{subsec:riemann_roch}

\begin{definition}[Riemann--Roch space] \label{def:app_riemann_roch_space}
For a divisor $D$ on $X$, the \emph{Riemann--Roch space} of $D$ is the $\F_q$-vector space
\begin{equation}
L(D) := \bigl\{ f \in F^\times : (f) + D \ge 0 \bigr\} \cup \{0\}.
\end{equation}
\end{definition}

If $D=\sum_P n_P P$, a function in $L(D)$ may have a pole of order at most $n_P$ at a place with $n_P>0$, must vanish to order at least $-n_P$ at a place with $n_P<0$, and has no other poles. The case that matters most below is $D=M P_\infty$ for a rational point $P_\infty$: then $L(MP_\infty)$ consists of the functions that are regular away from $P_\infty$ and have a pole of order at most $M$ there. On the projective line with $P_\infty$ the point at infinity these are exactly the polynomials of degree at most $M$, so $L(MP_\infty)$ is the analogue, on a general curve, of the space of message polynomials of a Reed--Solomon code. If $D_1\sim D_2$ with $D_1-D_2=(h)$, then multiplication by $h$ is an isomorphism $L(D_1)\to L(D_2)$.

\begin{theorem}[Riemann--Roch {\cite[Theorem~1.5.15]{Stichtenoth09}}] \label{thm:app_riemann_roch}
Let $K_X$ be a canonical divisor of $X$, that is, the divisor of a nonzero differential \textup{(}Section~\textup{\ref{subsec:app_weil_differentials})}. It has degree $2g-2$. For every divisor $D$ on $X$,
\begin{equation}
\dim L(D) - \dim L(K_X - D) = \deg(D) + 1 - g .
\end{equation}
\end{theorem}

Two consequences are used constantly~\cite[Corollary~1.4.12 and Theorem~1.5.17]{Stichtenoth09}. A nonzero function has as many zeros as poles, so $L(D)=0$ whenever $\deg(D)<0$. Applying this to $K_X-D$ shows that
\begin{equation} \label{eq:rr_exact}
\dim L(D) = \deg(D)+1-g \qquad\text{whenever } \deg(D) > 2g-2 .
\end{equation}

\begin{remark}[Line bundle notation] \label{rem:line_bundle_notation}
Geometers attach to a divisor $D$ a \emph{line bundle} (invertible sheaf) $\OO_X(D)$, whose sections over an open set are the functions satisfying the condition $(f)+D\ge 0$ on that open set. Its space of global sections is $H^0(X,\OO_X(D))=L(D)$. A reader who prefers function fields may read every $H^0(X,\OO_X(D))$ below as $L(D)$. The line bundle notation is used because it extends without change to the product varieties of Section~\ref{sec:product_geometry}. Three rules suffice for working with it on a curve. Linearly equivalent divisors give isomorphic line bundles, and every line bundle on $X$ is of the form $\OO_X(D)$. The \emph{degree} of the line bundle is $\deg(D)$. Tensor product of line bundles corresponds to addition of divisors, and for a line bundle $\cF$ one writes $\cF(D):=\cF\otimes\OO_X(D)$. Finally, the regular differentials on $X$ form a line bundle $\omega_X$, the \emph{canonical bundle}: if $\omega_0$ is a nonzero differential with divisor $K_X=(\omega_0)$, then $f\mapsto f\omega_0$ is an isomorphism $\OO_X(K_X)\xrightarrow{\ \sim\ }\omega_X$, because $(f\omega_0)=(f)+K_X$. The dictionary between divisors and line bundles is developed in~\cite[Sections~II.6 and~IV.1]{Hartshorne77}. A line bundle $\cF$ of degree $d\ge g-1$ is called \emph{nonspecial} if $H^1(X,\cF)=0$; by Riemann--Roch, $\cF$ is special exactly when $\omega_X\otimes\cF^{-1}$ has a nonzero section, that is, when $\cF\cong\omega_X(-F)$ for an effective divisor $F$ of degree $2g-2-d$. The special bundles of degree $d\ge g-1$ therefore lie in the image of the $(2g-2-d)$-fold symmetric power of $X$ in $\Pic^d(X)$, a closed subset of dimension at most $2g-2-d\le g-1<g=\dim\Pic^d(X)$; a general line bundle of degree $d\ge g-1$ is nonspecial.
\end{remark}

\subsection{Evaluation codes} \label{subsec:evaluation_codes}

\begin{definition}[Evaluation code {\cite{Goppa81}, \cite[Definition~2.2.1]{Stichtenoth09}}] \label{def:app_line_bundle_ev}
Let $P_1,\dots,P_n$ be distinct rational points of $X$, let $E=P_1+\dots+P_n$, and let $G$ be a divisor whose support is disjoint from $E$. The \emph{evaluation map} is the $\F_q$-linear map
\begin{equation}
\ev_E : L(G) \longrightarrow \F_q^n, \qquad f \longmapsto \bigl( f(P_1), \dots, f(P_n) \bigr),
\end{equation}
and the \emph{evaluation code} \textup{(}functional AG code\textup{)} is its image $C_L(E,G):=\ev_E\bigl(L(G)\bigr)\subseteq \F_q^n$. We use the same letter $E$ for the divisor and for its set of points.
\end{definition}

\begin{proposition}[Parameters of evaluation codes {\cite[Theorem~2.2.2 and Corollary~2.2.3]{Stichtenoth09}}] \label{prop:ag_parameters}
The kernel of $\ev_E$ is $L(G-E)$. Consequently, if $\deg(G)<n$ then $\ev_E$ is injective, and $C_L(E,G)$ has length $n$, dimension $\dim L(G)\ge \deg(G)+1-g$, and minimum distance at least $n-\deg(G)$. If moreover $\deg(G)>2g-2$, the dimension is exactly $\deg(G)+1-g$. Whether or not $\deg(G)<n$, a nonzero function in $L(G)$ vanishes at no more than $\deg(G)$ of the points $P_m$.
\end{proposition}

The proof is standard; a version which makes explicit where each hypothesis of Definition~\ref{def:app_line_bundle_ev} is used is given in Appendix~\ref{subsec:appendix_parameters_proof}.

\begin{remark}[Values of sections of a general line bundle] \label{rem:app_generator_matrices}
For a general line bundle $\cF$ in place of $\OO_X(G)$, the value of a section at a rational point $P$ lies in the one-dimensional fibre $\cF_P\otimes k(P)$, which has no preferred basis, so the evaluation code $\ev_E H^0(X,\cF)$ is defined only up to a rescaling of coordinates by nonzero scalars. Rescaling preserves length, dimension, the Hamming weight of every word, the Schur product structure up to rescaling, and every coding-theoretic quantity of this paper. In the two situations in which codes are defined here the basis is canonical: for $\OO_X(G)$ at a point outside $\Supp(G)$ it is the constant function $1$, and for differentials with at most a simple pole at $P$ the residue identifies the fibre with $k(P)$. These two choices are compatible by~\eqref{eq:residue_product}, which is why the duality of Proposition~\ref{prop:app_cauchy_duality} holds exactly. Appendix~\ref{subsec:appendix_trivialisation} gives the details. In Section~\ref{sec:localization} codes of general line bundles occur; there a trivialisation is fixed once at every grid point and all products are formed in it.
\end{remark}

\subsection{Differentials, residues and the dual code} \label{subsec:app_weil_differentials}

The dual of an evaluation code is described by differentials. A \emph{differential} on $X$ is an expression $\omega=f\,dt$ with $f,t\in F$ and $t$ a separating element. The differentials form a one-dimensional vector space over $F$. At a rational point $P$ with local parameter $t$, a differential has an expansion $\omega=\bigl(\sum_{m\ge m_0} a_m t^m\bigr)\,dt$ with $a_m\in\F_q$. The integer $v_P(\omega):=m_0$ (when $a_{m_0}\ne0$) and the \emph{residue} $\Res_P(\omega):=a_{-1}$ do not depend on the choice of $t$. A nonzero differential has a divisor $(\omega)=\sum_P v_P(\omega)P$, the divisors obtained in this way are the canonical divisors, and they all have degree $2g-2$~\cite[Corollary~1.5.16]{Stichtenoth09}. If $\omega$ has at most a simple pole at $P$ and $f$ is regular at $P$, then
\begin{equation} \label{eq:residue_product}
\Res_P(f\omega) = f(P)\,\Res_P(\omega).
\end{equation}
Stichtenoth's book defines differentials as Weil differentials, linear functionals on the adele ring of $F$; the two definitions agree~\cite[Theorem~4.3.2]{Stichtenoth09}. A treatment of residues on curves over an arbitrary field is given in~\cite{Tate68}.

For a divisor $D$ we write $H^0(X,\omega_X(D))=\{\omega : (\omega)+D\ge 0\}\cup\{0\}$ for the space of differentials with poles bounded by $D$; in the notation of~\cite{Stichtenoth09} this is $\Omega_F(-D)$. Multiplication by a fixed differential $\omega_0$ with divisor $K_X$ identifies $L(K_X+D)$ with $H^0(X,\omega_X(D))$.

\begin{theorem}[Residue theorem {\cite[Corollary~4.3.3]{Stichtenoth09}}] \label{thm:app_global_residue}
For every differential $\omega$ on $X$, the sum of the residues of $\omega$ over all places of $X$ is zero. \textup{(}At a place $P$ of degree greater than one the residue $\Res_P(\omega)\in k(P)$ is first mapped to $\F_q$ by the trace. This case does not occur below.\textup{)}
\end{theorem}

\begin{proposition}[The dual of an evaluation code {\cite[Theorem~2.2.8]{Stichtenoth09}}] \label{prop:app_cauchy_duality}
With $E=P_1+\dots+P_n$ and $G$ as in Definition~\textup{\ref{def:app_line_bundle_ev}}, let
\begin{equation}
\res_E : H^0\bigl(X,\omega_X(E-G)\bigr)\longrightarrow \F_q^n,\quad \omega\longmapsto \bigl(\Res_{P_1}(\omega),\dots,\Res_{P_n}(\omega)\bigr),
\end{equation}
and let $C_\Omega(E,G)$ be its image. Then $C_\Omega(E,G)=C_L(E,G)^{\perp}$. The kernel of $\res_E$ is $H^0(X,\omega_X(-G))$, so $\res_E$ is injective when $\deg(G)>2g-2$. If $2g-2<\deg(G)<n$, then $C_L(E,G)^\perp$ has dimension $n-\deg(G)+g-1$ and minimum distance at least $\deg(G)-(2g-2)$.
\end{proposition}

The proof, which combines the residue theorem with the Riemann--Roch theorem, is given in Appendix~\ref{subsec:appendix_dual_proof}. In words: the parity checks of $C_L(E,G)$ are the residue vectors of the differentials that have at most simple poles on $E$ and vanish along $G$. The zeros along $G$ cancel the poles that a codeword function $f\in L(G)$ is allowed to have, so that $f\omega$ has poles only on $E$, and the residue theorem then yields one linear relation among the values $f(P_m)$.

\subsection{Products of codes and multiplication of sections} \label{subsec:code_products}

For vectors $u,v\in\F_q^n$ let $u*v=(u_1v_1,\dots,u_nv_n)$ be their componentwise (Schur) product, and for linear codes $A,B\subseteq\F_q^n$ let $A*B$ be the linear span of the products $a*b$. We write $C^{*r}$ for the $r$-fold product of $C$ with itself and $C^{\langle r\rangle}$ for the same space when it is viewed as a code in its own right. Since evaluation is a ring homomorphism, $C_L(E,G)*C_L(E,G')\subseteq C_L(E,G+G')$. One fact about these products is used.

\begin{theorem}[Mumford {\cite{Mum69}}] \label{thm:mumford}
Let $X$ be a smooth projective geometrically irreducible curve of genus $g$ over a field, and let $\cF,\cG$ be line bundles on $X$ with $\deg\cF\ge2g+1$ and $\deg\cG\ge2g$. Then the multiplication map $H^0(X,\cF)\otimes H^0(X,\cG)\to H^0(X,\cF\otimes\cG)$ is surjective.
\end{theorem}

Mumford states the theorem over an algebraically closed field; since surjectivity of a linear map is the full rank of a matrix, and the formation of $H^0$ commutes with extension of the base field, the statement holds over $\F_q$. Iterating with $\cF=\cG=\OO_X(G)$ shows that $L(G)^{\otimes r}\to L(rG)$ is surjective for all $r\ge1$ when $\deg G\ge2g+1$.

\begin{proposition}[Schur fullness of evaluation codes] \label{prop:schur_fullness}
Let $C_L(E,G)$ be an evaluation code on $n$ rational points of a curve of genus $g$. If $\deg G\ge2g+1$ and $\deg(rG)-n>2g-2$, then $C_L(E,G)^{*r}=\F_q^n$.
\end{proposition}

\begin{proof}
By Theorem~\ref{thm:mumford}, $C_L(E,G)^{*r}=C_L(E,rG)$, the image of $L(rG)$ under $\ev_E$, whose kernel is $L(rG-E)$ by Proposition~\ref{prop:ag_parameters}. Since $\deg(rG)$ and $\deg(rG-E)=\deg(rG)-n$ both exceed $2g-2$, \eqref{eq:rr_exact} gives $\dim L(rG)-\dim L(rG-E)=n$, so the image is all of $\F_q^n$.
\end{proof}

\section{The code family} \label{sec:family}

\subsection{The Garcia--Stichtenoth tower} \label{subsec:tower_and_codes}

Let $F_1 = k_0(x_1)$ be the rational function field over $k_0=\F_{\ell^2}$, $\ell=27$. Following Garcia and Stichtenoth~\cite{GS95}, consider the sequence of function fields $F_1 \subset F_2 \subset F_3 \subset \cdots$ over $k_0$ in which $F_{i+1} = F_i(z_{i+1})$ is the Artin--Schreier extension defined by
\begin{equation}
z_{i+1}^\ell + z_{i+1} = x_i^{\ell + 1}, \qquad x_{i+1} := \frac{z_{i+1}}{x_i}. \label{eq:artin_schreier_tower}
\end{equation}
Let $X_i$ denote the smooth, projective, geometrically irreducible curve over $k_0$ with function field $F_i$, and let $g_i$ be its genus. Thus $X_1$ is the projective line and $X_2$ is the Hermitian curve over $\F_{729}$~\cite[Section~6.4]{Stichtenoth09}. The index $i$ is called the \emph{floor} of the tower.

\begin{proposition}[Properties of the tower {\cite{GS95}}] \label{prop:tower_properties}
For every floor $i \ge 1$ of the tower~\eqref{eq:artin_schreier_tower} the following hold.
\begin{enumerate}
    \item \emph{Total ramification at infinity.} The pole of $x_1$ in $F_1$ is totally ramified in $F_i/F_1$. Hence there is a unique point $P_{\infty,i}$ of $X_i$ above it, and this point is rational.
    \item \emph{Complete splitting of the evaluation points.} Every place $x_1=\xi$ of $F_1$ with $\xi \in k_0^*$ splits completely in $F_i/F_1$. The points of $X_i$ above these places are therefore rational and distinct from $P_{\infty,i}$. There are $\ell^2-1$ such places, each splits into $[F_i:F_1]=\ell^{\,i-1}$ points, and so the number of these points is
    \begin{equation}
    N_i = (\ell^2-1)\,\ell^{\,i-1} = 728 \cdot 27^{\,i-1} .
    \end{equation}
    We write $E_i$ for the sum of these $N_i$ points, the \emph{evaluation divisor} of floor $i$.
    \item \emph{Genus.} The genus of $X_i$ is
    \begin{equation} \label{eq:tower_genus}
    g_i=\begin{cases}
    \ell^{\,i}+\ell^{\,i-1}-\ell^{(i+1)/2}-2\ell^{(i-1)/2}+1, & i \text{ odd},\\[0.2em]
    \ell^{\,i}+\ell^{\,i-1}-\tfrac12\ell^{\,i/2+1}-\tfrac32\ell^{\,i/2}-\ell^{\,i/2-1}+1, & i \text{ even}.
    \end{cases}
    \end{equation}
    In particular $g_1=0$, $g_2=351$, and on every floor
    \begin{equation} \label{eq:genus_bound}
    g_i<\ell^{\,i-1}(\ell+1)=\frac{N_i}{\ell-1}=\frac{N_i}{26},
    \qquad
    \lim_{i \to \infty} \frac{g_i}{N_i} = \frac{1}{\ell - 1} = \frac{1}{26} =: \beta .
    \end{equation}
\end{enumerate}
\end{proposition}

The bound~\eqref{eq:genus_bound} follows from~\eqref{eq:tower_genus} because for $i\ge2$ the subtracted terms sum to more than the constant term $1$, and $g_1=0$. The limit expresses that the tower attains the Drinfeld--Vl\u{a}du\c{t} bound~\cite{DV83}; that this bound is attained over fields of square order was first shown by means of modular curves~\cite{Iha81,TVZ82}, and the towers of~\cite{GS95,GS96} are given by explicit equations. General references are~\cite[Chapter~7]{Stichtenoth09} and the survey~\cite{Beelen22}.

\subsection{The evaluation codes \texorpdfstring{$C_i$}{Ci}} \label{subsec:codes_Ci}

\begin{definition}[The code family] \label{def:ag_code_family}
For each floor $i \ge 1$ let $P_{\infty,i}$ and $E_i$ be as in Proposition~\textup{\ref{prop:tower_properties}}, and let $M_i := \lfloor N_i / 5 \rfloor$. The code of floor $i$ is the one-point evaluation code
\begin{equation}
C_i := C_L\bigl(E_i,\,M_iP_{\infty,i}\bigr) = \Bigl\{ \bigl( f(P) \bigr)_{P \in E_i} : f \in L(M_i P_{\infty,i}) \Bigr\} \subseteq k_0^{N_i}.
\end{equation}
The number $\theta:=\lim_{i\to\infty}M_i/N_i=1/5$ is the \emph{pole fraction} of the family.
\end{definition}

On the first floor $X_1=\PP^1$ the space $L(M_1P_{\infty,1})$ consists of the polynomials in $x_1$ of degree at most $M_1=145$, so $C_1$ is the Reed--Solomon code of length $728$ and dimension $146$ evaluated at the elements of $k_0^*$~\cite[Chapter~10]{MS77}, whereas for $i\ge 2$ the length $N_i\ge 19656$ exceeds $730$, the largest length of a Reed--Solomon code over $k_0$, so $C_i$ is a genuine AG code.

\paragraph{Computational meaning of explicit.}
Here explicit means that, given the floor $i$, one can construct a generator matrix in time polynomial in $N_i$, over the fixed field $k_0$. Fix a representation and ordering of $k_0$. Starting with $x_1\in k_0^*$, enumerate the completely split places recursively using the tower equations. Each lift requires solving a degree-$27$ equation over this fixed field. Next compute a basis of $L(M_iP_{\infty,i})$ by the function-field Riemann--Roch algorithm of Hess~\cite{Hess02}, with the polynomial-time implementation analysed in~\cite[Appendix~B, Theorem~B.9]{EH13}. The tower has extension degree $27^{i-1}=N_i/728$, genus $O(N_i)$ and divisor degree $M_i=O(N_i)$. Its recursive arithmetic and divisor data have polynomial size in $N_i$. Evaluating the basis at the enumerated places gives a generator matrix, and Gaussian elimination gives one for the dual. Since the constant field is fixed, finite-field factoring subroutines can be made deterministic in polynomial time. This assertion concerns construction of the codes, not discovery of the auxiliary geometric carriers or a polynomial-time filling algorithm.

\begin{proposition}[Parameters of $C_i$] \label{prop:Ci_parameters}
For every floor $i\ge1$ one has $2g_i-2<M_i<N_i$, and therefore $C_i$ has length $N_i$, dimension $\dim C_i = M_i+1-g_i$, minimum distance $d(C_i)\ge N_i-M_i\ge \tfrac45 N_i$, and dual distance $d(C_i^\perp)\ge M_i-2g_i+2$. In particular, on every floor,
\begin{equation}\label{eq:dual_params_every_floor}
\frac{d(C_i^\perp)}{N_i}>\frac{8}{65},\qquad \frac45-\frac1{N_i}\le\frac{\dim C_i^\perp}{N_i}<\frac{109}{130},
\end{equation}
and as $i\to\infty$ the rate of $C_i$ tends to $\theta-\beta=\tfrac{21}{130}$ and the rate of $C_i^\perp$ tends to $1-\theta+\beta=\tfrac{109}{130}$.
\end{proposition}

\begin{proof}
By~\eqref{eq:genus_bound}, $2g_i-2<\tfrac{N_i}{13}-2<\tfrac{N_i}{5}-1<M_i$, and $M_i\le N_i/5<N_i$. The assertions follow from Propositions~\ref{prop:ag_parameters} and~\ref{prop:app_cauchy_duality}.
\end{proof}

The value $\theta=\tfrac15$ lies in the window
\begin{equation} \label{eq:pole_window}
2\beta<\theta<\tfrac12-\beta, \qquad\text{that is,}\qquad \tfrac1{13}<\tfrac15<\tfrac{6}{13},
\end{equation}
and each inequality has a distinct purpose. The lower one makes $M_i$ exceed $2g_i-2$, which gives the exact dimension formula and a dual distance linear in $N_i$. The upper one says that $g_i+M_i$ stays below $N_i/2$ by a constant fraction of $N_i$, which is the hypothesis under which the two-code robustness theorem of Section~\ref{subsec:two_directions} applies to the pair $(C_i,C_i)$. The window is nonempty precisely when $\beta<\tfrac16$, that is, for towers over $\F_{\ell^2}$ with $\ell\ge8$. The window is what the two-direction input requires for the pair $(C_i,C_i)$ itself. The punctured pairs of Section~\ref{sec:transfer} and the support localisation of Section~\ref{sec:localization} are treated with explicit constants that impose narrower constraints on $\theta$ and $\beta$, and the choice $\ell=27$ is dictated by those constraints (Remark~\ref{rem:admissible_theta}).

\begin{remark}[Admissible pole fractions and the choice of the tower] \label{rem:admissible_theta}
Two numerical conditions of the proof constrain the pole fraction $\theta$ and the genus ratio $\beta$ beyond the window~\eqref{eq:pole_window}, and both come from constants that were fixed for convenience of verification rather than optimised. First, Proposition~\ref{prop:pair}, with the retained fraction $\tfrac34$ and $\epsilon=\tfrac13$ fixed in Section~\ref{subsec:one_constant}, needs $4+2g+2M<N/2$, that is, $\theta+\beta<\tfrac14-2/N$. Second, the strongest of the interpolation margins of Section~\ref{sec:localization}, $\mu_P>0$ in Lemma~\ref{lem:support_on_Delta}, needs $\theta+7\beta<\tfrac12-\tfrac5{1000}$ with a margin for the term $4u_1<4N/10^5$. At $\beta=\tfrac1{26}$ the two conditions allow $\theta<0.2115$ and $\theta<0.2257$ respectively, so the pole fraction $\tfrac15$ lies close to the edge of what the first of them permits. At $\theta=\tfrac15$ they allow $\beta<\tfrac1{20}$ and $\beta<0.0421$ respectively; the tower over $\F_{625}$, with $\beta=\tfrac1{24}$, satisfies both, but the explicit constants of Sections~\ref{sec:transfer}, \ref{sec:carriers} and~\ref{sec:localization} ($g\le N/25$, $g<N/26$) are stated for $\ell=27$ and would have to be re-verified for $\ell=25$. A sharper bound on the number of conditions imposed by the support of the shifted word is available (Remark~\ref{rem:few_conditions}), but it requires $\theta+20\beta<1-\tfrac{18}{1000}$ and hence $\ell\ge27$; it is not used here, and $\ell=27$ is retained because the constants are verified for it. The general criterion of Section~\ref{sec:general}, which treats the pair margin as a parameter, gives a uniform certificate for the tower codes at every pole fraction $\tfrac1{13}<\theta<\tfrac3{13}$ and for every $\ell\ge23$, with its own constants (Example~\ref{ex:tower_window}). Whether Theorem~\ref{thm:A} holds throughout the window~\eqref{eq:pole_window} is not decided by either argument (Remarks~\ref{rem:constants} and~\ref{rem:general_coefficients}).
\end{remark}

\begin{remark}[Notation of the central question] \label{rem:notation_match}
In Question~\ref{q:main} the ambient code is $C_n$, and the product expansion is required of its dual. On the tower we take $n=N_i$ and $C_n:=C_i^\perp$, so that $C_n^\perp=C_i$ is the evaluation code. All statements in the body of the paper are made for the evaluation codes $C_i$.
\end{remark}

\begin{remark}[Schur fullness and one-sidedness of the family] \label{rem:schur_family}
The ambient evaluation divisor of $C_i^\perp=C_L(E_i,E_i-M_iP_{\infty,i}+(\eta_i))$ (Proposition~\ref{prop:app_cauchy_duality} and Appendix~\ref{subsec:appendix_dual_proof}) has degree $N_i-M_i+2g_i-2\ge\tfrac45N_i-2\ge2g_i+1$, and three times this degree exceeds $N_i+2g_i-2$, so Proposition~\ref{prop:schur_fullness} gives $(C_i^\perp)^{*3}=k_0^{N_i}$ directly. This confirms, independently of Theorem~\ref{thm:A}, the necessary condition for three-fold expansion of $C_i$ recorded in Theorem~\ref{thm:central-schur-fullness}. In the opposite direction, $C_i^{*3}\subseteq C_L(E_i,3M_iP_{\infty,i})$ has dimension at most $3M_i+1<N_i$, so by Proposition~\ref{prop:diag} three copies of the ambient code $C_i^\perp$ have expansion at most $1/N_i$: the certificate of Theorem~\ref{thm:A} is a bound for three copies of the evaluation code and cannot hold for three copies of its dual. The same obstruction applies to $r$ copies of $C_i^\perp$ for every $r$ with $r\theta<1$, in particular at $r=2$.
\end{remark}

\section{Three-direction sum codes, filling mass and the certificate} \label{sec:sum_codes}

\subsection{The sum code and its line count} \label{subsec:sum_code}

Let $V_j=k_0^{n_j}$ for $j\in\{1,2,3\}$, and let $\Cj{j}\subseteq V_j$ be linear codes. An element of $V_1\otimes V_2\otimes V_3=k_0^{n_1\times n_2\times n_3}$ is a three-dimensional array $x=(x_{abc})$. A \emph{line in direction $1$} is the set of $n_1$ cells obtained by fixing the second and third coordinates and letting the first one vary; the restriction of $x$ to such a line is a vector in $V_1$. Lines in directions $2$ and $3$ are defined in the same way. Let $U_j$ be the space of arrays all of whose lines in direction $j$ are codewords of $\Cj{j}$, so that
\begin{equation}
U_1 = \Cj1 \otimes V_2 \otimes V_3, \qquad U_2 = V_1 \otimes \Cj2 \otimes V_3, \qquad U_3 = V_1 \otimes V_2 \otimes \Cj3 .
\end{equation}
The \emph{three-direction sum code} is
\begin{equation}
S\bigl(\Cj1,\Cj2,\Cj3\bigr) := U_1 + U_2 + U_3 \subseteq k_0^{n_1\times n_2\times n_3},
\end{equation}
and when the three codes are equal to a single code $C$ we write $S_3(C):=S(C,C,C)$. The sum code should not be confused with the tensor code $\Cj1\otimes\Cj2\otimes\Cj3=U_1\cap U_2\cap U_3$, in which every line in every direction is a codeword. The two are related by duality: since $U_1^\perp=(\Cj1)^\perp\otimes V_2\otimes V_3$, and similarly for the other directions,
\begin{equation} \label{eq:sum_code_dual}
S\bigl(\Cj1,\Cj2,\Cj3\bigr)^{\perp} = (\Cj1)^{\perp}\otimes(\Cj2)^{\perp}\otimes(\Cj3)^{\perp}.
\end{equation}
The sum code is therefore the code whose parity checks are the tensor products of parity checks of the three constituent codes. The sum code is monotone in each $\Cj{j}$.

\begin{definition}[Coded lines and line count] \label{def:lines_and_mass}
A \emph{coded line} in direction $j$ is a nonzero element of $U_j$ supported on a single line in direction $j$. Every $u_j\in U_j$ is the sum of its restrictions to the lines in direction $j$, and $\operatorname{lines}_j(u_j)$ denotes the number of these restrictions that are nonzero. A \emph{directional decomposition} of $x\in S(\Cj1,\Cj2,\Cj3)$ is an expression $x=u_1+u_2+u_3$ with $u_j\in U_j$, and the \emph{line count} of $x$ is
\begin{equation}
L(x) := \min\bigl\{ \operatorname{lines}_1(u_1)+\operatorname{lines}_2(u_2)+\operatorname{lines}_3(u_3) \;:\; x=u_1+u_2+u_3,\ u_j\in U_j \bigr\},
\end{equation}
the least number of coded lines whose sum is $x$. Any family of coded lines with sum $x$ can be grouped by direction into a directional decomposition with no more lines, because two coded lines on the same line in the same direction add up to a single coded line or to zero.
\end{definition}

In the notation of~\cite[Section~II]{KP25}, $\operatorname{lines}_j(u_j)=|u_j|_j$, and for codes of a common length $N$ the collection $(\Cj1,\Cj2,\Cj3)$ is $\rho$-product-expanding in the sense of~\cite[Definition~1]{KP25}, \cite[Section~2]{KP22} if and only if $\rho N L(x)\le\wt(x)$ for every $x\in S(\Cj1,\Cj2,\Cj3)$. A coded line in direction $j$ has weight at most $n_j$, and there are $n_1n_2n_3/n_j$ lines in direction $j$. In the cubic case $n_1=n_2=n_3=N$ this gives the trivial bounds
\begin{equation} \label{eq:trivial_bounds}
\wt(x)\le N\,L(x) \qquad\text{and}\qquad L(x)\le 3N^2 \qquad\text{for all } x\in S(\Cj1,\Cj2,\Cj3).
\end{equation}

\begin{definition}[Uniform $D = 3$ certificate] \label{def:uniform_certificate}
The family $\{C_i\}_{i \ge 1}$ has a \emph{uniform $D=3$ certificate} with constant $\rho>0$ if for every floor $i$ and every word $x \in S_3(C_i)$,
\begin{equation}
N_i\, L(x) \le \rho^{-1} \wt(x).
\end{equation}
\end{definition}

A certificate is a bound in the opposite direction to the first inequality in~\eqref{eq:trivial_bounds}: a word of small weight never requires many lines whose contributions cancel. Theorem~\ref{thm:A} asserts that the family of Definition~\ref{def:ag_code_family} has a uniform certificate.

\subsection{The two-direction input} \label{subsec:two_directions}

Let $C\subseteq k_0^n$ and $C'\subseteq k_0^{n'}$ be linear codes. We regard elements of $k_0^n\otimes k_0^{n'}$ as $n\times n'$ matrices, so that columns have length $n$ and rows have length $n'$. A matrix is \emph{column-coded} if all of its columns lie in $C$, and \emph{row-coded} if all of its rows lie in $C'$; thus $C\otimes C'$ is the tensor code and $C\otimes k_0^{n'}+k_0^n\otimes C'$ the two-direction sum code. When $n=n'$, let $\delta(w,w'):=\wt(w-w')/n^2$ be the normalised Hamming distance of two matrices and $\delta(w,T):=\min_{w'\in T}\delta(w,w')$ for a set $T$ of matrices.

\begin{externalinput}[Two-code robustness {\cite[Definition~1.1 and Corollary~6.3]{GargSengupta25}}] \label{ext:two_code_robustness}
For every $\epsilon\in(0,1)$ there is $\rhoGS(\epsilon)>0$ with the following property. Let $C=C_L(E,G)$ and $C'=C_L(\tilde E,\tilde G)$ be evaluation codes of the same length $n$ over a finite field $K$, on curves of genera $g$ and $g'$, with $\deg(G)\ge0$ and $\deg(\tilde G)\ge0$. If
\begin{equation} \label{eq:gs_margin}
4 + g + \deg(G) + g' + \deg(\tilde G) < (1 - \epsilon)\,n ,
\end{equation}
then the pair $(C,C')$ is $\rhoGS(\epsilon)$-robust: for every matrix $w\in K^{n\times n}$,
\begin{equation} \label{eq:robustness}
\rhoGS(\epsilon)\cdot\delta(w, \, C \otimes C') \le \tfrac12\bigl[\,\delta(w, \, C \otimes K^n) + \delta(w, \, K^n \otimes C')\,\bigr].
\end{equation}
\end{externalinput}

Corollary~6.3 of~\cite{GargSengupta25} is a reformulation of their Theorem~6.2, which is stated with a parameter $c_0>1$ in place of $\epsilon$; it gives $\rhoGS(\epsilon)=\epsilon^2/12$, and its only hypotheses are those of their definition of an AG code (distinct rational evaluation points, a rational divisor of non-negative degree with support disjoint from them, on a smooth geometrically irreducible curve) together with the margin~\eqref{eq:gs_margin}. Only the existence of $\rhoGS(\epsilon)$ is used below, and every constant is expressed through $\alpha:=\rhoGS(\tfrac13)$. Their Definition~1.1 places the first code along the rows rather than the columns; since~\eqref{eq:robustness} is symmetric in the two codes, this makes no difference. Robustness says that a matrix whose columns are on average close to $C$ and whose rows are on average close to $C'$ is close to the tensor code; the notion is due to Ben-Sasson and Sudan~\cite{BSS06}, and for Reed--Solomon codes the corresponding statement is the bivariate testing theorem of Polishchuk and Spielman~\cite{PS94}. What is needed below is the following consequence for decompositions of words of the two-direction sum code. The equivalence of robustness and expansion for two codes of linear distance is well known~\cite{KP22,KP25}; the short proof of the direction used here is included to keep track of the constant, and it is formulated for codes of possibly different lengths because pairs of unequal length occur in Section~\ref{sec:transfer}.

\begin{definition}[Two-direction expansion] \label{def:two_direction_expansion}
A pair $(C,C')$ of codes of lengths $n$ and $n'$ has \emph{expansion at least $\eta>0$} if every $y\in C\otimes k_0^{n'}+k_0^n\otimes C'$ can be written as $y=y_{\mathrm c}+y_{\mathrm r}$, with $y_{\mathrm c}$ column-coded with $u$ nonzero columns and $y_{\mathrm r}$ row-coded with $v$ nonzero rows, in such a way that
\begin{equation} \label{eq:two_direction_expansion}
\eta\,(n\,u+n'\,v)\le \wt(y).
\end{equation}
\end{definition}

For $n=n'$ the condition reads $\eta\,n\,(u+v)\le\wt(y)$, which is $\rho$-product-expansion of the pair with $\rho=\eta$ in the sense of~\cite{PK22,KP22}. The definition is symmetric: transposing matrices exchanges the roles of $C$ and $C'$.

\begin{lemma}[Robustness implies expansion] \label{lem:robust_to_expansion}
If $C,C'\subseteq k_0^n$ have the same length, $(C,C')$ is $\alpha$-robust, and both codes have minimum distance at least $\delta_0 n$, then $(C,C')$ has expansion at least $\alpha\delta_0/(1+\alpha)$.
\end{lemma}

\begin{proof}
Let $y=y_{\mathrm c}+y_{\mathrm r}$ with $y_{\mathrm c}$ column-coded and $y_{\mathrm r}$ row-coded. Apply~\eqref{eq:robustness} to $w=y_{\mathrm c}$. The first distance on the right is zero, and the second is at most $\delta(y_{\mathrm c},-y_{\mathrm r})=\wt(y)/n^2$ because $-y_{\mathrm r}$ is row-coded. Hence there is $c\in C\otimes C'$ with $\wt(y_{\mathrm c}-c)\le \wt(y)/(2\alpha)$. The matrix $y_{\mathrm c}-c$ is column-coded, the matrix $y_{\mathrm r}+c$ is row-coded, their sum is $y$, and $\wt(y_{\mathrm r}+c)\le\wt(y)+\wt(y_{\mathrm c}-c)\le\bigl(1+\tfrac1{2\alpha}\bigr)\wt(y)$. Every nonzero column of $y_{\mathrm c}-c$ is a nonzero codeword of $C$ and every nonzero row of $y_{\mathrm r}+c$ is a nonzero codeword of $C'$, so each has weight at least $\delta_0 n$. If the first matrix has $u$ nonzero columns and the second has $v$ nonzero rows, then
$\delta_0\,n\,(u+v)\le \wt(y_{\mathrm c}-c)+\wt(y_{\mathrm r}+c)\le\bigl(1+\tfrac1\alpha\bigr)\wt(y)$.
\end{proof}

\begin{corollary}[Two-direction expansion on the tower] \label{cor:tower_expansion}
On every floor $i\ge1$, the pair $(C_i,C_i)$ is $\alpha$-robust, where $\alpha:=\rhoGS(\tfrac13)$, and has expansion at least $\tfrac45\alpha/(1+\alpha)$. In particular it has expansion at least
\begin{equation} \label{eq:eta}
\eta:=\frac{22\,\alpha}{59\,(1+\alpha)},
\end{equation}
and this is the value of $\eta$ used throughout.
\end{corollary}

\begin{proof}
By~\eqref{eq:genus_bound} and $M_i\le N_i/5$, the left-hand side of~\eqref{eq:gs_margin} for $C=C'=C_i$ is $4+2g_i+2M_i<4+\tfrac{31}{65}N_i$, which is less than $\tfrac23N_i$ because $N_i\ge728>\tfrac{780}{37}$. So~\eqref{eq:gs_margin} holds with $\epsilon=\tfrac13$, and External Input~\ref{ext:two_code_robustness} gives robustness with constant $\alpha$. By Proposition~\ref{prop:Ci_parameters} the relative distance of $C_i$ is at least $\tfrac45$, and Lemma~\ref{lem:robust_to_expansion} gives expansion at least $\tfrac45\alpha/(1+\alpha)$, which exceeds $\eta$ because $\tfrac45>\tfrac{22}{59}$.
\end{proof}

The smaller constant $\eta$ is used because it also serves for the punctured pairs of Section~\ref{sec:transfer} (Proposition~\ref{prop:pair}). For orientation: with $\rhoGS(\epsilon)=\epsilon^2/12$ one has $\alpha=\tfrac1{108}$ and $\eta=\tfrac{22}{6431}>\tfrac1{400}$; no argument below depends on this value.

\subsection{The pair filling mass} \label{subsec:pair_filling}

The line count is a minimum over combinatorial data and is awkward to bound directly. The proof works instead with a quantity that measures cost by Hamming weight and sees two directions at a time. For $j\ne k$ let $U_{jk}:=U_j+U_k$. A \emph{slice} of an array is the two-dimensional array obtained by fixing one of the three coordinates; it is perpendicular to the direction of that coordinate. For $\{j,k,l\}=\{1,2,3\}$ and $h\in[n_l]$, we write $x|_{l=h}\in V_j\otimes V_k$ for the slice of $x$ on which the $l$-th coordinate equals $h$. Then
\begin{equation} \label{eq:Ujk_slices}
U_{jk}=\bigl\{x \;:\; x|_{l=h}\in \Cj{j}\otimes V_k+V_j\otimes\Cj{k}\ \text{ for all } h\in[n_l]\bigr\},
\end{equation}
since $U_{jk}=\bigl(\Cj{j}\otimes V_k+V_j\otimes\Cj{k}\bigr)\otimes V_l$ with the factors written in the order $j,k,l$, and an array lies in $A\otimes V_l$ exactly when all of its slices $x|_{l=h}$ lie in $A$. The third direction plays no role in the membership condition, and this is what allows the two-direction input to be applied slice by slice.

\begin{definition}[Pair filling mass] \label{def:pair_mass}
A \emph{pair filling} of $x\in S(\Cj1,\Cj2,\Cj3)$ is an expression $x=y_{12}+y_{13}+y_{23}$ with $y_{jk}\in U_{jk}$. The \emph{pair filling mass} of $x$ is
\begin{equation}
P(x) := \min\bigl\{ \wt(y_{12}) + \wt(y_{13}) + \wt(y_{23}) \;:\; x = y_{12} + y_{13} + y_{23}, \ y_{jk} \in U_{jk} \bigr\}.
\end{equation}
On floor $i$ of the tower, with $\Cj1=\Cj2=\Cj3=C_i$, we write $P_i(x)$ for this quantity.
\end{definition}

\begin{proposition}[Filling mass versus line count] \label{prop:sandwich}
Let $C\subseteq k_0^N$ be a code such that the pair $(C,C)$ has expansion at least $\eta$. Then every $x\in S_3(C)$ satisfies
\begin{equation} \label{eq:sandwich}
\wt(x)\;\le\;P(x)\;\le\;N\,L(x) \qquad\text{and}\qquad \eta\,N\,L(x)\;\le\;P(x).
\end{equation}
\end{proposition}

\begin{proof}
The inequality $\wt(x)\le P(x)$ is the subadditivity of the Hamming weight. For $P(x)\le N\,L(x)$, take a directional decomposition $x=u_1+u_2+u_3$ with $L(x)$ coded lines and use the pair filling $y_{12}=u_1+u_2$, $y_{13}=u_3$, $y_{23}=0$. For the last inequality, take a pair filling of weight $P(x)$. Each slice of $y_{jk}$ perpendicular to the remaining direction lies in the two-direction sum code of $(C,C)$, so by~\eqref{eq:two_direction_expansion} it is a sum of columns and rows that are codewords of $C$, their number being at most the weight of the slice divided by $\eta N$. Summing over the slices and the three pairs gives a directional decomposition of $x$ with at most $P(x)/(\eta N)$ coded lines.
\end{proof}

By~\eqref{eq:sandwich}, bounding the line count by a multiple of $\wt(x)/N$ is the same as bounding the filling ratio $P(x)/\wt(x)$. This is the reason for working with $P$: it is subadditive, it is measured in the same units as $\wt(x)$, and each of its three terms lives in a space $U_{jk}$ governed by the geometry of a product of only two curves (Section~\ref{subsec:app_product_bundles}).

\subsection{Consequences of a certificate} \label{subsec:consequences}

The following elementary facts explain the shape of Question~\ref{q:main}. They are formulated for general codes over a field $\F_q$ and use only the definition of product expansion. For a nonzero array $x$ let $\lm(x)$ be the largest number of nonzero entries of $x$ on a single line, in any of the three directions.

\begin{lemma}[Light lines force many coded lines] \label{lem:line_weight}
Every nonzero $x\in S(\Cj1,\Cj2,\Cj3)$ satisfies $L(x)\ge\wt(x)/\lm(x)$. If moreover $\Cj1=\Cj2=\Cj3=C\subseteq k_0^N$ and $(C,C)$ has expansion at least $\eta$, then
\begin{equation} \label{eq:line_weight}
\frac{N\,L(x)}{\wt(x)}\ \ge\ \frac{N}{\lm(x)}
\qquad\text{and}\qquad
\frac{P(x)}{\wt(x)}\ \ge\ \frac{\eta N}{\lm(x)} .
\end{equation}
In particular, a uniform certificate with constant $\rho$ forces $\lm(x)\ge\rho N_i$ for every nonzero $x\in S_3(C_i)$.
\end{lemma}

\begin{proof}
If $x$ is a sum of $L$ coded lines, every nonzero entry of $x$ lies on one of these lines, and each line contains at most $\lm(x)$ nonzero entries of $x$. Hence $\wt(x)\le L\,\lm(x)$. The inequalities~\eqref{eq:line_weight} follow from this and Proposition~\ref{prop:sandwich}.
\end{proof}

\begin{proposition}[Diagonal obstruction] \label{prop:diag}
Let $C_1,\dots,C_D\subseteq\F_q^n$ be linear codes with $C_1*\dots*C_D\ne\F_q^n$. Then the collection $(C_1^\perp,\dots,C_D^\perp)$ is not $\rho$-product-expanding for any $\rho>1/n$.
\end{proposition}

\begin{proof}
Take $z\ne0$ orthogonal to $C_1*\dots*C_D$ and let $\operatorname{diag}(z)$ be the array equal to $z_m$ at $(m,\dots,m)$ and zero elsewhere. Then $\langle\operatorname{diag}(z),c_1\otimes\dots\otimes c_D\rangle=\langle z,c_1*\dots*c_D\rangle=0$, so $\operatorname{diag}(z)\in(C_1\otimes\dots\otimes C_D)^\perp$, which is the sum code of the duals. A line contains at most one diagonal cell, so every decomposition into coded lines uses at least $\wt(z)$ lines, and the expansion is at most $\wt(z)/(n\wt(z))=1/n$.
\end{proof}

\begin{theorem}[Three consequences of dual expansion] \label{thm:central-schur-fullness}
Let $C\subsetneq\F_q^n$ be a linear code such that $(C^\perp,C^\perp,C^\perp)$ is $\rho_0$-product-expanding.
\begin{enumerate}[label=\textup{(\alph*)},leftmargin=2.4em]
\item \textup{(}dual distance\textup{)} $d(C^\perp)\ge\rho_0n$.
\item \textup{(}rates\textup{)} If moreover $d(C)\ge\delta n$, then $\rho_0-\tfrac1n\le\dim(C)/n\le1-\delta+\tfrac1n$; for a growing-length family with fixed $\delta,\rho_0>0$, the rates of $C$ and $C^\perp$ are bounded away from $0$ and $1$.
\item \textup{(}Schur fullness\textup{)} If $\rho_0>1/n$, then $C^{*3}=\F_q^n$.
\end{enumerate}
\end{theorem}

\begin{proof}
(a) A minimum-weight nonzero word of $C^\perp$ placed on a single line is a nonzero array of the sum code of the duals with $L=1$, so $\rho_0\le d(C^\perp)/n$. (b) Puncturing a code $A$ at $d(A)-1$ coordinates is injective, which gives the Singleton bound $\dim A\le n-d(A)+1$; apply it to $C$ and to $C^\perp$ and use $\dim C+\dim C^\perp=n$ together with (a). (c) is the contrapositive of Proposition~\ref{prop:diag}.
\end{proof}

\begin{corollary}[One-sidedness] \label{cor:onesided}
For a growing-length family of codes, three-fold product expansion of $(C,C,C)$ and of $(C^\perp,C^\perp,C^\perp)$ cannot both hold with constants independent of $n$ unless $C^{*3}=(C^\perp)^{*3}=\F_q^n$; for one-point AG codes of pole fraction $\theta<\tfrac13$, and in particular for the family of Definition~\textup{\ref{def:ag_code_family}}, three copies of the ambient code $C_i^\perp$ have expansion at most $1/N_i$ \textup{(}Remark~\textup{\ref{rem:schur_family})}.
\end{corollary}

\begin{remark}[Why the expansion is required of the duals; dense gate tensors] \label{rem:why_duals}
On a $D$-fold ambient code with factor codes $C_1,\dots,C_D$, a transversal $C^{D-1}Z$-type gate is carried by a tensor whose simplest instance is a vector $z\perp C_1*\dots*C_D$; for $D=3$ this is a $\mathrm{CCZ}$ gate and the trilinear form $\langle z,a*b*c\rangle$ lives on $[n]^3$~\cite{GG24,GL24,GG25b}. By the argument of Proposition~\ref{prop:diag} and Lemma~\ref{lem:line_weight}, a gate-carrying tensor of line multiplicity $\lm$ in the sum code of the duals forces $\rho(C_1^\perp,\dots,C_D^\perp)\le\lm/n$. On an ambient family satisfying (iii) of Question~\ref{q:main}, every gate-carrying tensor is therefore dense, $\lm\ge\rho_0 n$. The bounded-check codes of the qLDPC constructions~\cite{GL24,LLL26} sit at $\rho\le w/n$, outside this regime, since a parity check of weight $w$ placed on a line is a light word of the sum code of the duals. This is also why the codes of Theorem~\ref{thm:A} are not LDPC: every nonzero parity check of $C_i^\perp$ is a word of $C_i$ and has weight at least $\tfrac45N_i$, and every nonzero parity check of $C_i$ has weight at least $d(C_i^\perp)=\Omega(N_i)$.
\end{remark}

For the scope of the result it is useful to record the form in which a growing-component construction consumes product expansion.

\begin{theorem}[Subsystem product parameters {\cite{GG25b}}] \label{thm:subsystem_product}
Let $Q_1,\dots,Q_t$ be CSS codes $Q_i=(Q_{i,X},Q_{i,Z})$ of length $n$. If for every $i\le t$ the mixed collections
$\bigl(Q_{1,Z}^\perp,\dots,Q_{i-1,Z}^\perp,Q_{i,X}\bigr)$ and $\bigl(Q_{1,X}^\perp,\dots,Q_{i-1,X}^\perp,Q_{i,Z}\bigr)$
are $\rho$-product-expanding, then the subsystem product code has parameters $[[n^t,\ge k^t,\ge(\rho n)^t]]$ with check locality at most $tn$.
\end{theorem}

The collections consumed here are mixed, $t-1$ codes of one type and one larger code of the dual type, rather than the symmetric triple $(C,C,C)$ of Theorem~\ref{thm:A}; the transversal $\mathrm{CCZ}$ gate is supplied separately by Schur fullness $C^{*3}=\F_q^n$, which is why $t=2$ suffices there. Theorem~\ref{thm:A} establishes the symmetric statement for an explicit fixed-field family; it does not establish the mixed one.

\subsection{The targets} \label{subsec:main_theorem}

The two statements proved in Sections~\ref{sec:transfer}--\ref{sec:proof} are the following.

\begin{theorem}[Uniform filling bound] \label{thm:main_uniform_certificate}
There are $c,\lambda_0>0$ and $i_0$ such that $P_i(x)\le c^{-1}\wt(x)$ for all $i\ge i_0$ and all nonzero $x\in S_3(C_i)$ with $\wt(x)\le\lambda_0N_i^3$.
\end{theorem}

\begin{theorem}[Sharp form] \label{thm:main_sharp}
For every $\mu>0$ there are $\lambda_0>0$ and $i_0$ such that $P_i(x)\le(1+\mu)\wt(x)$ for all $i\ge i_0$ and all nonzero $x\in S_3(C_i)$ with $\wt(x)\le\lambda_0N_i^3$.
\end{theorem}

Theorem~\ref{thm:main_sharp} is Theorem~\ref{thm:B}, and it contains Theorem~\ref{thm:main_uniform_certificate} with $c^{-1}=1+\mu$. Both are proved in Section~\ref{sec:proof}, with explicit constants (Theorem~\ref{thm:sharp_filling}).

\begin{remark}[The uniform filling bound gives the certificate] \label{rem:light_words}
Write $\varrho(x):=N_iL(x)/\wt(x)$ for nonzero $x\in S_3(C_i)$, and assume Theorem~\ref{thm:main_uniform_certificate} with constants $c,\lambda_0,i_0$. For $i\ge i_0$ and $\wt(x)\le\lambda_0N_i^3$, Corollary~\ref{cor:tower_expansion} and Proposition~\ref{prop:sandwich} give $\eta N_iL(x)\le P_i(x)\le c^{-1}\wt(x)$, so $\varrho(x)\le(c\eta)^{-1}$. If $\wt(x)>\lambda_0N_i^3$, then $\varrho(x)<3/\lambda_0$ by~\eqref{eq:trivial_bounds}. On each of the finitely many floors $i<i_0$, $\varrho$ is bounded because $S_3(C_i)$ is finite. Hence the family has a uniform certificate, and one may take
\[
\rho^{-1}:=\max\Bigl\{(c\eta)^{-1},\ 3/\lambda_0,\ \max_{i<i_0}\max_{x}\varrho(x)\Bigr\}.
\]
This is the deduction of Theorem~\ref{thm:A} from Theorem~\ref{thm:B}; Corollary~\ref{cor:explicit_certificate} makes the constant explicit.
\end{remark}

Lemma~\ref{lem:line_weight} is the elementary obstruction that every proof of the targets has to respect: a word all of whose lines are light is expensive to fill, however small its weight. Section~\ref{subsec:rs_witness} shows that such words exist for Reed--Solomon codes satisfying all the two-direction hypotheses of this paper, so that the targets cannot be reached from those hypotheses alone; the proof in Sections~\ref{sec:generation}--\ref{sec:localization} excludes them on every floor of the tower, by showing that light words live on sparse carriers; Remark~\ref{rem:second_floor_witness} checks directly that the Reed--Solomon witness and its pullback are absent from the first two floors.

\subsection{Why the two-direction input does not suffice} \label{subsec:rs_witness}

The combinatorial part of the proof uses three hypotheses only: the two-direction input of Section~\ref{subsec:two_directions}, the linear distance of the codes, and the uniform expansion of punctured pairs proved in Section~\ref{sec:transfer}. The following Reed--Solomon example shows that families satisfying these hypotheses uniformly can fail the targets. The geometric part of the proof, Sections~\ref{sec:product_geometry}--\ref{sec:localization}, is therefore not dispensable. The example also identifies the property of the tower curves that the geometric argument exploits: their large gonality over the working field (Lemma~\ref{lem:gonality}).
\paragraph{The additive Reed--Solomon witness.}
Let $q=p^f$, let $T$ be $\F_q$ or $\F_q^*$, and let $RS_M\subseteq\F_q^{T}$ be the Reed--Solomon code of polynomials of degree at most $M$ evaluated on $T$, of length $n=|T|$. Consider
\begin{equation}
x_{\mathrm{add}}(u,v,w):=\begin{cases}1,& u+v+w=0,\\ 0,&\text{otherwise,}\end{cases}\qquad (u,v,w)\in T^3 .
\end{equation}
Every line of $x_{\mathrm{add}}$ carries at most one nonzero entry, and $\wt(x_{\mathrm{add}})$ is $n^2$ for $T=\F_q$ and $n(n-1)$ for $T=\F_q^*$. Let
\[
M^*_q:=\max\Bigl\{\min(e_1,e_2,e_3)\;:\;e_1+e_2+e_3=q-1,\ \tbinom{q-1}{e_1,\,e_2,\,e_3}\not\equiv0\pmod p\Bigr\}.
\]

\begin{proposition}[Additive witness] \label{prop:rs_witness}
\begin{enumerate}[label=(\roman*)]
\item $x_{\mathrm{add}}\in S_3(RS_M)$ if and only if $M\ge M^*_q$, for both choices of $T$.
\item $M^*_q=q/3-1$ if $p=3$, and $M^*_q=q/4-1$ if $p=2$ and $q\ge4$.
\item If $x_{\mathrm{add}}\in S_3(RS_M)$, then $L(x_{\mathrm{add}})\ge\wt(x_{\mathrm{add}})$, and if moreover $(RS_M,RS_M)$ has expansion at least $\eta'$, then $P(x_{\mathrm{add}})\ge\eta' n\,\wt(x_{\mathrm{add}})$.
\end{enumerate}
\end{proposition}

\begin{proof}
(i) The monomials $u^{e_1}v^{e_2}w^{e_3}$ with exponents at most $q-1$ (at most $q-2$ for $T=\F_q^*$) form a basis of the functions on $T^3$. The spaces $U_1$, $U_2$, $U_3$ are spanned by the monomials whose first, second or third exponent is at most $M$. Hence an array lies in $S_3(RS_M)$ if and only if every monomial in its expansion has some exponent at most $M$. On $\F_q^3$, $x_{\mathrm{add}}=1-(u+v+w)^{q-1}$, whose monomials are those with $e_1+e_2+e_3=q-1$ and nonzero multinomial coefficient. On $(\F_q^*)^3$ the same expression applies after the substitution $u^{q-1}=1$, which only turns the three monomials with an exponent $q-1$ into constants. In both cases the criterion is $M\ge M^*_q$.

(ii) By Lucas' theorem, the multinomial coefficient is nonzero modulo $p$ exactly when the base-$p$ digits of $e_1$, $e_2$, $e_3$ add up to the digits $p-1$ of $q-1$ without carries. For $p=3$ the triple $(q/3,q/3,q/3-1)$ is admissible. Conversely, in the top digit position the three digits add up to $2$, so one of $e_1,e_2,e_3$ has top digit $0$ and is at most $q/3-1$. For $p=2$ the triple $(q/2,q/4,q/4-1)$ is admissible. Conversely, the two top binary digits belong to at most two of $e_1,e_2,e_3$, and the third is at most $q/4-1$.
(iii) Every line contains at most one nonzero entry, so Lemma~\ref{lem:line_weight} applies with $\lm(x_{\mathrm{add}})=1$.
\end{proof}

\begin{corollary}[Linear distance and uniform punctured-pair expansion do not imply a certificate] \label{cor:H_insufficient}
Let $p\in\{2,3\}$, let $q$ range over the powers of $p$, and let $M_q$ be integers with $M_q/q\to\theta'$, where $\tfrac14<\theta'<\tfrac12$ if $p=2$ and $\tfrac13<\theta'<\tfrac12$ if $p=3$. Put $n:=q-1$ and $RS_q:=RS_{M_q}\subseteq\F_q^{\F_q^*}$. Then:
\begin{enumerate}[label=\textup{(\alph*)},leftmargin=2.4em]
\item the relative distance of $RS_q$ tends to $1-\theta'>\tfrac12$; moreover, for every $f$ with $2\theta'<f\le1$ there is $\eta'(f)>0$ such that, for all large $q$, every pair of codes obtained from $RS_q$ by retaining at least $fn$ coordinates in each has two-direction expansion at least $\eta'(f)$;\item for all large $q$, $x_{\mathrm{add}}\in S_3(RS_q)$ and $L(x_{\mathrm{add}})\ge\wt(x_{\mathrm{add}})=n(n-1)$, so that $(RS_q,RS_q,RS_q)$ is not $\rho$-product-expanding for any $\rho>1/n$, and $P(x_{\mathrm{add}})\ge\eta'(1)\,n\,\wt(x_{\mathrm{add}})$.
\end{enumerate}
In particular no analogue of Theorem~\ref{thm:main_uniform_certificate}, let alone of Theorem~\ref{thm:main_sharp}, holds for these families. The hypotheses of Sections~\ref{sec:sum_codes} and~\ref{sec:transfer} hold uniformly in $q$, while the filling ratio of a word of weight $n(n-1)$ grows linearly in $n$.\end{corollary}

\begin{proof}
Lemmas~\ref{lem:robust_to_expansion} and~\ref{lem:rectangle} and Proposition~\ref{prop:sandwich} were stated over $k_0$, but their proofs use no property of the field, and we apply them over $\F_q$.

(a) By Proposition~\ref{prop:ag_parameters} with $g=0$, $d(RS_q)\ge n-M_q$, so the relative distance tends to $1-\theta'$. Let $B,B'\subseteq\F_q^*$ with $|B|,|B'|\ge fn$, and put $\epsilon:=\tfrac12(1-2\theta'/f)>0$ and $\delta_0:=\tfrac12(1-\theta'/f)>0$. Suppose first that $|B|=|B'|=n_0$. The two punctured codes are evaluation codes of genus $0$ and degree $M_q$ on $n_0\ge fn$ points of $\PP^1$. Since $M_q/n_0\le M_q/(fn)\to\theta'/f$ and $2\theta'/f=1-2\epsilon<1-\epsilon$, the margin $4+2M_q<(1-\epsilon)n_0$ holds for all large $q$, and External Input~\ref{ext:two_code_robustness} gives $\rhoGS(\epsilon)$-robustness. Both codes have relative distance at least $1-M_q/n_0$, which exceeds $\delta_0$ for all large $q$, so Lemma~\ref{lem:robust_to_expansion} gives expansion at least $\eta_0:=\rhoGS(\epsilon)\delta_0/(1+\rhoGS(\epsilon))$. Suppose now that $|B|>|B'|$. The restriction of $RS_q|_B$ to any $I\subseteq B$ with $|I|=|B'|$ is injective, because $|B'|\ge fn>M_q$ for large $q$. Lemma~\ref{lem:rectangle} applies with $|B|/|B'|\le1/f$, $(|B|-|B'|)/d(RS_q|_{B'})\le(1/f-1)/\delta_0$ and $d(RS_q|_{B'})/|B'|\ge\delta_0$, and gives expansion at least\[
\eta'(f):=\min\Bigl\{\frac{\eta_0}{1/f+(1/f-1)/\delta_0},\ \delta_0\Bigr\}>0,
\]
and $\eta_0\ge\eta'(f)$, so the same constant serves in the square case.

(b) Since $M_q/q\to\theta'$ exceeds $\tfrac14$ if $p=2$ and $\tfrac13$ if $p=3$, Proposition~\ref{prop:rs_witness}(ii) gives $M_q\ge M^*_q$ for all large $q$, and then $x_{\mathrm{add}}\in S_3(RS_q)$ by Proposition~\ref{prop:rs_witness}(i). Every line of the array contains at most one nonzero entry, so $L(x_{\mathrm{add}})\ge\wt(x_{\mathrm{add}})$ by Lemma~\ref{lem:line_weight}, and $\rho\,n\,L(x_{\mathrm{add}})\le\wt(x_{\mathrm{add}})$ forces $\rho\le1/n$. The weight is $n(n-1)$: for each $u\in\F_q^*$ the values $v\in\F_q^*\setminus\{-u\}$ are exactly those with $w:=-u-v\in\F_q^*$. The case $f=1$ of~(a) and Proposition~\ref{prop:sandwich} give $P(x_{\mathrm{add}})\ge\eta'(1)\,n\,L(x_{\mathrm{add}})\ge\eta'(1)\,n\,\wt(x_{\mathrm{add}})$. For the last assertion, $\wt(x_{\mathrm{add}})=n(n-1)\le\lambda_0n^3$ for every fixed $\lambda_0>0$ once $n\ge1/\lambda_0$, while $P(x_{\mathrm{add}})/\wt(x_{\mathrm{add}})\ge\eta'(1)\,n\to\infty$.\end{proof}

The point where the proof of Theorem~\ref{thm:B} breaks down for these families is instructive, and it identifies the geometric input that the tower supplies. The support of $x_{\mathrm{add}}$ is the set of grid points of the surface $u+v+w=0$ in $(\PP^1)^{\times3}$, a divisor of axis degree $1$ in all three directions. The grid-cover lemma of Section~\ref{sec:localization} (Lemma~\ref{lem:grid_cover}) would cover it by $5N$ lines if the gonality of the curve over the working field exceeded $2$. But $\PP^1$ has gonality $1$, and indeed the four point divisors $F_{P,Q}$ in the proof of that lemma are linearly equivalent without being equal. On the tower curves the gonality over every working field is at least $N/730$ (Lemma~\ref{lem:gonality}), because the curves have $N+1$ rational points over the fixed field $\F_{729}$; this is what makes the cover possible. On the first floor of the tower, $C_1$ is the Reed--Solomon code of length $728$ on $\F_{729}^*$ with $M_1=145<242=M^*_{729}$, so $x_{\mathrm{add}}\notin S_3(C_1)$.
\begin{remark}[The additive witness on the second floor] \label{rem:second_floor_witness}
The witness pulls back along the tower. On the Hermitian curve $X_2$, with affine points $H=\{(x,y):y^\ell+y=x^{\ell+1}\}$ and evaluation set $E_2=\{(x,y)\in H:x\ne0\}$, consider the array $x_{\mathrm{add}}\circ x$, the indicator of $x_P+x_Q+x_R=0$ on $E_2^3$. It is a light-line word with $\lm=\ell$, since a point of $E_2$ with prescribed $x$-coordinate has $\ell$ choices of $y$. For $M<(\ell^2-1)\ell$ it lies in $S_3(C_L(E_2,MP_\infty))$ if and only if $M\ge\ell\,M^*_{\ell^2}$. Indeed, the monomials $x^ay^b$ with $a\le\ell^2-1$ and $b\le\ell-1$ span the functions on $H$, because $y^\ell=x^{\ell+1}-y$ and $x^{\ell^2}=x$ there; there are $|H|=\ell^3$ of them, so they form a basis. On $E_2$, where moreover $x^{\ell^2-1}=1$, those with $a\le\ell^2-2$ form a basis in the same way. The pole order of $x^ay^b$ at $P_\infty$ is $\ell a+(\ell+1)b$, and the Weierstrass semigroup at $P_\infty$ is generated by $\ell$ and $\ell+1$, so $L(MP_\infty)$ is spanned by the monomials with $b\le\ell-1$ and $\ell a+(\ell+1)b\le M$. For $M<(\ell^2-1)\ell$ these have $a\le\ell^2-2$, so $C_L(E_2,MP_\infty)$ is the span of a subset of the basis. The argument of Proposition~\ref{prop:rs_witness}(i) for $T=\F_{\ell^2}^*$ then applies verbatim, with the pole orders $\ell e_1,\ell e_2,\ell e_3$ in place of the exponents $e_1,e_2,e_3$. For $\ell=27$ the threshold is $27\cdot242=6534$, while $M_2=3931$, so $x_{\mathrm{add}}\circ x\notin S_3(C_2)$. On a floor $i\ge3$ the pullback of $x_{\mathrm{add}}$ has $\lm=N_i/728$, a constant fraction of $N_i$. By Lemma~\ref{lem:line_weight} such a word bounds $P/\wt$ below only by a constant, and does not conflict with Theorem~\ref{thm:B}.\end{remark}

\begin{remark}[Comparison with Kalachev's witness] \label{rem:kalachev}
The separation of Kalachev~\cite{Kal23} between product expansion and robust testability uses the primitive Reed--Solomon code of length $n=q-1$ and dimension $n/3$ over $\F_q$, $q=2^{2t}$, presented as the cyclic code with check polynomial $\prod_{j<n/3}(x-\omega^j)$. His witness is supported on the triples $(i,j,l)$ of cyclic coordinates with $i+j+l\equiv0\pmod n$, with entries twisted by a character of order three. In the evaluation coordinates $u=\omega^{-i}$, $v=\omega^{-j}$, $w=\omega^{-l}$ its support is the surface $uvw=1$. It is thus a light-line word adapted to the multiplicative group $\F_q^*$, and the twist is what places it in the sum code.

The monomial criterion in the proof of Proposition~\ref{prop:rs_witness} shows that no witness of this kind does better. On $(\F_q^*)^3$ one has $[uvw=1]=-\sum_{m=0}^{n-1}(uvw)^m$. A word $\varphi(u,v)\,[uvw=1]$ with $\varphi=\sum c_{\alpha\beta}u^\alpha v^\beta$ therefore expands into the distinct monomials $u^{m+\alpha}v^{m+\beta}w^m$ with $c_{\alpha\beta}\ne0$, exponents read modulo $n$. It lies in $S_3(RS_M)$ if and only if, for each such $(\alpha,\beta)$, the three intervals $[0,M]$, $[-\alpha,M-\alpha]$, $[-\beta,M-\beta]$ cover $\mathbb Z/n$. This requires $3(M+1)\ge n$, and $\alpha=-(M+1)$, $\beta=-2(M+1)$ attains it. Thus every nonzero word supported on $uvw=1$ needs $M\ge\lceil n/3\rceil-1$, which is Kalachev's rate $\tfrac13$. The witness $x_{\mathrm{add}}$ is adapted to the additive group of $\F_q$ instead, and by Proposition~\ref{prop:rs_witness}(ii) it lies in the three-direction sum code at every rate above $\tfrac14$ in characteristic $2$. Since the tensor square of a Reed--Solomon code is robustly testable at every rate below $\tfrac12$~\cite{PS94}, product expansion of three copies and robust testability part ways on the whole range of rates $(\tfrac14,\tfrac12)$ in characteristic $2$ and $(\tfrac13,\tfrac12)$ in characteristic $3$.\end{remark}

\section{Two-code expansion and restoration of punctures} \label{sec:transfer}

The proof does not work with the full array throughout. In the carrier constructions of Sections~\ref{sec:carriers} and~\ref{sec:localization}, a small set of coordinates is deleted from each of the three axes, the argument is carried out on the smaller array, and the result is brought back to the full array. Deleting coordinates is \emph{puncturing}, and bringing a bound back to the full array is what we call \emph{restoring the punctures}. This section supplies the two facts that make this possible. First, the two-direction expansion of Section~\ref{subsec:two_directions} persists, with the single constant $\eta$ of~\eqref{eq:eta}, for all pairs of codes obtained from $C_i$ by deleting at most a quarter of the coordinates (Lemma~\ref{lem:rectangle} and Proposition~\ref{prop:pair}). Second, the pair filling mass of a word is bounded by the filling mass of its punctured version and the weight that was deleted (Lemma~\ref{lem:transfer}). Everything in this section is linear algebra and coding theory, and the only external ingredient is External Input~\ref{ext:two_code_robustness}.

\subsection{From square pairs to rectangular pairs} \label{subsec:rectangles}

For a code $C\subseteq k_0^n$ and a set $B\subseteq[n]$ of retained coordinates, the \emph{punctured code} $C|_B\subseteq k_0^B$ consists of the restrictions of the codewords of $C$ to $B$. The restriction map $C\to C|_B$ is injective exactly when no nonzero codeword is supported on the set $A:=[n]\setminus B$ of deleted coordinates, that is, when $C\cap k_0^{A}=0$; this is the case whenever fewer than $d(C)$ coordinates are deleted~\cite[Chapter~1]{MS77}. A \emph{parity-check map} of a code $C'\subseteq k_0^{n'}$ is a linear map $H'$ defined on $k_0^{n'}$ whose kernel is $C'$.

\begin{lemma}[Rectangle lemma] \label{lem:rectangle}
Let $C\subseteq k_0^n$ and $C'\subseteq k_0^{n'}$ be codes with $n\ge n'$ and $C'\ne0$. Suppose that for some $I\subseteq[n]$ with $|I|=n'$ the restriction $C\to C|_I$ is injective and the square pair $(C|_I,C')$ has expansion at least $\eta_0$. Then $(C,C')$ has expansion at least
\begin{equation}\label{eq:rectangle}
 \min\left\{\frac{\eta_0}{\,n/n'+(n-n')/d(C')\,},\ \frac{d(C')}{n'}\right\}.
\end{equation}
\end{lemma}

\begin{proof}
Let $y$ be an $n\times n'$ matrix in the two-direction sum code of $(C,C')$, let $y_I$ be the $n'\times n'$ matrix formed by its rows with index in $I$, and let $y_{I^{\mathrm c}}$ be the matrix formed by the remaining $n-n'$ rows. Then $y_I$ lies in the two-direction sum code of $(C|_I,C')$, so by hypothesis $y_I=p+q$, where $p$ is column-coded with $u$ nonzero columns, $q$ is row-coded with $v$ nonzero rows, and
\begin{equation} \label{eq:rect_square}
 n'(u+v)\le\wt(y_I)/\eta_0 .
\end{equation}
Since $C\to C|_I$ is injective, every column of $p$ is the restriction of a unique codeword of $C$. Replacing each column by that codeword gives an $n\times n'$ matrix $\tilde p$ that is column-coded for $C$ and still has $u$ nonzero columns. Let $\tilde q$ be the $n\times n'$ matrix obtained from $q$ by adding zero rows outside $I$. It is row-coded with $v$ nonzero rows.

The matrix $z:=y-\tilde p-\tilde q$ lies in the two-direction sum code of $(C,C')$, and its rows with index in $I$ vanish. We claim that every row of $z$ lies in $C'$. Let $H'$ be a parity-check map of $C'$ and apply it to each row of $z$. This annihilates the row-coded part of $z$, so the result is the image of a column-coded matrix: each of its columns is a codeword of $C$. These codewords vanish on $I$, hence they are zero by injectivity. So $H'$ annihilates every row of $z$, which proves the claim.

Outside the rows of $I$ the matrix $\tilde q$ vanishes and $\tilde p$ has at most $u$ nonzero columns with $n-n'$ entries each, so $\wt(z)\le\wt(y_{I^{\mathrm c}})+(n-n')u$. Every nonzero row of $z$ is a nonzero codeword of $C'$ and has weight at least $d(C')$, so $z$ has at most $\wt(z)/d(C')$ nonzero rows. Now $y=\tilde p+(\tilde q+z)$ is a decomposition into a column-coded and a row-coded matrix, and its weighted line count is at most
\begin{equation}
 n u+n'v+\frac{n'}{d(C')}\bigl(\wt(y_{I^{\mathrm c}})+(n-n')u\bigr)
 \;\le\;\Bigl(\frac{n}{n'}+\frac{n-n'}{d(C')}\Bigr)\,n'(u+v)+\frac{n'}{d(C')}\wt(y_{I^{\mathrm c}}),
\end{equation}
where we used $n\ge n'$. By~\eqref{eq:rect_square} the right-hand side is at most
$\frac{n/n'+(n-n')/d(C')}{\eta_0}\,\wt(y_I)+\frac{n'}{d(C')}\,\wt(y_{I^{\mathrm c}})$,
and since $\wt(y_I)+\wt(y_{I^{\mathrm c}})=\wt(y)$ this is at most $\wt(y)$ divided by the quantity~\eqref{eq:rectangle}.
\end{proof}

\subsection{One constant for all punctured pairs on the tower} \label{subsec:one_constant}

We now fix a floor of the tower and drop it from the notation: $C=C_i\subseteq k_0^{N}$ is the code of that floor, its coordinates are indexed by the evaluation points $E$, the genus is $g$, and $M=\lfloor N/5\rfloor$. For a set $B\subseteq E$ of evaluation points, the punctured code $C|_B$ is the evaluation code $C_L(B,MP_\infty)$ on the same curve with the smaller evaluation set $B$. A nonzero function in $L(MP_\infty)$ has at most $M$ zeros. Hence, as soon as $|B|>M$, the restriction $C\to C|_B$ is injective and $d(C|_B)\ge|B|-M$. We call a puncturing \emph{safe} if it retains at least $3N/4$ of the $N$ evaluation points, and we call a pair $(C|_B,C|_{B'})$ of two safe puncturings a \emph{mixed pair}, since $B$ and $B'$ may differ in size. The fraction $\tfrac34$ is chosen so that~\eqref{eq:gs_margin} with $\epsilon=\tfrac13$ holds for every mixed pair, and so that the relative distance of the punctured codes stays above $\tfrac23$.

\begin{proposition}[One constant for safe mixed pairs]\label{prop:pair}
Consider a floor with $N>200$ and $g\le N/25$. For all sets $B,B'\subseteq E$ with $|B|\ge 3N/4$ and $|B'|\ge 3N/4$, the pair $(C|_B,C|_{B'})$ has expansion at least $\eta$. In particular, for $B=B'=E$, every $x\in S_3(C)$ satisfies $\eta\, N\,L(x)\le P(x)$.
\end{proposition}

The hypotheses hold on every floor of the tower, because $N_i\ge728$ and $g_i<N_i/26$ by~\eqref{eq:genus_bound}.

\begin{proof}
Suppose first that $|B|=|B'|=n_0$ with $n_0\ge3N/4$. Both codes are evaluation codes of length $n_0$ on the curve $X$ with divisor $MP_\infty$, and
$4+2g+2M\le 4+\frac{2N}{25}+\frac{2N}{5}=4+\frac{12N}{25}<\frac N2\le\frac23\,n_0$,
the strict inequality being equivalent to $N>200$. External Input~\ref{ext:two_code_robustness} with $\epsilon=\tfrac13$ shows that the pair is $\alpha$-robust. Both codes have minimum distance at least $n_0-M\ge n_0-\tfrac N5\ge\tfrac{11}{15}n_0>\tfrac23n_0$, because $N\le\tfrac43n_0$. By Lemma~\ref{lem:robust_to_expansion} with $\delta_0=\tfrac23$, the pair has expansion at least $\eta_0:=2\alpha/(3(1+\alpha))$.

Suppose now that the sizes differ. By the symmetry of Definition~\ref{def:two_direction_expansion} we may assume $n:=|B|>n':=|B'|$. Choose any $I\subseteq B$ with $|I|=n'$. The restriction $C|_B\to C|_I$ is injective because $n'>M$, and the square pair $(C|_I,C|_{B'})$ has expansion at least $\eta_0$ by the first part. With $d(C|_{B'})\ge\tfrac{11}{15}n'$ and $n\le N\le\tfrac43n'$ we get
$n/n'\le\tfrac43$, $(n-n')/d(C|_{B'})\le\tfrac{n'/3}{11n'/15}=\tfrac5{11}$ and $d(C|_{B'})/n'\ge\tfrac{11}{15}$,
so Lemma~\ref{lem:rectangle} gives expansion at least $\min\{\tfrac{33}{59}\eta_0,\tfrac{11}{15}\}=\tfrac{33}{59}\eta_0=\eta$, the minimum being the first term because $\eta_0<\tfrac23$. In the square case $\eta_0>\eta$. Finally, the last assertion is Proposition~\ref{prop:sandwich} for the pair $(C,C)$.
\end{proof}

\subsection{Puncturing transfer} \label{subsec:transfer}

We return to three arbitrary codes $\Cj{j}\subseteq V_j=k_0^{n_j}$, with the notation $U_j$, $U_{jk}$, $S(\Cj1,\Cj2,\Cj3)$ and $P$ of Section~\ref{sec:sum_codes}. Delete a set $A\subseteq[n_1]$ of $t$ coordinates of the first axis and retain its complement $B$, of size $n=n_1-t$. The part of the array with first coordinate in $A$ is the \emph{deleted slab} $A\times[n_2]\times[n_3]$. For an array $x$ we write $x|_A$ for its restriction to the deleted slab and $x'$ for its restriction to $B\times[n_2]\times[n_3]$. If $x$ lies in $S(\Cj1,\Cj2,\Cj3)$, then $x'$ lies in the sum code of the punctured triple $(\Cj1|_B,\Cj2,\Cj3)$, and we write $P'(x')$ for its pair filling mass with respect to that triple.

\begin{lemma}[Puncturing transfer]\label{lem:transfer}
With the notation above, suppose that the restriction $\Cj1\to\Cj1|_B$ is injective and that the two pairs $(\Cj1|_B,\Cj2)$ and $(\Cj1|_B,\Cj3)$ have expansion at least $\eta$. Then every $x\in S(\Cj1,\Cj2,\Cj3)$ satisfies
\begin{equation}\label{eq:transfer}
 P(x)\le\wt(x|_{A})
             +\left(1+\frac{2t}{\eta n}\right)P'(x').
\end{equation}
The same holds, with the corresponding hypotheses, for deletions in the second or the third coordinate.

Moreover, let $C\subseteq k_0^N$, let $B_1,B_2,B_3\subseteq[N]$ be sets of sizes $n_j=N-t_j$, all nonempty, and suppose that the restrictions $C\to C|_{B_j}$ are injective and that every pair $(C|_{B_j},C)$ and every pair $(C|_{B_j},C|_{B_k})$ with $j\ne k$ has expansion at least $\eta$. For $x\in S_3(C)$ let $x_B$ be the restriction of $x$ to $B_1\times B_2\times B_3$, and let $P_B$ be the pair filling mass with respect to the triple $(C|_{B_1},C|_{B_2},C|_{B_3})$. Then
\begin{equation}\label{eq:boxtransfer}
 P(x)\le\Gamma\,\bigl(\wt(x)-\wt(x_B)+P_B(x_B)\bigr),\qquad
 \Gamma:=\prod_{j=1}^3\left(1+\frac{2t_j}{\eta n_j}\right).
\end{equation}
If $t:=t_1+t_2+t_3<N$, then $\Gamma\le\exp\bigl(2t/(\eta(N-t))\bigr)$.
\end{lemma}

In words: the filling mass of $x$ exceeds that of its punctured version by at most the weight that was deleted, up to the factor $\Gamma$, and $\Gamma$ is close to $1$ when the number $t$ of deleted coordinates is small compared with $\eta N$. On a floor of the tower as in Proposition~\ref{prop:pair}, the hypotheses of the second part hold with the constant $\eta$ of~\eqref{eq:eta} whenever each $t_j\le N/4$.

\begin{proof}
Let $H_2$ and $H_3$ be parity-check maps of $\Cj2$ and $\Cj3$. The kernel of $H_2\otimes H_3$ on $V_2\otimes V_3$ is the two-direction sum code $\Cj2\otimes V_3+V_2\otimes\Cj3$. An array therefore lies in $U_{23}$ if and only if it is annihilated by $\mathrm{id}\otimes H_2\otimes H_3$, the map that applies $H_2\otimes H_3$ to every slice perpendicular to the first direction.

Take a pair filling $x'=y'_{12}+y'_{13}+y'_{23}$ of weight $P'(x')$ for the punctured triple. We extend its three pieces to arrays on the full index set. Extending $y'_{23}$ by zero on the deleted slab gives $Y_{23}\in U_{23}$ of the same weight. Every slice of $y'_{12}$ perpendicular to the third direction lies in the two-direction sum code of $(\Cj1|_B,\Cj2)$, an $n\times n_2$ matrix code. Applying the expansion of this pair to each slice writes $y'_{12}$ as the sum of an array whose lines in the first direction lie in $\Cj1|_B$ and an array whose lines in the second direction lie in $\Cj2$, and by~\eqref{eq:two_direction_expansion} the number $h_{12}$ of nonzero lines in the first direction satisfies $nh_{12}\le\wt(y'_{12})/\eta$. Extend each of these lines to the unique codeword of $\Cj1$ that restricts to it, and extend the lines in the second direction by zero. The result is an array $Y_{12}\in U_{12}$ that agrees with $y'_{12}$ outside the deleted slab and whose weight on the deleted slab is $\Delta_{12}\le t\,h_{12}\le\frac{t}{\eta n}\,\wt(y'_{12})$. The same construction gives $Y_{13}\in U_{13}$ with weight $\Delta_{13}\le\frac{t}{\eta n}\wt(y'_{13})$ on the deleted slab.

The array $z:=x-Y_{12}-Y_{13}-Y_{23}$ lies in $S(\Cj1,\Cj2,\Cj3)$ and is supported on the deleted slab. We claim that $z\in U_{23}$. Write $z=u_1+u_2+u_3$ with $u_j\in U_j$. The map $\mathrm{id}\otimes H_2\otimes H_3$ annihilates $u_2$ and $u_3$, so it sends $z$ to the image of $u_1$, which lies in $\Cj1\otimes\im H_2\otimes\im H_3$. On the other hand this image is supported on $A$ in the first coordinate, because $z$ is. So it lies in $(\Cj1\cap k_0^A)\otimes\im H_2\otimes\im H_3$, and $\Cj1\cap k_0^A=0$ by the injectivity of the restriction to $B$. The image is therefore zero, and the claim follows from the first paragraph.

Now $x=Y_{12}+Y_{13}+(Y_{23}+z)$ is a pair filling of $x$. On the deleted slab $Y_{23}$ vanishes, so $\wt(z)\le\wt(x|_A)+\Delta_{12}+\Delta_{13}$. The total weight of the filling is therefore at most
$\bigl(\wt(y'_{12})+\Delta_{12}\bigr)+\bigl(\wt(y'_{13})+\Delta_{13}\bigr)+\wt(y'_{23})+\wt(z)\le P'(x')+\wt(x|_A)+2(\Delta_{12}+\Delta_{13})$,
and $2(\Delta_{12}+\Delta_{13})\le\frac{2t}{\eta n}P'(x')$. This proves~\eqref{eq:transfer}. The argument is symmetric in the three coordinates.

For the second part, puncture the three coordinates one after the other. Let $x^{[0]}=x$, and let $x^{[j]}$ be the restriction of $x$ to the index set in which the first $j$ coordinates have been restricted to $B_1,\dots,B_j$. Thus $x^{[3]}=x_B$. At step $j$ the triple consists of the codes $C|_{B_1},\dots,C|_{B_{j-1}}$ followed by copies of $C$, and~\eqref{eq:transfer} applies to the deletion in the $j$-th coordinate, since the two pairs it requires are among those assumed to have expansion $\eta$. Let $w_j:=\wt(x^{[j-1]})-\wt(x^{[j]})$ be the weight of $x^{[j-1]}$ on the slab deleted at step $j$, and let $\gamma_j:=1+2t_j/(\eta n_j)$, so that $\Gamma=\gamma_1\gamma_2\gamma_3$. The three inequalities combine to
$P(x)\le w_1+\gamma_1w_2+\gamma_1\gamma_2w_3+\gamma_1\gamma_2\gamma_3\,P_B(x_B)\le\Gamma\,\bigl(w_1+w_2+w_3+P_B(x_B)\bigr)$,
and $w_1+w_2+w_3=\wt(x)-\wt(x_B)$. Finally, $1+\xi\le e^{\xi}$ and $n_j=N-t_j\ge N-t$ give $\Gamma\le\exp\bigl(2t/(\eta(N-t))\bigr)$.
\end{proof}

\section{The geometry of the triple product} \label{sec:product_geometry}

This section translates the sum code into geometry on the threefold $Y=X\times X\times X$ and collects the geometric tools used in Sections~\ref{sec:carriers}--\ref{sec:localization}: the description of the parity checks of a sum code as the sections of a line bundle $\LME$ (Lemma~\ref{lem:dual_evaluation_bundle}), the degree calculus for curves and $1$-cycles in $Y$, the theorem of the cube, and two vanishing lemmas (Lemmas~\ref{lem:kunneth_vanishing} and~\ref{lem:ambient_vanishing_D}). The second vanishing lemma, for a twist of $\omega_Y$ by a positive box product and an arbitrary effective divisor, is the geometric fact on which the support localisation of Section~\ref{sec:localization} rests; its proof uses only the product structure of $Y$ and needs no lifting hypothesis.

From now on a floor $i$ is fixed and dropped from the notation: $X:=X_i$ is the curve, $g$ its genus, $P_\infty:=P_{\infty,i}$ its point at infinity, $E:=E_i$ its evaluation divisor and $N:=N_i$. Let $\Xj1,\Xj2,\Xj3$ be three copies of $X$, and let
\begin{equation}
Y := \Xj1 \times \Xj2 \times \Xj3
\end{equation}
be their product, a smooth projective variety of dimension three. It comes with the coordinate projections $p_j:Y\to\Xj{j}$ and the pair projections $p_{jk}:Y\to\Xj{j}\times\Xj{k}$. A rational point of $Y$ is a triple $(P_1,P_2,P_3)$ of rational points of $X$, so the grid $E\times E\times E\subset Y(k_0)$ is the index set of the arrays in $k_0^{N\times N\times N}$. Under this identification a line of the array in direction $j$ is the set of grid points on a \emph{fibre curve} of $Y$, obtained by fixing the two coordinates other than the $j$-th, and a slice of the array is the set of grid points on a copy of the surface $\Xj{j}\times\Xj{k}$. The table summarises the dictionary.

\begin{center}
\renewcommand{\arraystretch}{1.25}
\begin{tabular}{@{}p{0.43\textwidth}p{0.52\textwidth}@{}}
\toprule
Coding theory & Geometry \\
\midrule
coordinate of a codeword & rational point $P\in E$ of the curve $X$ \\
codeword of $C_L(E,G)$ & function in $L(G)=H^0(X,\OO_X(G))$ \\
parity check of $C_L(E,G)$ & differential in $H^0(X,\omega_X(E-G))$, read through its residues \\
cell of a three-dimensional array & grid point $(P_1,P_2,P_3)\in E\times E\times E\subset Y$ \\
line in direction $j$; slice & fibre curve of $Y$; fibre surface $\Xj{j}\times\Xj{k}$ of $Y$ \\
word of the tensor code $C\otimes C\otimes C$ & section of the box product $\OO(G)\boxtimes\OO(G)\boxtimes\OO(G)$ on $Y$ \\
parity check of the sum code $S_3(C)$ & section of the line bundle $\LME$ on $Y$ \\
sparse family of prescribed lines & carrier: a divisor $J\subset Y$ containing the fibre curves of the lines that survive a few deletions (Section~\ref{sec:carriers}) \\
arrays supported on the carrier's grid points & sums of coded lines on the fibres in $J$ (Theorem~\ref{thm:exact_generation}) \\
\bottomrule
\end{tabular}
\end{center}

\subsection{Line bundles on \texorpdfstring{$Y$}{Y} and tensor codes} \label{subsec:app_product_bundles}

Divisors and line bundles on $Y$ are defined as on a curve~\cite[Section~II.6]{Hartshorne77}, with irreducible surfaces in $Y$ taking the place of points: a divisor $D$ on $Y$ is a formal integer combination of irreducible surfaces, a nonzero rational function $f$ on $Y$ has a divisor $(f)=\sum_S v_S(f)[S]$ of zeros and poles, where $v_S$ is the valuation of the discrete valuation ring $\OO_{Y,S}$, and the line bundle $\OO_Y(D)$ has global sections $\{f\in K(Y)^\times:(f)+D\ge0\}\cup\{0\}$. The simplest divisors on $Y$ come from the factors. If $D_j$ is a divisor on $\Xj{j}$, its pullback $p_j^*D_j$ is the divisor on $Y$ that imposes the condition $D_j$ on the $j$-th coordinate and no condition on the other two: for a point $P$, $p_j^{-1}(P)$ is the surface of points whose $j$-th coordinate is $P$. Correspondingly a line bundle $\AA_j$ on $\Xj{j}$ has a pullback $p_j^*\AA_j$ on $Y$, with $p_j^*\OO_{\Xj{j}}(D_j)=\OO_Y(p_j^*D_j)$.

\begin{definition}[Box product] \label{def:app_box_product}
The \emph{box product} of line bundles $\AA_j$ on $\Xj{j}$, $j=1,2,3$, is the line bundle $\AA_1 \boxtimes \AA_2 \boxtimes \AA_3 := p_1^* \AA_1 \otimes p_2^* \AA_2 \otimes p_3^* \AA_3$ on $Y$. The integers $\deg(\AA_j)$ are its \emph{factor degrees}. A line bundle on $Y$ isomorphic to a box product is said to have \emph{external class}. Box products on the pair surfaces $\Xj{j}\times\Xj{k}$ are defined in the same way with two factors.
\end{definition}

Box products are the geometric form of tensor products of codes. By the K\"unneth formula~\cite[Tag~0BED]{Stacks}, for coherent sheaves $\cF_j$ on $\Xj{j}$,
\begin{equation} \label{eq:kunneth}
H^r\bigl(Y,\cF_1\boxtimes\cF_2\boxtimes\cF_3\bigr)=\bigoplus_{r_1+r_2+r_3=r}H^{r_1}(\Xj1,\cF_1)\otimes H^{r_2}(\Xj2,\cF_2)\otimes H^{r_3}(\Xj3,\cF_3),
\end{equation}
and similarly for two factors, and for products in which one factor is a closed subscheme of a surface. For $r=0$ and $\cF_j=\OO(D_j)$ the identity says that every function on $Y$ with poles bounded by $\sum_j p_j^*D_j$ is a linear combination of products $f_1(P_1)f_2(P_2)f_3(P_3)$ with $f_j\in L(D_j)$, and evaluating such functions on a grid of rational points produces the tensor product of the three evaluation codes. When $g\ge1$ not every line bundle on $Y$ is a box product; the pullback of the diagonal of $\Xj1\times\Xj2$ defines one that is not. The canonical bundle of $Y$ is a box product, because the differentials of a product are the direct sum of the pullbacks of the differentials of the factors~\cite[Exercise~II.8.3]{Hartshorne77}:
\begin{equation} \label{eq:omega_Y}
\omega_Y=\omega_{\Xj1}\boxtimes\omega_{\Xj2}\boxtimes\omega_{\Xj3}.
\end{equation}
Since $p_j=\pi_j\circ p_{jk}$ with $\pi_j:\Xj{j}\times\Xj{k}\to\Xj{j}$ the projection, $p_j^*\AA_j\cong p_{jk}^*(\pi_j^*\AA_j)$: a box product involving only the directions $j$ and $k$ is the pullback of a line bundle on the pair surface. This is the counterpart of the description~\eqref{eq:Ujk_slices} of $U_{jk}$, in which the third direction is free.

\begin{lemma}[K\"unneth vanishing] \label{lem:kunneth_vanishing}
Let $\cF=\cF_1\boxtimes\cF_2\boxtimes\cF_3$ be a box product of line bundles on $Y$.
\begin{enumerate}[label=\textup{(\alph*)}]
\item If every $\cF_j$ is nonspecial, in particular if every factor degree exceeds $2g-2$, then $H^r(Y,\cF)=0$ for all $r>0$.
\item If at most one $\cF_j$ is special, then $H^2(Y,\cF)=H^3(Y,\cF)=0$. If some $\cF_j$ is nonspecial, then $H^3(Y,\cF)=0$.
\item If every factor degree is negative, then $H^0(Y,\cF)=H^1(Y,\cF)=0$.
\end{enumerate}
On a pair surface, a box product of two nonspecial line bundles has $H^r=0$ for every $r>0$. If at least one factor is nonspecial, then $H^2=0$. And if both factor degrees are negative, then $H^0=H^1=0$. One nonspecial factor need not force $H^1=0$.
\end{lemma}

\begin{proof}
On a curve the only cohomology groups are $H^0$ and $H^1$. In~\eqref{eq:kunneth} a summand with $r=\sum r_j$ contains $r$ factors $H^1(\Xj{j},\cF_j)$, which vanish for nonspecial $\cF_j$; this gives (a) and (b). If $\deg\cF_j<0$ then $H^0(\Xj{j},\cF_j)=0$, and a summand with $r\le1$ contains at least two factors $H^0$; this gives (c).
\end{proof}

\subsection{Parity checks of the sum code as sections of a line bundle} \label{subsec:dual_and_cohomology}

Let $E'_1,E'_2,E'_3\subseteq E$ be sets of evaluation points of cardinalities $n_j=|E'_j|\le N$, and let $G_1,G_2,G_3$ be divisors on $X$ with supports disjoint from $E'_1,E'_2,E'_3$ respectively. Put
\begin{equation}
C(E'_j,G_j):=C_L(E'_j,G_j)\subseteq k_0^{n_j},
\end{equation}
the evaluation code on the retained points of the $j$-th factor. The case of interest in Sections~\ref{sec:carriers}--\ref{sec:generation} is $G_j=MP_\infty$ for all $j$; in Section~\ref{sec:localization} divisors $G_j$ of larger degree occur, which is why the statement is made in this generality. Equation~\eqref{eq:sum_code_dual} says that the parity checks of the sum code $S(C(E'_1,G_1),C(E'_2,G_2),C(E'_3,G_3))$ are the tensor products of parity checks of the three codes, and Proposition~\ref{prop:app_cauchy_duality} says that the parity checks of $C(E'_j,G_j)$ are the residue vectors of the differentials in $H^0(\Xj{j},\omega_{\Xj{j}}(E'_j-G_j))$. Combining the two with the K\"unneth formula leads to the line bundle
\begin{equation} \label{eq:LM}
\LL_{G,E'} := \omega_{\Xj1}(E'_1 - G_1)\boxtimes\omega_{\Xj2}(E'_2 - G_2)\boxtimes\omega_{\Xj3}(E'_3 - G_3)
\end{equation}
on $Y$; when $G_j=MP_\infty$ for all $j$ we write $\LME$, and when the sets $E'_j$ are clear we write $\LL$. Its global sections are the linear combinations of products $\omega_1\otimes\omega_2\otimes\omega_3$ of differentials $\omega_j\in H^0(\Xj{j},\omega_{\Xj{j}}(E'_j-G_j))$. A section $\sigma$ has a value at each grid point $P=(P_1,P_2,P_3)\in E'_1\times E'_2\times E'_3$, defined on products by
\begin{equation} \label{eq:residue_value}
(\omega_1\otimes\omega_2\otimes\omega_3)(P):=\Res_{P_1}(\omega_1)\,\Res_{P_2}(\omega_2)\,\Res_{P_3}(\omega_3)
\end{equation}
and extended linearly. As explained in Remark~\ref{rem:app_generator_matrices}, the residue is the canonical way of turning the value of such a section into a scalar, so no choice is involved.

\begin{lemma}[Dual evaluation bundle] \label{lem:dual_evaluation_bundle}
Suppose that $\deg G_j>2g - 2$ for $j=1,2,3$.
\begin{enumerate}
\item For each $j$, the residue map $\res_{E'_j}$ is an isomorphism from $H^0\bigl(\Xj{j},\omega_{\Xj{j}}(E'_j - G_j)\bigr)$ onto the dual code $C(E'_j,G_j)^\perp$.
\item An array $z \in k_0^{n_1 \times n_2 \times n_3}$ belongs to the sum code $S\bigl(C(E'_1,G_1), C(E'_2,G_2), C(E'_3,G_3)\bigr)$ if and only if
\begin{equation}
\sum_{P \in E'_1 \times E'_2 \times E'_3} z_P \cdot \sigma(P) = 0 \qquad \text{for every global section } \sigma \in H^0(Y, \, \LL_{G,E'}). \label{eq:dual_orthogonality}
\end{equation}
\end{enumerate}
\end{lemma}

\begin{proof}
(1) By Proposition~\ref{prop:app_cauchy_duality}, applied with $E=E'_j$ and $G=G_j$, the image of $\res_{E'_j}$ is $C(E'_j,G_j)^\perp$, and the map is injective because $\deg G_j>2g-2$.
(2) By~\eqref{eq:kunneth}, $H^0(Y,\LL_{G,E'})$ is the tensor product of the three spaces of differentials, and by~\eqref{eq:residue_value} the map $\sigma\mapsto(\sigma(P))_P$ is the tensor product of the three residue maps. By part~(1) it is therefore an isomorphism of $H^0(Y,\LL_{G,E'})$ onto $C(E'_1,G_1)^\perp\otimes C(E'_2,G_2)^\perp\otimes C(E'_3,G_3)^\perp$, which is the dual of the sum code by~\eqref{eq:sum_code_dual}. Condition~\eqref{eq:dual_orthogonality} says that $z$ is orthogonal to this dual.
\end{proof}

\begin{remark}[Codes of general line bundles] \label{rem:shifted_codes}
Section~\ref{sec:localization} uses evaluation codes $\ev_{E'_j}H^0(X,\cG_j)$ of line bundles $\cG_j$ that are not given as $\OO(MP_\infty)$. Since the base field there is infinite, every line bundle on $X$ is isomorphic to $\OO(G_j)$ for a divisor $G_j$ disjoint from $E'_j$, and we fix such an isomorphism once; it trivialises $\cG_j$ at every point of $E'_j$ by the constant function $1$, and all evaluations, products and residues below are formed in these trivialisations, so that Lemma~\ref{lem:dual_evaluation_bundle} applies verbatim to $\ev_{E'_j}H^0(X,\cG_j)=C(E'_j,G_j)$. A different choice rescales the coordinates of all codes involved by the same nonzero scalars and changes neither weights nor the sum-code membership of an array rescaled in the same way.
\end{remark}

\subsection{Degrees, projection degrees and positivity} \label{subsec:positivity}

By a \emph{curve in $Y$} we mean an integral curve: a closed subvariety $\Sigma\subset Y$ of dimension one, reduced and irreducible. Fibre curves are examples, but a curve in $Y$ may project onto two or all three factors, and it may be singular. More generally a \emph{subcurve} of a closed subscheme of $Y$ is a closed subscheme $V$ of pure dimension one, possibly reducible and nonreduced; its \emph{cycle} is $[V]=\sum_\Sigma m_\Sigma[\Sigma]$, the sum of its irreducible components with their multiplicities $m_\Sigma=\length(\OO_{V,\eta_\Sigma})$ at the generic points. For a line bundle $\cF$ on $Y$ and a subcurve $V$, we put
\begin{equation} \label{eq:curve_degree}
\deg(\cF|_V):=\chi(V,\cF|_V)-\chi(V,\OO_V).
\end{equation}
This is additive in $\cF$, and by Snapper's theorem~\cite{Kleiman66} it equals $\sum_\Sigma m_\Sigma\deg(\cF|_\Sigma)$, where for an integral $\Sigma$ the degree is computed on the normalisation $\tilde\Sigma\to\Sigma$; we also write $\cF\cdot V$ for it. The intersection number $\cF\cdot V$ depends only on the algebraic equivalence class of $\cF$, and if $\cF=\OO_Y(D)$ for an effective divisor $D$ containing no component of $V$, then $\cF\cdot V=\length(D\cap V)$. The \emph{projection degrees} of $V$ are
\begin{equation} \label{eq:projection_degrees}
d_j(V):=\deg\bigl(p_j^*\OO_X(P)\big|_V\bigr)=\sum_\Sigma m_\Sigma\deg(p_j|_\Sigma:\Sigma\to\Xj{j})\ \ge 0,
\end{equation}
where $P$ is any point of $X$ and the degree of a constant map is $0$. A curve cannot be contracted by all three projections, so $d_1(V)+d_2(V)+d_3(V)>0$ for every nonzero subcurve. The same definitions apply on the pair surfaces.

\begin{lemma}[Degrees of box products on subcurves] \label{lem:box_degree}
For a box product $\AA=\AA_1\boxtimes\AA_2\boxtimes\AA_3$ and a subcurve $V$ of $Y$,
\begin{equation}
\AA\cdot V = \deg(\AA_1)\,d_1(V) + \deg(\AA_2)\,d_2(V) + \deg(\AA_3)\,d_3(V). \label{eq:box_intersection_formula}
\end{equation}
In particular $\AA\cdot V=\deg(\AA_j)$ for a fibre curve in direction $j$, and $\AA\cdot V\ge0$ for every $V$ when all factor degrees are non-negative.
\end{lemma}

\begin{proof}
Degrees add under tensor products and are additive over the cycle, so it suffices to show that $\deg((p_j^*\AA_j)|_\Sigma)=d_j(\Sigma)\deg(\AA_j)$ for an integral $\Sigma$. The bundle is the pullback of $\AA_j$ under the composite $\tilde\Sigma\to\Sigma\to\Xj{j}$, a morphism of smooth projective curves that is either constant or finite of degree $d_j(\Sigma)$. In the first case the pullback is trivial; in the second, pulling back multiplies degrees by $d_j(\Sigma)$~\cite[Proposition~II.6.9]{Hartshorne77}.
\end{proof}

\begin{definition}[Nef line bundles] \label{def:app_nef}
A line bundle $\NN$ on $Y$, or on a pair surface, is \emph{nef} if $\NN\cdot\Sigma\ge0$ for every integral curve $\Sigma$, equivalently for every subcurve.
\end{definition}

Three properties follow directly from the definition. A tensor product of nef line bundles is nef. The pullback of a nef line bundle under a morphism is nef, because its degree on a curve $\Sigma$ is zero if $\Sigma$ is contracted and otherwise a positive multiple of the degree on the image. A box product with non-negative factor degrees is nef, by Lemma~\ref{lem:box_degree}. The following criterion supplies nef bundles that are not box products.

\begin{lemma}[Systems without fixed curves are nef] \label{lem:nef_system}
Let $S$ be a smooth projective surface and $W\subseteq H^0(S,\MM)$ a nonzero space of sections of a line bundle $\MM$ such that no curve of $S$ lies in the zero locus of every section in $W$. Then $\MM$ is nef.
\end{lemma}

\begin{proof}
For an integral curve $\Sigma$, choose $\sigma\in W$ not vanishing identically on $\Sigma$; then $\sigma|_\Sigma$ is a nonzero section of $\MM|_\Sigma$, whose degree is therefore non-negative.
\end{proof}

\begin{proposition}[Theorem of the cube on $Y$ {\cite[Chapter~I, Theorem~5.1]{Milne08}}] \label{prop:cube}
Let $\MM$ be a line bundle on $Y$ and $(P_1,P_2,P_3)$ a rational point. Let $\MM_j:=\MM|_{\ell_j}$ be the restriction of $\MM$ to the fibre curve $\ell_j\cong X$ through the point in direction $j$, and $\MM_{jk}:=\MM|_{S_{jk}}$ its restriction to the fibre surface $S_{jk}\cong\Xj{j}\times\Xj{k}$ through the point. Then
\begin{equation} \label{eq:cube}
\MM\otimes p_1^*\MM_1\otimes p_2^*\MM_2\otimes p_3^*\MM_3\;\cong\;p_{12}^*\MM_{12}\otimes p_{13}^*\MM_{13}\otimes p_{23}^*\MM_{23}.
\end{equation}
Consequently, for rational points $P,P_0$ of $\Xj1$ and $Q,Q_0$ of $\Xj2$ and the fibre curves $\ell_{P,Q}:=\{P\}\times\{Q\}\times\Xj3$,
\begin{equation} \label{eq:cube_fibres}
\MM|_{\ell_{P,Q}}\otimes\MM|_{\ell_{P_0,Q_0}}\;\cong\;\MM|_{\ell_{P,Q_0}}\otimes\MM|_{\ell_{P_0,Q}}\qquad\text{as line bundles on } X .
\end{equation}
\end{proposition}

\begin{proof}
The theorem of the cube states that the line bundle $\NN:=\MM\otimes\bigotimes_jp_j^*\MM_j\otimes\bigotimes_{j<k}p_{jk}^*\MM_{jk}^{-1}$ on the product of three complete varieties is trivial as soon as its restrictions to the three fibre surfaces $S_{jk}$ through a point are trivial, and they are: on $S_{12}$, for instance, $p_{12}^*\MM_{12}$ restricts to $\MM_{12}$, $p_{13}^*\MM_{13}$ and $p_{23}^*\MM_{23}$ restrict to $\pi_1^*\MM_1$ and $\pi_2^*\MM_2$, and $p_3^*\MM_3$ restricts to the trivial bundle, so the restriction of $\NN$ is $\MM_{12}\otimes\pi_1^*\MM_1\otimes\pi_2^*\MM_2\otimes\pi_1^*\MM_1^{-1}\otimes\pi_2^*\MM_2^{-1}\otimes\MM_{12}^{-1}$, which is trivial. For~\eqref{eq:cube_fibres}, apply~\eqref{eq:cube} with base point $(P_0,Q_0,R_0)$ and restrict to $\ell_{P,Q}$: the pullbacks $p_1^*$, $p_2^*$ and $p_{12}^*$ become trivial, $p_{13}^*\MM_{13}$ restricts to $\MM|_{\ell_{P,Q_0}}$, $p_{23}^*\MM_{23}$ to $\MM|_{\ell_{P_0,Q}}$, and $p_3^*\MM_3$ to $\MM|_{\ell_{P_0,Q_0}}$.

The theorem of the cube is stated in~\cite{Milne08} over an algebraically closed field. It holds over any field $K$ over which $Y$, $\MM$ and the base point are defined, and this is the form used in Lemma~\ref{lem:grid_cover}: if a line bundle $\NN$ on $Y_K$ becomes trivial over an algebraic closure $\Omega$, then $h^0(Y_K,\NN)=h^0(Y_\Omega,\NN_\Omega)=1$ by flat base change, and a nonzero section of $\NN$ over $K$ vanishes nowhere, because its base change is a nonzero section of the trivial bundle on the proper integral variety $Y_\Omega$, hence a nonzero constant, so $\NN\cong\OO_{Y_K}$.
\end{proof}

\subsection{Cohomology and interpolation; two vanishing lemmas} \label{subsec:vanishing}

Besides its global sections, a line bundle $\cF$ on $Y$ has higher cohomology groups $H^r(Y,\cF)$ for $r=1,2,3$, finite-dimensional vector spaces~\cite[Chapter~III]{Hartshorne77} whose role in this paper is that of obstructions to interpolation. For a closed subscheme $Z\subset Y$ with ideal sheaf $\II_Z$, there is an exact sequence
\begin{equation} \label{eq:restriction_sequence}
0\to H^0\bigl(Y,\II_Z\cF\bigr)\to H^0(Y,\cF)\to H^0\bigl(Z,\cF|_Z\bigr)\to H^1\bigl(Y,\II_Z\cF\bigr),
\end{equation}
in which the middle arrow is restriction to $Z$; for an effective divisor $D$ one has $\II_D\cF=\cF(-D)$. If $H^1(Y,\II_Z\cF)=0$, then every section of $\cF$ defined on $Z$ is the restriction of a global section. Applied to $\cF=\LME$, whose global sections are the parity checks of the sum code, a vanishing statement of this kind produces global parity checks with prescribed behaviour along $Z$. Serre duality on $Y$ reads $H^r(Y,\cF)^\vee\cong H^{3-r}(Y,\omega_Y\otimes\cF^{-1})$.

\begin{lemma}[Vanishing with an arbitrary effective divisor] \label{lem:ambient_vanishing_D}
Let $\cF=\cF_1\boxtimes\cF_2\boxtimes\cF_3$ be a box product with positive factor degrees, and let $D$ be an effective divisor on $Y$. Then
\begin{equation} \label{eq:vanishing_D}
H^2\bigl(Y,\,\omega_Y\otimes\cF\otimes\OO_Y(D)\bigr)=0 ,\qquad\text{equivalently}\qquad H^1\bigl(Y,\,\cF^{-1}(-D)\bigr)=0 .
\end{equation}
\end{lemma}

\begin{proof}
The two statements are Serre dual. Extending scalars to an algebraic closure changes neither, so we may assume that the base field is algebraically closed, hence perfect. The divisor sequence $0\to\cF^{-1}(-D)\to\cF^{-1}\to\cF^{-1}|_D\to0$ gives an exact sequence
\[
H^0\bigl(D,\cF^{-1}|_D\bigr)\to H^1\bigl(Y,\cF^{-1}(-D)\bigr)\to H^1\bigl(Y,\cF^{-1}\bigr),
\]
and the last group vanishes by Lemma~\ref{lem:kunneth_vanishing}(c). It remains to show that $H^0(D,\cF^{-1}|_D)=0$.

We first treat a prime divisor $Z$ and an arbitrary effective divisor $D'$, and show that $H^0(Z,\cF^{-1}(-D')|_Z)=0$. Write $D'=mZ+G$ with $m\ge0$ and $G$ effective not containing $Z$. Since $Z$ is integral and $G$ does not contain it, the local equation of $G$ is a nonzerodivisor on $\OO_Z$, so $\OO_Z(-G)\hookrightarrow\OO_Z$ and it suffices to treat $D'=mZ$. Since the base field is perfect, $Z$ is generically smooth, and on its smooth locus the conormal sequence $0\to\OO_Z(-Z)\to\Omega_Y|_Z\to\Omega_Z\to0$ is exact on the left. As $\Omega_Y=\bigoplus_jp_j^*\omega_{\Xj{j}}$ by~\eqref{eq:omega_Y}, at least one component $\OO_Z(-Z)\to p_j^*\omega_{\Xj{j}}|_Z$ of the conormal map is nonzero at the generic point of $Z$, and a nonzero map between line bundles on an integral scheme is injective. Iterating, $\OO_Z(-mZ)\hookrightarrow p_j^*\omega_{\Xj{j}}^{\otimes m}|_Z$, and hence
\[
\cF^{-1}(-mZ)|_Z\hookrightarrow\bigl(p_j^*\omega_{\Xj{j}}^{\otimes m}\bigr)|_Z\otimes\cF^{-1}|_Z .
\]
We claim that the target has no nonzero section. If the projection $p_j|_Z$ is constant, then $Z=\{P\}\times\Xj{k}\times\Xj{l}$, the factor $p_j^*\omega^{\otimes m}$ restricts to the trivial bundle, and $\cF^{-1}|_Z=\cF_k^{-1}\boxtimes\cF_l^{-1}$ has no sections by Lemma~\ref{lem:kunneth_vanishing}(c) on the pair surface. If $p_j|_Z$ is dominant, consider the generic fibre $Z_\xi$ of $p_j|_Z:Z\to\Xj{j}$, an integral projective curve over the function field $\kappa=K(\Xj{j})$, contained in $(\Xj{k}\times\Xj{l})_\kappa$. On it $p_j^*\omega^{\otimes m}$ is trivial and $\cF|_{Z_\xi}$ is the restriction of $\cF_k\boxtimes\cF_l$; the projections of $Z_\xi$ to $(\Xj{k})_\kappa$ and $(\Xj{l})_\kappa$ are not both constant, each is either constant or finite, and pulling back multiplies degrees by the degree of the map, so $\deg(\cF|_{Z_\xi})>0$ by the two-factor form of Lemma~\ref{lem:box_degree}. A line bundle of negative degree on an integral projective curve over any field has no nonzero section, and a nonzero section of a line bundle on the integral scheme $Z$ restricts to a nonzero section on the dense generic fibre. This proves the claim.

Now let $D$ be an arbitrary effective divisor and argue by induction on the number of prime components of $D$ counted with multiplicity. Write $D=Z+D'$ with $Z$ prime. The standard exact sequence $0\to\OO_Z(-D')\to\OO_{Z+D'}\to\OO_{D'}\to0$ for the sum of two effective Cartier divisors, tensored with $\cF^{-1}$, gives
\[
0\to H^0\bigl(Z,\cF^{-1}(-D')|_Z\bigr)\to H^0\bigl(D,\cF^{-1}|_D\bigr)\to H^0\bigl(D',\cF^{-1}|_{D'}\bigr),
\]
and both outer groups vanish, the first by the previous paragraph and the second by induction (for $D'=0$ it is zero). Hence $H^0(D,\cF^{-1}|_D)=0$.
\end{proof}

\begin{remark}[No lifting hypothesis is needed] \label{rem:no_DI}
Over the complex numbers the Kodaira vanishing theorem gives $H^r(Y,\omega_Y\otimes\AA)=0$ for $r>0$ and ample $\AA$, and in positive characteristic this holds for varieties of dimension at most $p$ that lift modulo $p^2$, by the theorem of Deligne, Illusie and Raynaud~\cite{DeligneIllusie87}; the threefold $Y$ over $\F_{3^6}$ satisfies both hypotheses. Neither hypothesis is needed here. The pair divisors of a carrier have external class (Proposition~\ref{prop:carrier}(i)), so $\LME(-J)$ is a box product and the vanishing $H^1(Y,\LME(-J))=0$ on a carrier $J$ (Corollary~\ref{cor:carrier_vanishing}) follows from Lemma~\ref{lem:kunneth_vanishing} alone. The only non-K\"unneth vanishing statement used in this paper is Lemma~\ref{lem:ambient_vanishing_D}, which concerns $H^2$ rather than $H^1$, and which is proved above from the product structure of $Y$ alone. For a product of four or more curves the analogous argument produces $H^{\dim Y-1}$-vanishing only, which is why the method of this paper is specific to three directions (Remark~\ref{rem:four_directions}).
\end{remark}

\begin{lemma}[Koszul sequences for complete-intersection curves] \label{lem:koszul}
Let $\MM_1,\MM_2$ be line bundles on $Y$ with sections $\sigma_1,\sigma_2$ whose zero loci have no common component, and let $\Sigma_{12}:=Z(\sigma_1)\cap Z(\sigma_2)$ be the complete-intersection curve. For every line bundle $\cF$ on $Y$ there is an exact sequence
\begin{equation} \label{eq:koszul}
0\to\cF\otimes\MM_1^{-1}\otimes\MM_2^{-1}\to\cF\otimes\MM_1^{-1}\oplus\cF\otimes\MM_2^{-1}\to\cF\to\cF|_{\Sigma_{12}}\to0 .
\end{equation}
Consequently: if $H^1(Y,\cF\otimes\MM_1^{-1})=H^1(Y,\cF\otimes\MM_2^{-1})=0$ and $H^2(Y,\cF\otimes\MM_1^{-1}\otimes\MM_2^{-1})=0$, then $H^0(Y,\cF)\to H^0(\Sigma_{12},\cF|_{\Sigma_{12}})$ is surjective; if moreover $H^1(Y,\cF)=0$, $H^2(Y,\cF\otimes\MM_a^{-1})=0$ for $a=1,2$ and $H^3(Y,\cF\otimes\MM_1^{-1}\otimes\MM_2^{-1})=0$, then $H^1(\Sigma_{12},\cF|_{\Sigma_{12}})=0$. The curve $\Sigma_{12}$ is a local complete intersection of codimension two in the smooth threefold $Y$, hence Gorenstein, with dualizing sheaf $\omega_{\Sigma_{12}}=(\omega_Y\otimes\MM_1\otimes\MM_2)|_{\Sigma_{12}}$.
\end{lemma}

\begin{proof}
The Koszul complex of the regular sequence $(\sigma_1,\sigma_2)$ is exact because the two zero loci have no common component~\cite[Section~III.7]{Hartshorne77}; split~\eqref{eq:koszul} into two short exact sequences and chase the long exact cohomology sequences. The last assertion is adjunction for local complete intersections~\cite[Theorem~III.7.11]{Hartshorne77}.
\end{proof}

\section{Sparse carriers}\label{sec:carriers}

A \emph{carrier} is a divisor $J$ on $Y$ adapted to a prescribed set of lines of the array: after a few coordinates have been deleted, it contains the fibre curves of $Y$ on which the remaining prescribed lines lie, its pair divisors have external class of small factor degrees and no axis components, and each of them meets every row and column of the retained pair grid in few points. This section constructs carriers for sparse families of lines (Proposition~\ref{prop:carrier}). The family is measured by an integer $s$ whose square exceeds the number $m$ of prescribed lines. The carrier then costs a pole order of $a=g+s-1$ in each factor of each pair divisor, each pair divisor contains fewer than $s$ points on every row and column of the corresponding pair grid, and its construction requires the deletion of at most $2m/s$ coordinate indices. As long as the retained sets remain safe in the sense of Section~\ref{subsec:one_constant}, Lemma~\ref{lem:transfer} bounds the cost of these deletions.

\subsection{Conventions: base field and terminology} \label{subsec:carrier_conventions}

\paragraph{Extension of scalars.}
The construction uses auxiliary line bundles and sections that in general do not exist over $k_0$. Remark~\ref{rem:mds_obstruction} gives the reason. We construct the auxiliary bundles, sections and divisors over a working field $K=k_0(V)$ of the form specified in Section~\ref{subsec:fields}, and write $\Omega$ for an algebraic closure of the current $K$. In particular, $K$ need not be algebraic over $k_0$. The curve $X$, the pair surfaces and the threefold $Y$ may be base changed to either field, with base-change subscripts suppressed when the field is clear. The evaluation points remain the $k_0$-rational points of $E$. Cohomology, geometric components and subcurve inequalities may be checked over $\Omega$. The arithmetic gonality bound and the grid-cover argument are applied over $K$ to divisors defined over $K$.

The genus and the degrees of line bundles are unchanged by field extension, and flat base change for coherent cohomology preserves the dimensions and vanishing assertions used below. Thus the geometric arguments may be verified after passage to $\Omega$. The original codes and input arrays are defined over $k_0$, but the shifted arrays and auxiliary codes of Section~\ref{sec:localization} may be defined only over $K$. For decompositions of the original arrays in the scalar extensions of the original codes, the following lemma returns the coding-theoretic conclusions to $k_0$.

\begin{lemma}[Extension of scalars] \label{lem:scalars}
Let $K$ be an extension field of $k_0$, and for a code $C\subseteq k_0^n$ let $C_K\subseteq K^n$ be its $K$-linear span. Then $C_K\cap k_0^n=C$. For every $x\in S(\Cj1,\Cj2,\Cj3)$, the line count and the pair filling mass of $x$ computed over $K$ with respect to the triple $(\Cj1_K,\Cj2_K,\Cj3_K)$ are equal to $L(x)$ and $P(x)$. If $x\in k_0^{n_1\times n_2\times n_3}$ is a $K$-linear combination of coded lines of the codes $\Cj{j}_K$ on a given set of lines, then it is a $k_0$-linear combination of coded lines of the codes $\Cj{j}$ on the same lines.
\end{lemma}

\begin{proof}
Choose a $k_0$-linear map $\varphi:K\to k_0$ with $\varphi(1)=1$ and apply it entry by entry to vectors and arrays. The resulting map is $k_0$-linear, fixes every vector or array with entries in $k_0$, and does not enlarge supports. It maps $C_K$ into $C$, since it sends $\sum_t\lambda_tc_t$ with $\lambda_t\in K$ and $c_t\in C$ to $\sum_t\varphi(\lambda_t)c_t$. This gives the first assertion. The spaces $U_j$ and $U_{jk}$ formed with the codes $\Cj{j}_K$ are the $K$-spans of the original ones, so for the same reason the map sends them into $U_j$ and $U_{jk}$, and it maps a coded line over $K$ to a coded line on the same line of the array or to zero. Applied to a directional decomposition, a pair filling, or a decomposition into coded lines on a given set of lines over $K$, it therefore produces one over $k_0$ with no more lines, no more weight, and the same set of lines. The opposite inequalities are trivial, because a decomposition over $k_0$ is also one over $K$.
\end{proof}

The restriction on the working fields is essential: finite generation alone does not imply the gonality bound used in Section~\ref{sec:localization}. Generic Picard parameters and generic coefficients can be adjoined within the class $K=k_0(V)$, as explained in Section~\ref{subsec:fields}. The divisor constructions below preserve definition over the current $K$. Passing to $\Omega$ to verify their geometric properties does not replace $K$ in the arithmetic argument.

\paragraph{Terminology.}
Throughout this section $E_1^0,E_2^0,E_3^0\subseteq E$ are sets of evaluation points, one for each factor of $Y$; they play the role of the retained sets $E'_j$ of Section~\ref{subsec:dual_and_cohomology}. The \emph{grid} is $E_1^0\times E_2^0\times E_3^0$, and for $j<k$ the \emph{pair grid} is the set $E_j^0\times E_k^0$ of points of the pair surface $\Xj{j}\times\Xj{k}$. A fibre curve of $Y$ through grid points, such as $\Xj1\times\{b\}\times\{c\}$ with $b\in E_2^0$ and $c\in E_3^0$, is an \emph{evaluation coordinate fibre}, and its intersection with the grid is its \emph{grid line}; this is a line of the array in the sense of Section~\ref{subsec:sum_code}. On a pair surface $\Xj{j}\times\Xj{k}$, an \emph{axis} is a curve $\{b\}\times\Xj{k}$ or $\Xj{j}\times\{c\}$ on which one of the two projections is constant. It is an \emph{evaluation axis} if $b\in E_j^0$, respectively $c\in E_k^0$, and its points in the pair grid then form a \emph{row}, respectively a \emph{column}, of the pair grid. A \emph{coordinate-axis component} of a divisor on a pair surface is a component that is an axis. The two \emph{coordinate degrees} of an effective divisor $D$ on a pair surface are its intersection numbers with $\{b\}\times\Xj{k}$ and with $\Xj{j}\times\{c\}$; by Lemma~\ref{lem:box_degree} they are the projection degrees $d_j(D)$ and $d_k(D)$, and if $\OO(D)\cong\cF_j\boxtimes\cF_k$ has external class they are $\deg\cF_k$ and $\deg\cF_j$ respectively. The \emph{footprint} of a divisor on a pair surface or on $Y$ is the set of pair-grid points, respectively grid points, that lie on it. We write $h^0(\BB):=\dim H^0(X,\BB)$ for a line bundle $\BB$ on $X$.

\subsection{Line bundles for which every \texorpdfstring{$s$}{s} points form an information set} \label{subsec:mds_bundles}

For each integer $d$, the line bundles of degree $d$ on $X$ are the points of an irreducible projective variety $\Pic^d(X)$ of dimension $g$, the Picard variety, which exists as a $k_0$-variety because $X$ has a rational point~\cite{Milne86,Kleiman05}; tensoring with a fixed line bundle of degree $d'$ is an isomorphism $\Pic^d(X)\to\Pic^{d+d'}(X)$, and $\Pic^0(X)$ is an abelian variety, so every $\Pic^d(X)$ is smooth and geometrically integral with a rational point. The line bundles of degree $g-1$ that have a nonzero section are those of the form $\OO(Q_1+\dots+Q_{g-1})$; for $g\ge1$ they form a closed subset $W_{g-1}\subset\Pic^{g-1}(X)$, the image of the $(g-1)$-fold symmetric product of $X$, of dimension at most $g-1$.

\begin{lemma}[Finite-grid information sets]\label{lem:MDSbundle}
Let $X$ be a smooth projective curve of genus $g$ over an algebraically closed field, let $E^0$ be a set of $n$ distinct points of $X$, and let $1\le s\le n$. A general line bundle $\BB$ on $X$ of degree $a=g+s-1$ satisfies $h^0(\BB)=s$ and the following property: for every $s$-element subset $A\subseteq E^0$, evaluation at the points of $A$ is an isomorphism from $H^0(X,\BB)$ onto the direct sum of the fibres of $\BB$ at these points. Consequently a section of $\BB$ that vanishes at $s$ points of $E^0$ is zero, and evaluation at any $e\le s$ points of $E^0$ is surjective. Here ``general'' means: outside a proper closed subset of $\Pic^a(X)$ determined by $E^0$ and $s$.
\end{lemma}

\begin{proof}
For genus zero, $\BB=\OO((s-1)Q)$ has the required property, since its sections are the polynomials of degree at most $s-1$ in a coordinate with pole at $Q$, and a nonzero section has exactly $s-1$ zeros. Let $g\ge1$. For an $s$-element subset $A\subseteq E^0$, the kernel of the evaluation map at $A$ is $H^0(X,\BB(-A))$, and $\BB(-A)$ has degree $g-1$. This kernel is nonzero exactly when $\BB(-A)\in W_{g-1}$, that is, when $\BB$ lies in the translate $\OO(A)\otimes W_{g-1}$, a closed subset of $\Pic^{g+s-1}(X)$ of dimension at most $g-1$. There are finitely many subsets $A$, and a finite union of proper closed subsets cannot cover the irreducible variety $\Pic^{g+s-1}(X)$. For $\BB$ outside this union, evaluation at every $s$-element subset is injective, so $h^0(\BB)\le s$, while Riemann--Roch gives $h^0(\BB)\ge a+1-g=s$. Hence $h^0(\BB)=s$ and each of these evaluation maps is an isomorphism. The two consequences follow because every set of at most $s$ points of $E^0$ is contained in an $s$-element subset.
\end{proof}

\begin{remark}[Why the base field has to be enlarged] \label{rem:mds_obstruction}
In coding terms, the lemma says that the evaluation code of $\BB$ on $E^0$ is an MDS code of length $n$ and dimension $s$. An MDS code of dimension $s\ge2$ over $\F_q$ has length at most $q+s-1$~\cite[Chapter~11]{MS77}: shortening it $s-2$ times gives an MDS code of dimension two, whose coordinates are distinct points of the projective line over $\F_q$. Hence for $s\ge2$ and $n>729+s-1$ no bundle $\BB$ as in the lemma can be defined over $k_0=\F_{729}$. This is the situation of interest: in all applications $s<N/2$ and $n\ge3N/4$, so $n>729+s-1$ as soon as $3N/4>729+N/2$, that is, $N>2916$, which holds on every floor beyond the first. The lemma, and with it the whole construction of this section, takes place over an extension of $k_0$. In the proof the enlarged field is used to find a point of $\Pic^{g+s-1}(X)$ outside a proper closed subset, and the same is true of the choice of a general section in the proof of Proposition~\ref{prop:carrier}.
\end{remark}

\subsection{The carrier construction} \label{subsec:carrier_construction}

The lines that a carrier has to contain are prescribed as follows. For each pair of directions $j<k$, with remaining direction $l$, a set $F_{jk}\subseteq E_j^0\times E_k^0$ of pair-grid points is given, and each point $(b,c)\in F_{jk}$ stands for the evaluation coordinate fibre in direction $l$ whose $j$-th and $k$-th coordinates are $b$ and $c$. A divisor on the pair surface that passes through $(b,c)$ pulls back under $p_{jk}$ to a divisor on $Y$ that contains this fibre.

\begin{proposition}[Carrier construction]\label{prop:carrier}
Let $F_{jk}\subseteq E_j^0\times E_k^0$ for $1\le j<k\le3$, and put $m=|F_{12}|+|F_{13}|+|F_{23}|$. Let $s$ be an integer with $s^2>m$ and $1\le s\le\min_j|E_j^0|$, and set $a=g+s-1$. Then, over a suitable extension field, there are subsets $E_j^1\subseteq E_j^0$, the \emph{retained sets}, obtained by deleting at most $2m/s<2s$ coordinate indices in total from the three sets $E_j^0$, and effective divisors $D_{jk}$ on the pair surfaces $\Xj{j}\times\Xj{k}$, such that:
\begin{enumerate}[label=(\roman*),leftmargin=2.4em]
\item $D_{jk}$ has no coordinate-axis components and external class: $\OO(D_{jk})\cong\cF^{(jk)}_j\boxtimes\cF^{(jk)}_k$ with $0\le\deg\cF^{(jk)}_j,\deg\cF^{(jk)}_k\le a$; in particular both coordinate degrees of $D_{jk}$ are at most $a$.
\item $D_{jk}$ contains every \emph{surviving} point of $F_{jk}$, that is, every point of $F_{jk}$ that lies in the retained pair grid $E_j^1\times E_k^1$.
\item Every row and every column of the retained pair grid $E_j^1\times E_k^1$ contains at most $s-1$ points of $D_{jk}$.
\end{enumerate}
Consequently the divisor $J:=p_{12}^*D_{12}+p_{13}^*D_{13}+p_{23}^*D_{23}$ on $Y$, the \emph{carrier} of the family, has external class $\OO_Y(J)\cong\LL_1\boxtimes\LL_2\boxtimes\LL_3$ with $0\le\deg\LL_j\le2a$. The footprint of $J$ on the retained grid $E_1^1\times E_2^1\times E_3^1$ is the union of the retained parts of the grid lines of at most $3N(s-1)$ evaluation coordinate fibres contained in $J$, namely the fibres in the remaining direction through the points of the divisors $D_{jk}$ in the retained pair grids. Each of these fibres meets the fibres of the other two directions among them in at most $\nu=2(s-1)$ retained grid points. If $m=0$ one may take all $D_{jk}=0$, $J=0$ and no deletions.
\end{proposition}

\begin{proof}
\emph{Step 1: the linear systems.} For each $j$ choose a line bundle $\BB_j$ on $\Xj{j}$ by Lemma~\ref{lem:MDSbundle}, applied to the point set $E_j^0$ and the integer $s$, and put $R_j:=H^0(\Xj{j},\BB_j)$, a space of dimension $s$. By the K\"unneth formula, $R_j\otimes R_k$ is the space of global sections of $\BB_j\boxtimes\BB_k$ on the pair surface, of dimension $s^2$. Let $\Lambda_{jk}\subseteq R_j\otimes R_k$ be the subspace of sections that vanish at all points of $F_{jk}$. Each point imposes one linear condition, so the codimension $c_{jk}$ of $\Lambda_{jk}$ is at most $|F_{jk}|\le m<s^2$, and $\Lambda_{jk}\ne0$.

\emph{Step 2: forced axes and deletions.} Call an evaluation axis \emph{forced} for the pair $(j,k)$ if every section in $\Lambda_{jk}$ vanishes identically on it. Suppose that the axes $\{b_1\}\times\Xj{k},\dots,\{b_e\}\times\Xj{k}$ are forced, with $e\le s$. Restricting a section of $\BB_j\boxtimes\BB_k$ to these $e$ axes amounts to evaluating its first tensor factor at $b_1,\dots,b_e$. By Lemma~\ref{lem:MDSbundle} this evaluation is surjective on $R_j$, so the restriction map on $R_j\otimes R_k$ has rank $es$, and its kernel, which contains $\Lambda_{jk}$, has codimension $es$. Hence $es\le c_{jk}$. Since $c_{jk}<s^2$, there cannot be $s$ forced axes of this kind, and therefore their number is at most $c_{jk}/s\le|F_{jk}|/s$. The same holds for the forced axes $\Xj{j}\times\{c\}$. Delete from $E_j^0$ every index whose axis is forced for one of the two pairs that involve the direction $j$, and let $E_j^1$ be the set of indices that remain. An index, a row, a column or a grid point is \emph{retained} if it is not affected by these deletions. The number of deleted indices is at most $(|F_{jk}|+|F_{jl}|)/s$ for the direction $j$, and at most $2m/s<2s$ in total.

\emph{Step 3: the divisors $D_{jk}$.} Let $\Phi_{jk}\subseteq E_j^0\times E_k^0$ be the set of pair-grid points at which every section in $\Lambda_{jk}$ vanishes. It contains $F_{jk}$. Choose a nonzero $\sigma\in\Lambda_{jk}$ that does not vanish identically on any evaluation axis that is not forced, and does not vanish at any pair-grid point outside $\Phi_{jk}$. Each of these finitely many requirements, including $\sigma\ne0$, excludes a proper subspace of $\Lambda_{jk}$, and a vector space over an infinite field is not a finite union of proper subspaces, so such a section exists over any infinite field over which $\Lambda_{jk}$ is defined. Its divisor of zeros is an effective divisor with $\OO(\operatorname{div}\sigma)\cong\BB_j\boxtimes\BB_k$. We remove all geometric axis components without enlarging the field of definition. Suppose $\sigma$ is defined over the current working field $K$. Let
\[
U_j:=\im\bigl(R_k^\vee\longrightarrow R_j,\ \varphi\longmapsto(\mathrm{id}\otimes\varphi)(\sigma)\bigr),\qquad
U_k:=\im\bigl(R_j^\vee\longrightarrow R_k,\ \psi\longmapsto(\psi\otimes\mathrm{id})(\sigma)\bigr).
\]
The nonzero coefficient spaces $U_j,U_k$ are defined over $K$. On the smooth curve $X_K$, their base ideals are invertible, say $\OO(-A_j)$ and $\OO(-A_k)$ for effective $K$-divisors $A_j,A_k$. These ideals are the images of the evaluation maps $U_j\otimes_K\BB_j^{-1}\to\OO_{X_K}$ and $U_k\otimes_K\BB_k^{-1}\to\OO_{X_K}$, so they commute with extension of the ground field. Dividing $\sigma$ by the canonical sections of the pullbacks of $A_j$ and $A_k$ gives a section of $\BB_j(-A_j)\boxtimes\BB_k(-A_k)$ defined over $K$.

Over $\Omega$, the multiplicity of a vertical axis in $\operatorname{div}(\sigma)$ is the minimum order of vanishing of the coefficient sections in $U_j$, hence is exactly the multiplicity recorded by $A_j$. The horizontal case is identical. Thus, writing $Z_{jk}:=\pi_j^*A_j+\pi_k^*A_k$ and $D_{jk}:=\operatorname{div}(\sigma)-Z_{jk}$, where $\pi_j,\pi_k$ are the projections of the pair surface, $Z_{jk}$ collects all geometric axis components with their multiplicities, and $D_{jk}$ is an effective divisor defined over $K$ with no geometric coordinate-axis components. In particular,
\[
\OO(D_{jk})\;\cong\;\bigl(\BB_j(-A_j)\bigr)\boxtimes\bigl(\BB_k(-A_k)\bigr)
\]
has external class with factor degrees $a-\deg A_j$ and $a-\deg A_k$, both at most $a$. These factor degrees are the coordinate degrees of the effective divisor $D_{jk}$ without axis components, hence non-negative. This proves~(i).

The section $\sigma$ vanishes identically on every axis in $Z_{jk}$. If such an axis is an evaluation axis, it is forced by the choice of $\sigma$, and its index has been deleted. If it is not an evaluation axis, it contains no pair-grid point. Thus no retained pair-grid point lies on $Z_{jk}$, and a retained pair-grid point lies on $D_{jk}$ if and only if $\sigma$ vanishes there, which by the choice of $\sigma$ happens if and only if the point belongs to $\Phi_{jk}$. The points of $D_{jk}$ in the retained pair grid are therefore exactly the retained points of $\Phi_{jk}$, and they include the surviving points of $F_{jk}$. This proves~(ii).

\emph{Step 4: rows and columns.} Suppose that a retained row, with first coordinate $b$, contained $s$ points $(b,c_1),\dots,(b,c_s)$ of $\Phi_{jk}$. The restriction of any section in $\Lambda_{jk}$ to the axis $\{b\}\times\Xj{k}$ is a section of $\BB_k$ that vanishes at the $s$ points $c_1,\dots,c_s$ of $E_k^0$, and is zero by Lemma~\ref{lem:MDSbundle}. The axis would then be forced and $b$ would have been deleted. The same argument applies to columns. Since the points of $D_{jk}$ in the retained pair grid all belong to $\Phi_{jk}$ by Step~3, this proves~(iii).

\emph{Step 5: the carrier.} Each direction belongs to two of the three pairs, so $\OO_Y(J)=\bigotimes_{j<k}p_{jk}^*\OO(D_{jk})$ is a box product whose $j$-th factor is the product of the two factors in direction $j$, of degree between $0$ and $2a$. A retained grid point lies on $p_{jk}^*D_{jk}$ if and only if its $j$-th and $k$-th coordinates form a point of $D_{jk}$, and in that case the whole evaluation coordinate fibre in the remaining direction through it lies in $p_{jk}^*D_{jk}$. So the footprint of $J$ on the retained grid is the union of the retained parts of the grid lines of these fibres. By~(iii) the divisor $D_{jk}$ has at most $N(s-1)$ points in the retained pair grid, which gives at most $3N(s-1)$ fibres. Finally, consider one of these fibres, say in direction $l$ through the retained point $(b,c)$ of $D_{jk}$. A fibre of the family in direction $k$ lies over a retained point, with $j$-th coordinate $b'$ and $l$-th coordinate $e$, of the divisor of the pair $\{j,l\}$, and it meets the given fibre only if $b'=b$, in the single grid point with coordinates $b,c,e$. The number of such $e$ is the number of retained points of that divisor with $j$-th coordinate $b$, at most $s-1$ by~(iii). In the same way the fibres in direction $j$ meet it in at most $s-1$ grid points, which gives the bound $\nu=2(s-1)$.
\end{proof}

\begin{remark}[Prescribed family and footprint]\label{rem:footprint}
The family of fibres prescribed by the sets $F_{jk}$ and the footprint of the carrier are different objects, and the footprint can be larger: the divisor $D_{jk}$ passes through the surviving points of $F_{jk}$, but it may contain further points of the pair grid. Any use of the carrier must handle the whole footprint, not only the prescribed family; Proposition~\ref{prop:carrier} controls the footprint only through $a$, $\nu$ and the count $3N(s-1)$, not through $m$. Theorem~\ref{thm:exact_generation} is formulated for the whole footprint.
\end{remark}

\begin{corollary}[Vanishing for carriers] \label{cor:carrier_vanishing}
Let $J$ be a carrier as in Proposition~\textup{\ref{prop:carrier}}, and let $\LME$ be the line bundle~\eqref{eq:LM} for sets $E'_j\subseteq E$ of sizes $n_j$, for instance the retained sets $E_j^1$. If $n_j>M+2a$ for $j=1,2,3$, then $H^r(Y,\LME(-J))=0$ for all $r>0$.
\end{corollary}

\begin{proof}
By~\eqref{eq:LM} and Proposition~\ref{prop:carrier}, $\LME(-J)$ is the box product with factors $\omega_{\Xj{j}}(E'_j-MP_\infty)\otimes\LL_j^{-1}$, of degrees $2g-2+n_j-M-\deg\LL_j\ge2g-2+n_j-M-2a>2g-2$. Lemma~\ref{lem:kunneth_vanishing}(a) applies.
\end{proof}

Since $a=g+s-1$, the condition $n_j>M+2a$ reads $s<(n_j-M-2g+2)/2$. Together with the requirement $s^2>m$ this delimits the families of lines to which the construction applies. The next proposition fixes a regime in which all the requirements of Sections~\ref{sec:transfer} and~\ref{sec:carriers} hold at once; Section~\ref{sec:localization} works in a narrower regime still.

\begin{proposition}[Parameter regime] \label{prop:carrier_regime}
Consider a floor with $N>200$ and $g\le N/25$, take sets $E_j^0\subseteq E$ with $|E_j^0|\ge N-N/8$, and let $s$ be an integer with $s^2>m$ and $1\le s\le N/16$. Let $t_j$ be the number of indices that Proposition~\ref{prop:carrier} deletes from axis $j$, $t=t_1+t_2+t_3$, and $n_j=|E_j^1|$. Then:
\begin{enumerate}[label=(\alph*)]
\item $t<2s\le N/8$, so each retained set $E_j^1$ has at least $3N/4$ points and is a safe puncturing in the sense of Section~\ref{subsec:one_constant};
\item $n_j>M+2a$ for $j=1,2,3$, so Corollary~\ref{cor:carrier_vanishing} applies with $E'_j=E_j^1$;
\item the factor $\Gamma$ of Lemma~\ref{lem:transfer} for the total deletions from $E$ to $E^1$ satisfies $\Gamma<\exp\bigl(2q/(\eta(N-q))\bigr)$, where $q=\sum_j(N-n_j)$ is the total number of indices deleted from $E$.
\end{enumerate}
\end{proposition}

\begin{proof}
(a) Proposition~\ref{prop:carrier} gives $t\le 2m/s<2s\le N/8$, so each retained set has more than $N-N/8-N/8=\tfrac34N$ points. (b) $n_j\ge\tfrac34N$, while $M+2a=M+2g+2s-2\le\tfrac{N}{5}+\tfrac{2N}{25}+\tfrac{N}{8}-2<\tfrac{3N}{4}$. (c) By (a) the retained sets are safe, so Proposition~\ref{prop:pair} supplies the hypotheses of the second part of Lemma~\ref{lem:transfer}, which gives the bound.
\end{proof}

\section{Exact generation on a carrier}\label{sec:generation}

This section proves the central geometric statement of the paper: an array supported on the footprint of a carrier belongs to the sum code if and only if it is a sum of codewords placed on the coordinate fibres contained in the carrier (Theorem~\ref{thm:exact_generation}). Combined with the support localisation of Section~\ref{sec:localization}, which places every light word on such a footprint, and with the crossing bound of Proposition~\ref{prop:carrier}, this reduces the filling problem to the elementary count of Lemma~\ref{lem:weighted_filling}.

\subsection{Setting} \label{subsec:gen_setting}

A floor is fixed. For the geometric arguments of this section, base change to an algebraic closure $\Omega$ of a field of definition of the auxiliary divisors, and write $\kappa:=\Omega$ throughout this section. The resulting decompositions of the original arrays descend to $k_0$ by Lemma~\ref{lem:scalars}. Let $E'_1,E'_2,E'_3\subseteq E$ be sets of evaluation points of sizes $n_j=|E'_j|$, put $n_{\min}:=\min_jn_j$, let $G$ be a divisor on $X$ of degree $M$ whose support is disjoint from $E$, and let
\begin{equation}
C_j:=C_L(E'_j,G)\subseteq k_0^{n_j},\qquad S_B:=S(C_1,C_2,C_3)
\end{equation}
be the three punctured codes and their sum code. Put $\LL_j:=\omega_X(E'_j-G)$ and $\LL:=\LL_1\boxtimes\LL_2\boxtimes\LL_3$, the bundle~\eqref{eq:LM}. On the tower, $G=MP_\infty$ and the $E'_j$ are the retained sets of a carrier.

Let $D_{12},D_{13},D_{23}$ be effective divisors on the pair surfaces that have no coordinate-axis components and external class with factor degrees at most an integer $a\ge0$: $\OO(D_{jk})\cong\cF^{(jk)}_j\boxtimes\cF^{(jk)}_k$ with $0\le\deg\cF^{(jk)}_j,\deg\cF^{(jk)}_k\le a$. The three \emph{cylinders} are
\begin{equation}
\Pi_1:=p_{12}^*D_{12},\qquad \Pi_2:=p_{13}^*D_{13},\qquad \Pi_3:=p_{23}^*D_{23},\qquad J:=\Pi_1+\Pi_2+\Pi_3,
\end{equation}
with remaining directions $3$, $2$ and $1$ respectively; $\Pi_i$ is the product of $D_{jk}$ with the third curve. For a carrier, these are the data of Proposition~\ref{prop:carrier} with $a=g+s-1$, and $J$ is the carrier. Let $R_i$ be the reduced union of the evaluation coordinate fibres contained in $\Pi_i$, let $R:=R_1\cup R_2\cup R_3$, and let $\Phi\subseteq E'_1\times E'_2\times E'_3$ be the footprint of $J$, the set of grid points on $J$. Since a fibre of direction $l$ lies in $\Pi_i=p_{jk}^*D_{jk}$ exactly when its $(j,k)$-coordinates form a point of $D_{jk}$, $\Phi$ is the set of grid points on $R$. For a fibre $\ell\subset R$ in direction $j$, let $V_\ell\subseteq k_0^{\Phi}$ be the space of codewords of $C_j$ written along the grid line of $\ell$, extended by zero.

\begin{lemma}[Elementary properties of the configuration] \label{lem:config}
\begin{enumerate}[label=\textup{(\alph*)}]
\item No fibre of $R_i$ is contained in another cylinder $\Pi_j$, $j\ne i$. Two fibres in the same direction are disjoint; two fibres in different directions meet in at most one point, transversally.
\item The three cylinders have pairwise no common component, so each pairwise intersection $C_{ij}:=\Pi_i\cap\Pi_j$ is a complete-intersection curve, Gorenstein by Lemma~\textup{\ref{lem:koszul}}. Every coordinate projection $C_{ij}\to\Xj{r}$ is finite, and so is its restriction to every subcurve.
\item $\LL(-J)$ is a box product whose factor degrees are at least $2g-2+n_j-M-2a$.
\end{enumerate}
\end{lemma}

\begin{proof}
(a) A direction-$3$ fibre $\{b\}\times\{c\}\times\Xj3$ lies in $\Pi_2=p_{13}^*D_{13}$ only if $(b,e)\in D_{13}$ for every $e$, that is, if the axis $\{b\}\times\Xj3$ is a component of $D_{13}$, which is excluded; the other cases are identical. The assertions about pairs of fibres are clear. (b) A common component of $p_{12}^*D_{12}$ and $p_{13}^*D_{13}$ would be a surface $\Gamma\times\Xj3=\{(b,c,e):(b,c)\in\Gamma\}$ with $\Gamma$ a component of $D_{12}$ that also has the form $\{(b,c,e):(b,e)\in\Gamma'\}$; this forces $\Gamma=\{b\}\times\Xj2$, an axis. For finiteness, fix the first coordinate $b$ of a point of $C_{23}=p_{13}^*D_{13}\cap p_{23}^*D_{23}$: the third coordinate $e$ satisfies $(b,e)\in D_{13}$, which leaves finitely many $e$ because $D_{13}$ contains no axis, and then $(c,e)\in D_{23}$ leaves finitely many $c$. Fixing the second or third coordinate is similar, and the other curves are treated in the same way; a projective morphism with finite fibres is finite. (c) Each direction belongs to two of the three pairs, so $\OO_Y(-J)$ is a box product with factor degrees at least $-2a$, and $\LL$ has factor degrees $2g-2+n_j-M$.
\end{proof}

\paragraph{The grid ideal.}
For each $j$ let $q_j$ denote a local equation of the divisor $p_j^*E'_j$, the union of the surfaces $p_j^{-1}(P)$ with $P\in E'_j$; thus $q_j$ is the pullback of a uniformizer at $P$ near such a surface and a unit elsewhere. The \emph{grid ideal} is the ideal sheaf
\begin{equation}\label{eq:grid_ideal}
\mathfrak j:=\OO_Y(-p_1^*E'_1-p_2^*E'_2)+\OO_Y(-p_1^*E'_1-p_3^*E'_3)+\OO_Y(-p_2^*E'_2-p_3^*E'_3)\;\subseteq\;\OO_Y,
\end{equation}
that is, locally $\mathfrak j=(q_1q_2,q_1q_3,q_2q_3)$.
Recall that a subscheme is \emph{reduced} if its ideal is radical, and that the \emph{reduced union} of a family of fibres is their set-theoretic union with the reduced structure.

\begin{lemma}[The grid ideal]\label{lem:grid_ideal}
$\mathfrak j$ is the ideal of the reduced union of all evaluation coordinate fibres. At a grid point it is the ideal of the three fibres through that point. In particular $\mathfrak j\subseteq\II_R$.
\end{lemma}

\begin{proof}
Set-theoretically, $\mathfrak j$ vanishes exactly where at least two of $q_1,q_2,q_3$ vanish. The locus $q_1=q_2=0$ consists of the points whose first coordinate lies in $E'_1$ and whose second lies in $E'_2$, the union of the evaluation coordinate fibres in direction~$3$; the other two pairs give the fibres in directions $2$ and~$1$. For the scheme structure, work at a point $(P_1,P_2,P_3)$ of $Y$, with $x_j$ the pullback of a uniformizer at $P_j$. The points of $E'_j$ are distinct, so $p_j^*E'_j$ has multiplicity one along each of its surfaces, and $q_j$ is $x_j$ up to a unit if $P_j\in E'_j$ and a unit otherwise. If all three $P_j$ are retained, so that the point is a grid point, then
\[
\mathfrak j=(x_1x_2,x_1x_3,x_2x_3)=(x_1,x_2)\cap(x_1,x_3)\cap(x_2,x_3),
\]
since $(x_1,x_2)\cap(x_1,x_3)=(x_1,x_2x_3)$ and intersecting with $(x_2,x_3)$ gives $(x_1x_2,x_1x_3,x_2x_3)$. This is the ideal of the three fibres through the point, and as an intersection of prime ideals it is radical. If exactly two are retained, say $P_1$ and $P_2$, then $q_3$ is a unit and $\mathfrak j=(x_1x_2,x_1,x_2)=(x_1,x_2)$, the prime ideal of the single fibre through the point. If at most one is retained, two of the three generators are units and $\mathfrak j$ is the unit ideal. In every case $\mathfrak j$ is radical with the stated zero set. The last assertion holds because $R$ is a reduced subscheme of the reduced union of all fibres.
\end{proof}

\subsection{Tools: interpolation on Gorenstein curves and gluing on crossings}\label{subsec:gen_tools}

Let $Z$ be a projective curve over $\kappa$, that is, a projective scheme of pure dimension one. $Z$ is \emph{Cohen--Macaulay} if it has no embedded points and \emph{Gorenstein} if moreover its dualizing sheaf $\omega_Z$ is invertible; local complete intersections in smooth varieties are Gorenstein. A \emph{CM subcurve} of $Z$ is a closed subscheme $V\subseteq Z$ of pure dimension one without embedded points. For a finite subscheme $T\subset Z$, the scheme-theoretic intersection $T\cap V$ has ideal $\II_T+\II_V$, and $\length(T\cap V)$ is the dimension of its ring of functions; for a reduced $T$ this is the number of points of $T$ on $V$. The \emph{arithmetic genus} of $V$ is $p_a(V):=1-\chi(\OO_V)$, and $\deg(\cF|_V)$ is defined by~\eqref{eq:curve_degree}.

\begin{lemma}[Interpolation criterion on Gorenstein curves]\label{lem:fs_adjunction}
Let $Z$ be a projective Gorenstein curve over $\kappa$, $T\subset Z$ a finite subscheme, and $\cF$ a line bundle on $Z$. If, for every nonzero CM subcurve $V\subseteq Z$,
\begin{equation}\label{eq:gen1}
\deg(\cF|_V)>2p_a(V)-2+\length(T\cap V),
\end{equation}
then $H^1(Z,\II_{T,Z}\cF)=0$; equivalently, $H^0(Z,\cF)\to H^0(T,\cF|_T)$ is surjective and $H^1(Z,\cF)=0$.
\end{lemma}

The lemma is proved in Appendix~\ref{sec:appendix_curves}; it is a form of the automatic adjunction of Catanese, Franciosi, Hulek and Reid~\cite{CFHR99}. It extends the familiar statement that on a smooth curve of genus $g$ one has $H^1(\cF(-T))=0$ when $\deg\cF-\#T>2g-2$. On a reducible or nonreduced curve the condition must hold for every subcurve, because the obstruction can be concentrated on one piece. When $Z$ is an effective Cartier divisor on a smooth surface $\Xi$, its CM subcurves are exactly the effective divisors $V\le Z$ on $\Xi$, and adjunction gives $2p_a(V)-2=V\cdot(V+K_\Xi)$~\cite[Exercise~V.1.3]{Hartshorne77}.

The second tool describes sections of a line bundle on the reduced union $R$ of fibres.

\begin{lemma}[Gluing on reduced crossings]\label{lem:gluing}
Let $\cG$ be a line bundle on $Y$. A global section of $\cG|_R$ is the same thing as a family $(\tau_\ell)_{\ell\subset R}$ of sections $\tau_\ell\in H^0(\ell,\cG|_\ell)$, one for each fibre $\ell$ of $R$, such that $\tau_\ell(P)=\tau_{\ell'}(P)$ in the fibre $\cG(P)$ whenever two fibres $\ell\ne\ell'$ meet at a point $P$.
\end{lemma}

\begin{proof}
Restriction gives a map from $H^0(R,\cG|_R)$ to the compatible families, and it is injective because $R$ is reduced and the fibres cover it. For surjectivity it suffices to show that the natural map $\OO_R\to\bigoplus_\ell\OO_\ell$ identifies $\OO_R$ with the sheaf of families of functions agreeing at common points; the statement for $\cG|_R$ follows by tensoring with the locally free sheaf $\cG$. This is a local question at the points of $R$. Away from the crossings $R$ is a single fibre and there is nothing to prove. At a point $P$ where exactly the fibres in directions $j$ and $k$ meet, Lemma~\ref{lem:grid_ideal} and Lemma~\ref{lem:config}(a) show that the completed local ring of $R$ is $\kappa[[u,v]]/(uv)$, with $u,v$ the local coordinates along the two fibres; it maps injectively to $\kappa[[u]]\times\kappa[[v]]$, with image the pairs of power series with equal constant term. At a point where three fibres meet the completed local ring is $\kappa[[u,v,w]]/(uv,uw,vw)$, which maps injectively to $\kappa[[u]]\times\kappa[[v]]\times\kappa[[w]]$ with image the triples with a common constant term. In both cases the image is exactly the set of compatible families.
\end{proof}

\subsection{The exact generation theorem}\label{subsec:exact_generation}

\begin{theorem}[Exact generation on a carrier]\label{thm:exact_generation}
In the setting of Section~\textup{\ref{subsec:gen_setting}}, suppose that
\begin{equation}\label{eq:gen_hyp}
M\ge0,\qquad M>2g-2,\qquad n_{\min}>3M+4a .
\end{equation}
Then the restriction map $H^0(Y,\LL)\to H^0(R,\LL|_R)$ is surjective, and
\begin{equation}\label{eq:exact_generation}
S_B\cap k_0^{\Phi}=\sum_{\ell\subset R}V_\ell ,
\end{equation}
where $k_0^\Phi$ is identified with the arrays supported on $\Phi$. In words: an array supported on the footprint of $J$ lies in the sum code if and only if it is a sum of codewords placed on the fibres contained in $J$. No reducedness or transversality assumption is made on the pair divisors or their intersections.
\end{theorem}

The proof occupies the rest of the section. We work over $\kappa$ and descend at the end. The strategy is to extend sections of $\LL$ from $R$ to $J$ and then to $Y$. The second extension is Corollary~\ref{cor:carrier_vanishing}. The first is delicate at the crossings of the cylinders, and it is carried out in two steps, through an auxiliary curve $K$ that contains the pairwise intersections $C_{ij}$ with a canonical scheme structure.

\paragraph{The curve $K$.}
At a point of $Y$ let $x,y,z$ be local equations of $\Pi_1,\Pi_2,\Pi_3$; by Lemma~\ref{lem:config}(b) they are pairwise coprime in the regular local ring $\OO_{Y,P}$, which is a unique factorisation domain. Define the ideal sheaf $\II_K$ by
\begin{equation}\label{eq:IK}
\II_K:=(xy,\,xz,\,yz)\quad\text{locally},\qquad\text{that is,}\qquad \II_K=\OO_Y(-\Pi_1-\Pi_2)+\OO_Y(-\Pi_1-\Pi_3)+\OO_Y(-\Pi_2-\Pi_3).
\end{equation}
Its zero set is $C_{12}\cup C_{13}\cup C_{23}$, and $K$ denotes the closed subscheme it defines (possibly empty).

\begin{lemma}[$K$ is Cohen--Macaulay; the ideal of $K$ in $J$]\label{lem:K_CM}
\begin{enumerate}[label=\textup{(\alph*)}]
\item $K$ is a Cohen--Macaulay curve.
\item Multiplication by the pair products gives an isomorphism of $\OO_J$-modules
\begin{equation}\label{eq:IKJ}
\OO_{\Pi_1}(-\Pi_2-\Pi_3)\oplus\OO_{\Pi_2}(-\Pi_1-\Pi_3)\oplus\OO_{\Pi_3}(-\Pi_1-\Pi_2)\;\xrightarrow{\ \sim\ }\;\II_{K,J}:=\II_K/\II_J ,
\end{equation}
and, with $\cF$ any line bundle on $Y$, an isomorphism
\begin{equation}\label{eq:IRKJ}
\bigoplus_{i=1}^3\II_{R_i,\Pi_i}\,\cF(-\Pi_j-\Pi_k)|_{\Pi_i}\;\xrightarrow{\ \sim\ }\;\II_{R\cup K,J}\,\cF ,\qquad\{i,j,k\}=\{1,2,3\},
\end{equation}
where $R\cup K$ is the scheme-theoretic union, with ideal $\II_R\cap\II_K$.
\item The following colon identities hold locally: $(\II_K:x)=(y,z)$ and $\bigl((x,yz):y\bigr)=(x,z)$.
\end{enumerate}
\end{lemma}

\begin{proof}
(a) The ideal $(xy,xz,yz)$ is the ideal of $2\times2$ minors of the $3\times2$ matrix with rows $(x,0)$, $(0,y)$, $(-z,z)$, and it has height two: its generators have no common prime factor since $x,y,z$ are pairwise coprime, and it is contained in the height-two prime ideals of the components of the $C_{ij}$. In a Cohen--Macaulay local ring height equals grade, so the Hilbert--Burch theorem~\cite[Theorem~20.15]{Eisenbud95} gives a resolution $0\to\OO^2\to\OO^3\to\II_K\to0$ and shows that $\OO_K$ is Cohen--Macaulay of codimension two, that is, a Cohen--Macaulay curve.

(b) The map $\OO_Y(-\Pi_2-\Pi_3)\hookrightarrow\OO_Y\to\OO_J$ has image $(yz)\cdot\OO_J$ and kernel $\{f\cdot yz: f\cdot yz\in(xyz)\}=\OO_Y(-J)$, because $x\mid f$ whenever $x\mid f\cdot yz$. It therefore factors through an injection $\OO_{\Pi_1}(-\Pi_2-\Pi_3)=\OO_Y(-\Pi_2-\Pi_3)/\OO_Y(-J)\hookrightarrow\OO_J$, and similarly for the other two summands; the three images generate $\II_{K,J}$. The sum is direct: if $\alpha yz+\beta xz+\gamma xy\in(xyz)$, then reducing modulo $x$ gives $x\mid\alpha yz$, hence $x\mid\alpha$ by coprimeness, so $\alpha yz\in(xyz)$; likewise $\beta xz,\gamma xy\in(xyz)$. This proves~\eqref{eq:IKJ}. For~\eqref{eq:IRKJ}, a local section $\alpha yz+\beta xz+\gamma xy$ of $\II_{K,J}$ vanishes on $R$ if and only if it vanishes on each fibre $\ell\subset R$. On a fibre $\ell\subset R_1\subset\Pi_1$ we have $x|_\ell=0$, so the section restricts to $\alpha|_\ell\cdot(yz)|_\ell$, and $(yz)|_\ell$ is a nonzero function on the integral curve $\ell$ by Lemma~\ref{lem:config}(a), hence a nonzerodivisor; so the section vanishes on $\ell$ if and only if $\alpha|_\ell=0$. Running over the fibres of $R_1$, and treating $R_2$, $R_3$ in the same way, the sections of $\II_{K,J}$ vanishing on $R$ are exactly those with $\alpha\in\II_{R_1,\Pi_1}$, $\beta\in\II_{R_2,\Pi_2}$, $\gamma\in\II_{R_3,\Pi_3}$; conversely such a section vanishes on $R_1$ because $\alpha$ does and $x$ does, and on $R_2,R_3$ likewise. Since $R$ is reduced, vanishing on $R$ is membership in $\II_R$, and $\II_{R\cup K}=\II_R\cap\II_K$. Tensoring with $\cF$ gives~\eqref{eq:IRKJ}.

(c) If $tx=\alpha xy+\beta xz+\gamma yz$, then $x\mid\gamma$, so $t\in(y,z)$; the reverse inclusion is clear. If $ty=\alpha x+\beta yz$, then $y\mid\alpha$, so $t-\beta z\in(x)$ and $t\in(x,z)$; the reverse inclusion is clear.
\end{proof}

\paragraph{Step 1: extension from $R\cup K$ to $J$.}

\begin{lemma}\label{lem:step1}
Under~\eqref{eq:gen_hyp}, $H^1(J,\II_{R\cup K,J}\LL)=0$. Consequently $H^0(J,\LL|_J)\to H^0(R\cup K,\LL|_{R\cup K})$ is surjective.
\end{lemma}

\begin{proof}
By~\eqref{eq:IRKJ} it suffices to show that $H^1\bigl(\Pi_i,\II_{R_i,\Pi_i}\LL(-\Pi_j-\Pi_k)|_{\Pi_i}\bigr)=0$ for each $i$; take $i=1$, $(j,k)=(2,3)$, so that $\Pi_1=D_{12}\times\Xj3$. The subscheme $R_1\subset\Pi_1$ is $Z\times\Xj3$, where $Z\subset D_{12}$ is the reduced set of pair-grid points of $E'_1\times E'_2$ lying on $D_{12}$, so $\II_{R_1,\Pi_1}=\II_{Z,D_{12}}\boxtimes\OO_{\Xj3}$. The line bundle $\LL(-\Pi_2-\Pi_3)$ restricted to $\Pi_1$ is a box product: its $(1,2)$-part is $\cF_{12}:=\bigl(\LL_1\otimes\cF^{(13)}_1{}^{-1}\bigr)\boxtimes\bigl(\LL_2\otimes\cF^{(23)}_2{}^{-1}\bigr)$ restricted to $D_{12}$, since $\Pi_2=p_{13}^*D_{13}$ contributes the factor $\cF^{(13)}_1{}^{-1}$ in direction $1$ and $\Pi_3=p_{23}^*D_{23}$ the factor $\cF^{(23)}_2{}^{-1}$ in direction $2$; and its $\Xj3$-part is $\cG_3:=\LL_3\otimes\cF^{(13)}_3{}^{-1}\otimes\cF^{(23)}_3{}^{-1}$, of degree at least $2g-2+n_3-M-2a>2g-2$, hence with $H^1(\Xj3,\cG_3)=0$. By the K\"unneth formula on $D_{12}\times\Xj3$,
\[
H^1\bigl(\Pi_1,(\II_{Z,D_{12}}\cF_{12})\boxtimes\cG_3\bigr)=H^1(D_{12},\II_{Z,D_{12}}\cF_{12})\otimes H^0(\Xj3,\cG_3)\;\oplus\;H^0(D_{12},\II_{Z,D_{12}}\cF_{12})\otimes H^1(\Xj3,\cG_3),
\]
so it remains to prove $H^1(D_{12},\II_{Z,D_{12}}\cF_{12})=0$, and for this we apply Lemma~\ref{lem:fs_adjunction} on the surface $\Xj1\times\Xj2$ to the Gorenstein curve $D_{12}$. Let $V\le D_{12}$ be a nonzero effective subdivisor with coordinate degrees $u=d_1(V)$ and $v=d_2(V)$; since $V\le D_{12}$ and $D_{12}$ has no axis components, $0\le u,v\le a$. Hodge index on the product surface gives $V^2\le2uv$: with $f_1,f_2$ the classes of the two kinds of axes, the class $W:=V-vf_1-uf_2$ satisfies $W\cdot f_1=W\cdot f_2=0$, hence $W\cdot(f_1+f_2)=0$ with $f_1+f_2$ ample, so $W^2\le0$, and $W^2=V^2-2uv$. Adjunction gives $2p_a(V)-2=V^2+(2g-2)(u+v)\le2uv+(2g-2)(u+v)\le a(u+v)+(2g-2)(u+v)$, using $u,v\le a$. The reduced cluster $Z\cap V$ consists of points of $V$ with first coordinate in $E'_1$; over each such coordinate the fibre of $V\to\Xj1$ is a finite scheme of length $u$, so $\length(Z\cap V)\le n_1u$, and likewise $\le n_2v$, hence $\length(Z\cap V)\le\min(n_1u,n_2v)\le\tfrac12(n_1u+n_2v)$. Finally $\deg(\cF_{12}|_V)\ge(2g-2+n_1-M-a)u+(2g-2+n_2-M-a)v$ by Lemma~\ref{lem:box_degree}. Therefore
\begin{multline*}
\deg(\cF_{12}|_V)-(2p_a(V)-2)-\length(Z\cap V)\;\ge\;\Bigl(\frac{n_1}2-M-2a\Bigr)u+\Bigl(\frac{n_2}2-M-2a\Bigr)v\\
\ge\;\Bigl(\frac{n_{\min}}2-M-2a\Bigr)(u+v)>0,
\end{multline*}
since $n_{\min}>3M+4a\ge2M+4a$ and $u+v>0$. Lemma~\ref{lem:fs_adjunction} applies. The consequence follows from~\eqref{eq:restriction_sequence} on $J$.
\end{proof}

\paragraph{Step 2: extension from $T=R\cap K$ to $K$.}
Let $T:=R\cap K$ be the scheme-theoretic intersection, with ideal $\II_T=\II_R+\II_K$; it is a finite subscheme of $K$, since $R$ meets each $C_{ij}$ in finitely many points by Lemma~\ref{lem:config}(b). Write $H:=\II_R$, $\mathcal P:=H+\II_K$ and $\mathcal P_A:=H+(x,yz)$ locally.

\begin{lemma}[Filtration of $\II_{T,K}\LL$]\label{lem:filtration}
There is a filtration $0\subseteq\cG_1\subseteq\cG_2\subseteq\II_{T,K}\LL$ by subsheaves with
\begin{equation}\label{eq:layers}
\cG_1\cong\II_{Z_{23},C_{23}}\LL(-\Pi_1)|_{C_{23}},\qquad \cG_2/\cG_1\cong\II_{Z_{13},C_{13}}\LL(-\Pi_2)|_{C_{13}},\qquad \II_{T,K}\LL/\cG_2\cong\II_{T_{12},C_{12}}\LL|_{C_{12}},
\end{equation}
where $Z_{23}\subset C_{23}$, $Z_{13}\subset C_{13}$ and $T_{12}\subset C_{12}$ are the finite subschemes defined locally by the ideals
\begin{equation}\label{eq:layer_ideals}
\frac{(\mathcal P:x)}{(y,z)},\qquad \frac{(\mathcal P_A:y)}{(x,z)},\qquad \frac{H+(x,y)}{(x,y)}
\end{equation}
respectively, and these ideals contain the images of $(q_3,q_1q_2)$, $(q_2,q_1q_3)$ and $(q_1q_2,q_1q_3,q_2q_3)$ respectively.
\end{lemma}

\begin{proof}
Let $A_1\subset Y$ be the subscheme with ideal $(x,yz)=\OO_Y(-\Pi_1)+\OO_Y(-\Pi_2-\Pi_3)$, the intersection of $\Pi_1$ with $\Pi_2+\Pi_3$. By Lemma~\ref{lem:K_CM}(c), the kernel of $\OO_K\to\OO_{A_1}$ is $(x,yz)/\II_K=x\cdot\OO/(\II_K:x)=x\cdot\OO_{C_{23}}$, and the kernel of $\OO_{A_1}\to\OO_{C_{12}}$ is $(x,y)/(x,yz)=y\cdot\OO/((x,yz):y)=y\cdot\OO_{C_{13}}$; globally, multiplication by the equations of $\Pi_1$ and $\Pi_2$ gives exact sequences
\begin{equation}\label{eq:OK_filtration}
0\to\OO_{C_{23}}(-\Pi_1)\to\OO_K\to\OO_{A_1}\to0,\qquad 0\to\OO_{C_{13}}(-\Pi_2)\to\OO_{A_1}\to\OO_{C_{12}}\to0 .
\end{equation}
Now intersect the ideal $\II_{T,K}=\mathcal P/\II_K\subseteq\OO_K$ with the two kernels. First, $\II_{T,K}\cap\bigl((x)+\II_K\bigr)/\II_K=\bigl((\mathcal P\cap(x))+\II_K\bigr)/\II_K\cong(\mathcal P\cap(x))/(\II_K\cap(x))=x(\mathcal P:x)/x(\II_K:x)$, and multiplication by $x$ identifies this with $(\mathcal P:x)/(y,z)$, an ideal of $\OO_{C_{23}}$, twisted by $-\Pi_1$. This is $\cG_1$, and the quotient $\II_{T,K}/\cG_1$ is the image of $\mathcal P$ in $\OO_{A_1}$, namely $\mathcal P_A/(x,yz)$ since $\mathcal P+(x,yz)=H+(x,yz)=\mathcal P_A$. Second, intersecting with the kernel of $\OO_{A_1}\to\OO_{C_{12}}$ gives $(\mathcal P_A\cap(y))/((x,yz)\cap(y))=y(\mathcal P_A:y)/y(x,z)\cong(\mathcal P_A:y)/(x,z)$, twisted by $-\Pi_2$; this is $\cG_2/\cG_1$, and the final quotient is the image of $\mathcal P_A$ in $\OO_{C_{12}}$, namely $(H+(x,y))/(x,y)$. These are exact kernel computations in the local rings and glue to sheaves on $K$, $A_1$ and the $C_{ij}$; tensoring with the line bundle $\LL$ preserves exactness.

For the containments: $t\in(H:x)$ means that $tx$ vanishes on $R$, which is automatic on $R_1\subset\Pi_1$ and, on a fibre $\ell$ of $R_2$ or $R_3$, is equivalent to $t|_\ell=0$ because $x|_\ell$ is a nonzero function on the integral curve $\ell$ (Lemma~\ref{lem:config}(a)); as $R$ is reduced, $(H:x)=\II_{R_2\cup R_3}$. The function $q_3$ vanishes on every fibre of $R_2$ (direction $2$, third coordinate in $E'_3$) and of $R_3$ (direction $1$, third coordinate in $E'_3$), and $q_1q_2$ vanishes on $R_2$ (first coordinate in $E'_1$) and on $R_3$ (second coordinate in $E'_2$); hence $(q_3,q_1q_2)\subseteq\II_{R_2\cup R_3}=(H:x)\subseteq(\mathcal P:x)$. Likewise $(H:y)=\II_{R_1\cup R_3}\supseteq(q_2,q_1q_3)$ and $(\mathcal P_A:y)\supseteq(H:y)$. Finally $H\supseteq\mathfrak j=(q_1q_2,q_1q_3,q_2q_3)$ by Lemma~\ref{lem:grid_ideal}. The first two layer ideals contain the Cartier-section ideals $(q_3)$ and $(q_2)$, respectively, and the third contains $(q_1q_2)$; these sections are nonzerodivisors on the respective curves by Lemma~\ref{lem:config}(b). Thus the subschemes $Z_{23},Z_{13},T_{12}$ are contained in finite Cartier sections and are finite.
\end{proof}

\begin{lemma}[Lengths of the layer clusters]\label{lem:layer_lengths}
Let $V$ be a CM subcurve of $C_{23}$, $C_{13}$ or $C_{12}$, let $d_j:=d_j(V)$ be its projection degrees, and put $D_V:=d_1+d_2+d_3$ and $S_V:=n_1d_1+n_2d_2+n_3d_3$. Then
\begin{equation}\label{eq:layer_lengths}
\length(Z_{23}\cap V)\le\tfrac12S_V,\qquad \length(Z_{13}\cap V)\le\tfrac12S_V,\qquad \length(T_{12}\cap V)\le\tfrac23S_V .
\end{equation}
\end{lemma}

\begin{proof}
By Lemma~\ref{lem:config}(b) every coordinate projection of $V$ is finite, so no component of $V$ lies in a surface $p_j^{-1}(P)$, and since $V$ has no embedded points each $q_j$ is a nonzerodivisor on $\OO_V$. The subscheme of $V$ cut out by $q_j$ is therefore an effective Cartier divisor on $V$ of length $\deg(p_j^*\OO(E'_j)|_V)=n_jd_j$ (Lemma~\ref{lem:box_degree}), and the one cut out by $q_jq_k$ has length $n_jd_j+n_kd_k$. Since the ideal of $Z_{23}$ contains $q_3$ and $q_1q_2$, the subscheme $Z_{23}\cap V$ is contained in each of these Cartier divisors, so $\length(Z_{23}\cap V)\le\min(n_3d_3,\,n_1d_1+n_2d_2)\le\tfrac12S_V$. The other two bounds are proved in the same way: $\length(Z_{13}\cap V)\le\min(n_2d_2,n_1d_1+n_3d_3)$ and $\length(T_{12}\cap V)\le\min_{j<k}(n_jd_j+n_kd_k)\le\tfrac23S_V$, the latter because the three sums $n_jd_j+n_kd_k$ add up to $2S_V$.
\end{proof}

\begin{lemma}[Genus bound for subcurves of the $C_{ij}$]\label{lem:genus_Cij}
For every nonzero CM subcurve $V$ of $C_{12}$, $C_{13}$ or $C_{23}$, with $D_V$ as above and $d_{\max}:=\max_jd_j(V)$,
\begin{equation}\label{eq:genus_Cij}
2p_a(V)-2\le(2g-2)D_V+2a\,(D_V-d_{\max}).
\end{equation}
\end{lemma}

\begin{proof}
For $r\in\{1,2,3\}$ choose a box product $\AA^{(r)}=\AA^{(r)}_1\boxtimes\AA^{(r)}_2\boxtimes\AA^{(r)}_3$ with $\deg\AA^{(r)}_r=g-1$ and $\deg\AA^{(r)}_j=g-1+a$ for $j\ne r$, such that every factor $\AA^{(r)}_j$ and every twist $\AA^{(r)}_j\otimes\NN$ of degree at least $g-1$, where $\NN$ runs over the finitely many line bundles $\cF^{(jk)}_j{}^{-1}$ and their pairwise products that occur as $j$-th factors of $\OO(-\Pi_i)$ and $\OO(-\Pi_i-\Pi_{i'})$, is nonspecial. This is possible because, for each such $\NN$, the bundles $\AA$ with $\AA\otimes\NN$ special form a proper closed subset of the Picard variety (Remark~\ref{rem:line_bundle_notation}), and finitely many proper closed subsets do not cover it. We claim that $H^1(C_{ij},\AA^{(r)}|_{C_{ij}})=0$. By Lemma~\ref{lem:koszul} for the complete intersection $C_{ij}=\Pi_i\cap\Pi_j$ it suffices that $H^1(Y,\AA^{(r)})=0$, $H^2(Y,\AA^{(r)}(-\Pi_i))=H^2(Y,\AA^{(r)}(-\Pi_j))=0$ and $H^3(Y,\AA^{(r)}(-\Pi_i-\Pi_j))=0$. All these bundles are box products. The first vanishing holds by Lemma~\ref{lem:kunneth_vanishing}(a), all factors being nonspecial. In $\AA^{(r)}(-\Pi_i)$ the two directions of the pair of $\Pi_i$ are twisted, each by a bundle of degree at least $-a$; an enhanced factor, of degree $g-1+a$, stays at degree at least $g-1$ and is nonspecial by the choice, and the untwisted factors are nonspecial; so at most the factor in direction $r$, when $r$ belongs to the pair, can be special, and Lemma~\ref{lem:kunneth_vanishing}(b) gives the $H^2$-vanishing. In $\AA^{(r)}(-\Pi_i-\Pi_j)$ the direction shared by the two cylinders is twisted twice and the other two directions once; of the latter at least one differs from $r$, so it is an enhanced factor twisted by a bundle of degree at least $-a$, hence nonspecial, and Lemma~\ref{lem:kunneth_vanishing}(b) gives $H^3=0$. This proves the claim. The vanishing passes to every subcurve $V\subseteq C_{ij}$: the kernel of $\OO_{C_{ij}}\to\OO_V$ is supported in dimension at most one, so its $H^2$ vanishes and $H^1(C_{ij},\AA^{(r)})\to H^1(V,\AA^{(r)}|_V)$ is surjective. Riemann--Roch on $V$ then gives $0\le h^0(\AA^{(r)}|_V)=\deg(\AA^{(r)}|_V)+1-p_a(V)$, that is, $p_a(V)-1\le(g-1)D_V+a(D_V-d_r)$ by Lemma~\ref{lem:box_degree}. Choosing $r$ with $d_r=d_{\max}$ and doubling gives~\eqref{eq:genus_Cij}.
\end{proof}

\begin{lemma}\label{lem:step2}
Under~\eqref{eq:gen_hyp}, $H^1(K,\II_{T,K}\LL)=0$. Consequently $H^0(K,\LL|_K)\to H^0(T,\LL|_T)$ is surjective.
\end{lemma}

\begin{proof}
By Lemma~\ref{lem:filtration} and the long exact cohomology sequences, it suffices to show that the three layers in~\eqref{eq:layers} have vanishing $H^1$, and for each we apply Lemma~\ref{lem:fs_adjunction} to the Gorenstein curve $C_{23}$, $C_{13}$ or $C_{12}$ (Lemma~\ref{lem:config}(b)). Let $V$ be a nonzero CM subcurve of the relevant curve, with $d_j$, $D_V$, $S_V$, $d_{\max}$ as above; note $S_V\ge n_{\min}D_V$ and $D_V>0$.

\emph{First layer} ($\cF=\LL(-\Pi_1)|_{C_{23}}$, cluster $Z_{23}$). The cylinder $\Pi_1=p_{12}^*D_{12}$ has external class with factor degrees at most $a$ in directions $1,2$ and $0$ in direction $3$, so $\Pi_1\cdot V\le a(d_1+d_2)=a(D_V-d_3)$, and $\deg(\LL|_V)=\sum_j(2g-2+n_j-M)d_j=(2g-2)D_V+S_V-MD_V$. With~\eqref{eq:genus_Cij} and~\eqref{eq:layer_lengths},
\begin{multline*}
\deg(\cF|_V)-(2p_a(V)-2)-\length(Z_{23}\cap V)\;\ge\;S_V-MD_V-a(D_V-d_3)-2a(D_V-d_{\max})-\tfrac12S_V\\
\ge\;\tfrac12S_V-(M+2a)D_V ,
\end{multline*}
where the last step uses $d_3+2d_{\max}\ge D_V$. This is at least $(\tfrac12n_{\min}-M-2a)D_V>0$.

\emph{Second layer} ($\cF=\LL(-\Pi_2)|_{C_{13}}$, cluster $Z_{13}$). Identical, with $\Pi_2=p_{13}^*D_{13}$, $\Pi_2\cdot V\le a(D_V-d_2)$ and $d_2+2d_{\max}\ge D_V$.

\emph{Third layer} ($\cF=\LL|_{C_{12}}$, cluster $T_{12}$). Here
\begin{multline*}
\deg(\cF|_V)-(2p_a(V)-2)-\length(T_{12}\cap V)\;\ge\;S_V-MD_V-2a(D_V-d_{\max})-\tfrac23S_V\\
\ge\;\tfrac13S_V-MD_V-\tfrac43aD_V ,
\end{multline*}
using $d_{\max}\ge D_V/3$; this is at least $\tfrac13(n_{\min}-3M-4a)D_V>0$.

In all three cases~\eqref{eq:gen1} holds, so $H^1=0$. The consequence follows from~\eqref{eq:restriction_sequence} on $K$.
\end{proof}

\paragraph{Step 3: assembly and descent.}

\begin{proof}[Proof of Theorem~\ref{thm:exact_generation}]
\emph{Surjectivity.} Let $\tau\in H^0(R,\LL|_R)$. By Lemma~\ref{lem:step2} there is $\tau_K\in H^0(K,\LL|_K)$ with $\tau_K|_T=\tau|_T$. The scheme-theoretic union $R\cup K$ sits in the exact sequence $0\to\OO_{R\cup K}\to\OO_R\oplus\OO_K\to\OO_T\to0$, obtained from $\II_R\cap\II_K$, $\II_R$, $\II_K$ and $\II_R+\II_K$; tensoring with $\LL$ and taking global sections shows that the pair $(\tau,\tau_K)$ comes from a section $\tau'\in H^0(R\cup K,\LL|_{R\cup K})$. By Lemma~\ref{lem:step1}, $\tau'$ is the restriction of a section $\tau''\in H^0(J,\LL|_J)$, and since $\LL(-J)$ is a box product with factor degrees at least $2g-2+n_{\min}-M-2a>2g-2$ (Lemma~\ref{lem:config}(c) and~\eqref{eq:gen_hyp}), Lemma~\ref{lem:kunneth_vanishing}(a) gives $H^1(Y,\LL(-J))=0$, so by~\eqref{eq:restriction_sequence} $\tau''$ is the restriction of a global section of $\LL$. Hence $H^0(Y,\LL)\to H^0(R,\LL|_R)$ is surjective; if $K=\varnothing$ the middle step is void.

\emph{The code identity.} Put $A_\Phi:=\sum_{\ell\subset R}V_\ell\subseteq\kappa^\Phi$ and let $z\in\kappa^\Phi$. For a fibre $\ell\subset R$ in direction $j$ through the grid point coordinates $(b,c)$ in the other two directions, $z$ is orthogonal to $V_\ell$ if and only if the restriction of $z$ to the grid line of $\ell$ is a parity check of $C_j$, that is, by Lemma~\ref{lem:dual_evaluation_bundle}(1) and $M>2g-2$, the residue vector of a unique differential $\omega_\ell\in H^0(X,\LL_j)$. Now $\LL|_\ell\cong\LL_j$ canonically: the two factors of $\LL$ in the fixed directions are $\omega_X(E'_k-G)$ evaluated at $b$ and at $c$, points of $E'_k$ where these bundles have the residue trivialisation of Appendix~\ref{subsec:appendix_trivialisation}. In this identification the value of $\omega_\ell$, viewed as a section of $\LL|_\ell$, at a grid point $P$ of $\ell$ is $z_P$ by~\eqref{eq:residue_value}. Hence $z\in A_\Phi^\perp$ if and only if the family $(\omega_\ell)_{\ell\subset R}$ consists of sections of $\LL|_\ell$ whose values at every grid point are the entries of $z$; at a crossing point $P$ of two fibres the values of the two sections are both $z_P$, so the family is compatible and, by Lemma~\ref{lem:gluing}, defines a section $\tau_z\in H^0(R,\LL|_R)$ with $\tau_z(P)=z_P$ for all $P\in\Phi$. Conversely every section of $\LL|_R$ arises in this way from its values on $\Phi$, because a section of $\LL|_\ell$ is determined by its residue vector. By the surjectivity just proved, $\tau_z$ is the restriction of a global section $\sigma\in H^0(Y,\LL)$, whose values $(\sigma(P))_{P\in E'_1\times E'_2\times E'_3}$ form a parity check of $S_B$ by Lemma~\ref{lem:dual_evaluation_bundle}(2), and $\sigma(P)=z_P$ on $\Phi$. Thus $A_\Phi^\perp$ is the set of restrictions to $\Phi$ of parity checks of $S_B$: $A_\Phi^\perp=\res_\Phi(S_B^\perp)$, where the orthogonal complements are taken in $\kappa^\Phi$ (the inclusion $\supseteq$ holds because the restriction of a parity check of $S_B$ to $\Phi$ is orthogonal to every $V_\ell\subseteq S_B$). Taking orthogonal complements once more, $A_\Phi=\{z\in\kappa^\Phi:\langle z,\res_\Phi\sigma\rangle=0\ \forall\sigma\in S_B^\perp\}=S_B\cap\kappa^\Phi$, since for $z$ supported on $\Phi$ the pairing with $\sigma$ only sees $\res_\Phi\sigma$. This is~\eqref{eq:exact_generation} over $\kappa$.

\emph{Descent.} The pair divisors may have been chosen over an extension of $k_0$, but the fibres of $R$ are $k_0$-rational fibres, the codes $C_j$ are defined over $k_0$, and both sides of~\eqref{eq:exact_generation} over $\kappa$ are the $\kappa$-spans of the corresponding $k_0$-spaces. By Lemma~\ref{lem:scalars}, a $k_0$-array in $S_B\cap k_0^\Phi$ that is a $\kappa$-linear combination of coded lines on the fibres of $R$ is a $k_0$-linear combination of such coded lines. Hence~\eqref{eq:exact_generation} holds over $k_0$.
\end{proof}

\begin{remark}[Shared components and unequal multiplicities] \label{rem:shared_components}
The proof makes no use of transversality, and the three equations $x,y,z$ need not form a regular sequence. In $\kappa[[u,v,w]]$ take $x=(u-v)^p$, $y=(u-w)^q$, $z=(v-w)^r$ with $p,q,r>0$: the three cylinders share the diagonal curve, $\II_K=(xy,xz,yz)$ defines a generically non-Gorenstein curve for $p=q=r=1$ (its generic Artinian ring $\kappa[[\alpha,\beta]]/(\alpha,\beta)^2$ has a two-dimensional socle), and the layer clusters have lengths $q+r-1$, $r+\max(p-q,0)$ and $p+q+\min(p,q)-1$, which add up to $2(p+q+r)-2=\length(T)$ for $H=(uv,uw,vw)$. The filtration of Lemma~\ref{lem:filtration} therefore retains the full nilpotent structure of $T$, and the criterion of Lemma~\ref{lem:fs_adjunction} is only ever applied to the complete-intersection curves $C_{ij}$ and their CM subcurves, never to $K$ itself. These calculations are valid in every characteristic, including characteristic $3$.
\end{remark}

\section{Support localisation}\label{sec:localization}

This section proves that the support of a light word of $S_3(C_i)$ lies, after the deletion of few coordinates, on the footprint of a carrier whose pair divisors have small factor degrees. The precise statement is Proposition~\ref{prop:localization}. The argument has four stages. The lines of the grid carrying more than half of their entries in the support are few, and a first carrier absorbs them. Multiplying by the equation of that carrier produces a word $y$ of a shifted sum code all of whose lines are light. The support of $y$ lies on the fixed divisor $\Delta$ of a large linear system, and $\Delta$ has tiny degree on the coordinate axes. Finally, the grid points of such a divisor are covered by few lines, which a second carrier absorbs. Throughout, a floor $i\ge4$ is fixed, so that $N=N_i>10^6$, and $g<N/26$, $M=\lfloor N/5\rfloor$, $C=C_i$.
\subsection{Fields and gonality}\label{subsec:fields}

\begin{definition}[Gonality over a field]\label{def:gonality}
For a field $K\supseteq k_0$ put $X_K:=X\times_{k_0}K$. The \emph{gonality} $\gamma_K(X)$ is the least degree of a nonconstant $K$-morphism $X_K\to\PP^1_K$, equivalently the least degree of a nonconstant element of the function field $K(X_K)$.
\end{definition}

\begin{lemma}[Gonality bounds]\label{lem:gonality}
\begin{enumerate}[label=\textup{(\alph*)}]
\item $\gamma_{k_0}(X)\ge(N+1)/730\ge N/730$.
\item Let $V$ be a smooth, geometrically integral $k_0$-variety with a $k_0$-rational point, and $K:=k_0(V)$ its function field. Then $\gamma_K(X)=\gamma_{k_0}(X)$.
\end{enumerate}
\end{lemma}

\begin{proof}
(a) A $k_0$-morphism $X\to\PP^1$ of degree $d$ maps the $N+1$ rational points $E\cup\{P_\infty\}$ of $X$ into the $730$ rational points of $\PP^1_{k_0}$, and each fibre contains at most $d$ points; hence $N+1\le730d$.

(b) The inequality $\gamma_K(X)\le\gamma_{k_0}(X)$ holds by base change. For the reverse inequality, fix $d\ge1$. We construct a proper parameter scheme of pencils of degree $d$. Since $X$ has a rational point, the Picard variety $\Pic^d(X)$ represents the line bundles of degree $d$ and carries a Poincar\'e bundle $\mathcal P$ on $X\times\Pic^d(X)$~\cite{Kleiman05}. Let $T_0:=(2g-1)P_\infty$. For every line bundle $\mathcal L$ of degree $d$ the bundle $\mathcal L(T_0)$ has degree $d+2g-1>2g-2$, so by~\eqref{eq:rr_exact} and cohomology and base change, $\mathcal V:=\pi_*\bigl(\mathcal P\otimes\pi_X^*\OO(T_0)\bigr)$ is a vector bundle of rank $d+g$ on $\Pic^d(X)$, whose fibre at $\mathcal L$ is $H^0(X,\mathcal L(T_0))$. Let $\mathrm{Gr}\to\Pic^d(X)$ be the Grassmannian bundle of two-dimensional subspaces of $\mathcal V$, a proper $k_0$-scheme. Let $\Sigma_d\subseteq\mathrm{Gr}$ be the closed subscheme of those subspaces on which the bundle map $\mathcal V\to\pi_*\bigl(\mathcal P\otimes\pi_X^*\OO(T_0)|_{T_0\times\Pic^d}\bigr)$ vanishes. It is closed because the vanishing of a map on the tautological subbundle is a closed condition, and it is proper. A $K$-point of $\Sigma_d$ is a pair $(\mathcal L,W)$ with $\mathcal L$ a line bundle of degree $d$ on $X_K$ and $W\subseteq H^0(X_K,\mathcal L(T_0))$ a two-dimensional space of sections vanishing on $T_0$, that is, a two-dimensional subspace of $H^0(X_K,\mathcal L)$.

Now let $f:X_K\to\PP^1_K$ be a nonconstant $K$-morphism of degree $d$. Then $\mathcal L:=f^*\OO(1)$ has degree $d$ and $W:=f^*\langle x_0,x_1\rangle$ is a two-dimensional subspace of $H^0(X_K,\mathcal L)$, so $\Sigma_d(K)\ne\varnothing$. The $K$-point gives a rational map $V\dashrightarrow\Sigma_d$ defined over $k_0$. Since $V$ has a smooth $k_0$-point and $\Sigma_d$ is proper, the Lang--Nishimura lemma~\cite[Theorem~3.6.11]{Poonen17} gives $\Sigma_d(k_0)\ne\varnothing$. A $k_0$-point $(\mathcal L_0,W_0)$ gives two linearly independent sections of $\mathcal L_0$ over $k_0$, hence a nonconstant rational map $X\dashrightarrow\PP^1_{k_0}$. After removal of the base locus of $W_0$ this is a morphism of degree at most $d$. Hence $\gamma_{k_0}(X)\le d$.
\end{proof}
\paragraph{The working field.}
The auxiliary bundles and sections needed over the working field in this section and in Section~\ref{sec:carriers} are chosen \emph{generically}. A line bundle of degree $e$ is taken at the generic point of $\Pic^e(X)$, and a section with algebraically independent coefficients in a basis of its space of allowed sections over the current field. If the current field is $K_1=k_0(V_1)$, adjoining a generic point of $\Pic^e(X)_{K_1}$ gives the function field of $V_1\times\Pic^e(X)$, and adjoining generic section coefficients gives a purely transcendental extension. Equivalently, over an open parameter locus on which a space of sections is a vector bundle, choose a generic trivialisation. Thus the field $K$ remains the function field of a finite product $V$ of Picard varieties of $X$ and affine spaces over $k_0$. Each such $V$ is smooth and geometrically integral and has a $k_0$-rational point: $\OO_X(eP_\infty)$ supplies one on each $\Pic^e(X)$. The open loci imposing genericity need not themselves contain a $k_0$-point. Lemma~\ref{lem:gonality} applies to the smooth model $V$ and gives
\begin{equation}\label{eq:gonality_K}
\gamma_K(X)\ge N/730 .
\end{equation}

A generic Picard point avoids every proper closed subset defined over the current field, including the base changes of the exceptional loci defined over $k_0$ in Lemmas~\ref{lem:MDSbundle} and~\ref{lem:short_jets}. Fix an algebraic closure $\Omega$ of the current $K$. Dimension counts, cohomology, geometric components and subcurve inequalities may be checked after base change to $\Omega$, and auxiliary points and sections used only in those checks need not be adjoined to $K$. The removal of axis components in Proposition~\ref{prop:carrier} is defined over $K$ by its coefficient base ideals. The fixed divisor $\Delta$ below is also defined over $K$, and its formation commutes with base change, as shown in Section~\ref{subsec:fixed_divisor}. Thus Lemma~\ref{lem:grid_cover} will be applied over $K$ with~\eqref{eq:gonality_K}. No corresponding gonality lower bound over $\Omega$ is asserted. The original input array and codes remain defined over $k_0$, whereas the shifted array and auxiliary codes may be defined only over $K$. Lemma~\ref{lem:scalars} is applied to the final decompositions of the original array in the original scalar-extended code spaces.
\begin{lemma}[Uniform short jets]\label{lem:short_jets}
Let $s\ge2$, $a=g+s-1$ and $t=\lfloor s/2\rfloor$. There is a nonempty open subset of $\Pic^a(X)$ defined over $k_0$ such that every line bundle $\AA$ represented by a $K$-point of this open subset satisfies $h^0(X_K,\AA)=s$ and $H^1(X_\Omega,\AA_\Omega(-Z))=0$ for every effective divisor $Z$ on $X_\Omega$ of degree at most $t$, nonreduced divisors included. Consequently $h^0(X_\Omega,\AA_\Omega(-Z))=s-\deg Z$ for $\deg Z\le t$, $h^0(X_\Omega,\AA_\Omega(-Z))\le s-t$ for every effective $Z$ of degree at least $t$, and $\AA$ is base point free.
\end{lemma}

\begin{proof}
Work over $\Omega$ and suppress base-change subscripts. The exceptional locus described below is defined over $k_0$. If $H^1(\AA(-Z))\ne0$ for some effective $Z$ of degree $t$, then by Serre duality $\omega_X\otimes\AA^{-1}(Z)$ has a nonzero section, so $\AA\cong\omega_X(Z-F)$ for an effective divisor $F$ of degree $2g-2-a+t=g-s-1+t$. If this number is negative there is no such $\AA$. Otherwise all such $\AA$ lie in the image of the morphism $X^{(t)}\times X^{(g-s-1+t)}\to\Pic^a(X)$, $(Z,F)\mapsto\omega_X(Z-F)$. This image is a closed subset defined over $k_0$ of dimension at most $g-s-1+2t<g$, because $2t\le s$. For $\AA$ outside this image and $Z$ of degree $t$, the vanishing holds. For $Z$ of smaller degree, enlarge $Z$ to a divisor $Z'\ge Z$ of degree $t$ and use the surjection $H^1(\AA(-Z'))\to H^1(\AA(-Z))$. Riemann--Roch then gives $h^0(\AA(-Z))=a-\deg Z+1-g=s-\deg Z$ for $\deg Z\le t$, including $h^0(\AA)=s$ for $Z=0$. For $\deg Z\ge t$, choose $Z'\le Z$ of degree $t$, so that $h^0(\AA(-Z))\le h^0(\AA(-Z'))=s-t$. Base point freeness is the case $Z=P$, a point, which has degree $1\le t$ since $s\ge2$: $h^0(\AA(-P))=s-1<h^0(\AA)$.
\end{proof}

\subsection{Heavy lines, the first carrier and the shifted word}\label{subsec:heavy_lines}

Fix a real number $\lambda$ with $0<\lambda\le7\cdot10^{-14}$ and a nonzero array $x\in S_3(C)$ with
\begin{equation}\label{eq:light_hyp}
w:=\wt(x)\le\lambda N^3 .
\end{equation}
Call a grid line \emph{heavy} if it contains more than $N/2$ points of $\supp(x)$. The total number of incidences between points of $\supp(x)$ and lines in the three directions is $3w$, so the number $m_1$ of heavy lines satisfies $m_1\cdot N/2<3w$, that is,
\begin{equation}\label{eq:m1}
m_1<6w/N\le6\lambda N^2 .
\end{equation}
Put
\begin{equation}\label{eq:u1}
u_1:=\lfloor\sqrt{m_1}\rfloor+1,\qquad b_1:=g+u_1-1,\qquad\text{so that}\qquad u_1^2>m_1\quad\text{and}\quad u_1<\sqrt{6\lambda}\,N+1<\frac{N}{10^5},
\end{equation}
the last inequality because $\lambda\le7\cdot10^{-14}$ and $N>10^6$. Apply Proposition~\ref{prop:carrier} with $E_j^0=E$, with the heavy lines as prescribed family ($m=m_1$) and with $s=u_1$. It produces retained sets $B^{(1)}_j\subseteq E$ with fewer than $2u_1$ indices deleted in total, a box $B_1:=B^{(1)}_1\times B^{(1)}_2\times B^{(1)}_3$ of minimal side length
\begin{equation}\label{eq:nstar}
n_*:=\min_j|B^{(1)}_j|>N-2u_1 ,
\end{equation}
and a carrier $J_1$ with pair divisors $D^{(1)}_{jk}$ of external class with factor degrees at most $b_1$; by Proposition~\ref{prop:carrier}, $\OO_Y(J_1)\cong\LL_1\boxtimes\LL_2\boxtimes\LL_3$ with $0\le\deg\LL_j\le2b_1$. Let $s_1\in H^0(Y,\OO_Y(J_1))$ be the canonical section, with divisor $J_1$. By Remark~\ref{rem:shifted_codes} choose divisors $G'_j$ on $X$, disjoint from $E$, with $\OO_X(MP_\infty)\otimes\LL_j\cong\OO_X(G'_j)$, and fix these isomorphisms; then
\begin{equation}\label{eq:Mprime}
\deg G'_j\le M':=M+2b_1 .
\end{equation}
Let $C'_j:=C_L(B^{(1)}_j,G'_j)$ be the shifted punctured codes, and define the \emph{shifted word}
\begin{equation}\label{eq:y_def}
y:=s_1\cdot x_{B_1},\qquad y_P:=s_1(P)\,x_P\quad\text{for }P\in B_1 ,
\end{equation}
where $s_1(P)$ is the value of $s_1$ at the grid point $P$ in the fixed trivialisations.

\begin{lemma}[The shifted word]\label{lem:shifted_word}
\begin{enumerate}[label=\textup{(\alph*)}]
\item $y\in S(C'_1,C'_2,C'_3)$, and $\wt(y)\le w$.
\item Every grid line of $B_1$ contains at most $N/2$ points of $T:=\supp(y)$.
\item If $P\in B_1$ and $x_P\ne0$, then either $P$ lies on the footprint of $J_1$ or $P\in T$.
\end{enumerate}
\end{lemma}

\begin{proof}
(a) By the K\"unneth formula, $s_1=\sum_\alpha f^{(1)}_\alpha\otimes f^{(2)}_\alpha\otimes f^{(3)}_\alpha$ with $f^{(j)}_\alpha\in H^0(X,\LL_j)$. The restriction $x_{B_1}$ lies in the sum code of the punctured triple $(C|_{B^{(1)}_j})_j$; write $x_{B_1}=u_1+u_2+u_3$ accordingly. A line of $(f^{(1)}_\alpha\otimes f^{(2)}_\alpha\otimes f^{(3)}_\alpha)\cdot u_1$ in direction $1$, through the coordinates $(b,c)$ in directions $2,3$, equals $f^{(2)}_\alpha(b)f^{(3)}_\alpha(c)\cdot\bigl(f^{(1)}_\alpha(P)\,h(P)\bigr)_{P\in B^{(1)}_1}$ for some $h\in L(MP_\infty)$, and $f^{(1)}_\alpha h$ is a section of $\OO(MP_\infty)\otimes\LL_1\cong\OO(G'_1)$, so this line lies in $C'_1$. Hence $s_1\cdot u_1$ has all its direction-$1$ lines in $C'_1$, and similarly for the other directions; summing over $\alpha$ and over the three directions gives $y\in S(C'_1,C'_2,C'_3)$. The weight bound is clear. (b) and (c): a heavy line of $x$ whose two fixed coordinates were retained survives in $B_1$, and by Proposition~\ref{prop:carrier}(ii) its fibre lies in $J_1$, so $s_1$ vanishes on it and the corresponding line of $y$ is zero; a line that is not heavy contains at most $N/2$ points of $\supp(x)\supseteq\supp(y)$. For (c), $x_P\ne0$ and $P\notin J_1$ give $s_1(P)\ne0$ and $y_P\ne0$.
\end{proof}

If $y=0$ the second carrier below is empty and Proposition~\ref{prop:localization} holds with $u_2=1$; from now on assume $y\ne0$, so that $T\ne\varnothing$. By Lemma~\ref{lem:dual_evaluation_bundle}, applied with $E'_j=B^{(1)}_j$ and $G_j=G'_j$, which is legitimate because $\deg G'_j\ge M>2g-2$, the parity checks of $S(C'_1,C'_2,C'_3)$ are the values of the sections of
\begin{equation}\label{eq:Lambda}
\Lambda:=\boxprod\omega_{\Xj{j}}\bigl(B^{(1)}_j-G'_j\bigr),\qquad \deg\Lambda_j=2g-2+|B^{(1)}_j|-\deg G'_j\ge2g-2+n_*-M' ,
\end{equation}
where $\Lambda_j$ denotes the $j$-th factor.

\subsection{The fixed divisor of the auxiliary system}\label{subsec:fixed_divisor}

Put
\begin{equation}\label{eq:aux_params}
s:=\lceil N/1000\rceil,\qquad a:=g+s-1,\qquad t:=\lfloor s/2\rfloor,
\end{equation}
and choose, over the working field, line bundles $\AA_1,\AA_2,\AA_3$ of degree $a$ satisfying Lemma~\ref{lem:short_jets}; put $\AA:=\AA_1\boxtimes\AA_2\boxtimes\AA_3$. Since $g<N/26$ and $s\le N/1000+1$,
\begin{equation}\label{eq:a_bound}
a<\frac{N}{26}+\frac{N}{1000}=\frac{513}{13000}N .
\end{equation}
Let
\begin{equation}\label{eq:W_def}
W:=\bigl\{\sigma\in H^0(Y,\AA):\sigma(P)=0\ \text{for all }P\in T\bigr\},\qquad r:=\codim W=\dim H^0(Y,\AA)-\dim W ,
\end{equation}
and let $\mathsf B\subseteq K^T$ be the image of the evaluation map $H^0(Y_K,\AA)\to K^T$, $\sigma\mapsto(\sigma(P))_{P\in T}$, a code of length $|T|$ and dimension $r$ over the working field $K$. Then
\begin{equation}\label{eq:trivial_r}
r=\dim\mathsf B\le|T|=\wt(y)\le w ,
\end{equation}
and this trivial bound is all that the proof uses.

\begin{remark}[A sharper bound on the number of conditions]\label{rem:few_conditions}
The bound $r\le|T|$ of~\eqref{eq:trivial_r} can be improved to $r<|T|/18$, as follows. Since each $\AA_j$ is base point free, $\mathsf B$ has full support. Put $d_{18}:=n_*-M'-18a$ and suppose $d_{18}>18$. Let $\mathsf B^{\langle18\rangle}\subseteq K^T$ be the span of the $18$-fold Schur products of elements of $\mathsf B$. A product of $18$ sections of $\AA$ is a section of $\AA^{\otimes18}$, a box product with factor degrees $18a$, and by the argument of Lemma~\ref{lem:shifted_word}(a), applied to $y\in S(C'_1,C'_2,C'_3)$ and such a section, the array $y\cdot(\sigma_1\cdots\sigma_{18})|_T$ lies in the sum code $S(C''_1,C''_2,C''_3)$ of the codes $C''_j:=C_L(B^{(1)}_j,G''_j)$ with $\OO(G''_j)\cong\OO(G'_j)\otimes\AA_j^{\otimes18}$ and $\deg G''_j\le M'+18a$. By linearity, $y\cdot v\in S(C''_1,C''_2,C''_3)$ for every $v\in\mathsf B^{\langle18\rangle}$. The minimum distance of a sum code is at least the minimum distance of its factor codes: if $0\ne z$ lies in $S(C''_1,C''_2,C''_3)$ and $H_j$ is a parity-check map of $C''_j$, then $(H_1\otimes H_2\otimes H_3)z=0$. If $z$ were supported on a set whose projections to the three axes each had fewer than $\min_jd(C''_j)$ elements, each $H_j$ would be injective on the coordinates of that projection, their tensor product would be injective on the arrays supported on the product of the projections, and $z$ would vanish. Since $d(C''_j)\ge|B^{(1)}_j|-\deg G''_j\ge n_*-M'-18a=d_{18}$ by Proposition~\ref{prop:ag_parameters}, every nonzero array of $S(C''_1,C''_2,C''_3)$ has weight at least $d_{18}$. As $y$ has full support on $T$, multiplication by $y$ preserves weights, so every nonzero vector of $\mathsf B^{\langle18\rangle}$ has weight at least $d_{18}$. The product Singleton bound of Randriambololona~\cite[Theorem~2 and Remark~3(a)]{Randriam13}, applied with $t=18$ to the code $\mathsf B$ of full support (its proof uses only parity-check matrices, basis exchange and a pigeonhole count, and is valid over any field), provides codewords $c_1,\dots,c_{18}\in\mathsf B$ with $1\le\wt(c_1*\cdots*c_{18})\le\max\{17,\,|T|-18(r-1)\}$. The product is a nonzero element of $\mathsf B^{\langle18\rangle}$, so its weight is at least $d_{18}>18>17$, whence $|T|-18(r-1)\ge d_{18}>18$ and $r<|T|/18$. The factor $18$ improves the density threshold $\lambda$ of Proposition~\ref{prop:localization} by a factor of about $14$ and the constant $\rho$ of Corollary~\ref{cor:explicit_certificate} by a factor of about $18$, at the price of an external input and of the condition $d_{18}>18$, which at $\beta=\tfrac1{26}$ requires $\theta+20\beta<1-\tfrac{18}{1000}$ and is the reason the sharper bound would force $\ell\ge27$. Since the constants of this paper are not optimised, we do not use the sharper bound. The argument of this remark also shows that every nonzero word of a sum code has weight at least the smallest minimum distance of its three factor codes, which is used in Remark~\ref{rem:first_floors}.
\end{remark}

Let $\Delta$ be the \emph{fixed divisor} of the linear system $|W|$: the largest effective divisor on $Y_K$ along which every section of $W$ vanishes, with multiplicities. Here $W\ne0$ because $\dim W=s^3-r>0$. Its definition over $K$ and compatibility with geometric base change can be seen directly from the base ideal
\[
\mathfrak b:=\im\bigl(W\otimes_K\AA^{-1}\longrightarrow\OO_{Y_K}\bigr).
\]
Since $Y_K$ is regular, the codimension-one part of $\mathfrak b$ is an effective Cartier divisor $\Delta$ defined over $K$, and
\[
\mathfrak b=\OO_{Y_K}(-\Delta)\mathfrak b'
\]
for an ideal $\mathfrak b'$ whose zero locus has codimension at least two. The base ideal commutes with extension of the ground field, and the zero locus of $\mathfrak b'_\Omega$ still has codimension at least two. Consequently $\Delta_\Omega$ is exactly the fixed divisor of $W\otimes_K\Omega$, including all geometric multiplicities. In particular, removing the fixed divisor introduces no algebraic extension of $K$.

Let $\delta_j:=\Delta\cdot\ell$ for a fibre curve $\ell$ in direction $j$, the \emph{axis degree} of $\Delta$ in direction $j$, which does not depend on $\ell$ because all fibre curves in one direction are algebraically equivalent, and put $\delta:=\max_j\delta_j$.

\begin{lemma}[Small axis degrees]\label{lem:axis_degrees}
$s^2\min(\delta_j,t)\le r$ for $j=1,2,3$. In particular, if $r<ts^2$, then $\delta\le r/s^2$.
\end{lemma}

\begin{proof}
Base change to $\Omega$ for this proof. The dimensions $s,r$ and the axis degrees are unchanged. The information points chosen below are used only for this geometric verification and do not enlarge the arithmetic working field $K$. Fix $j$ and let $k,l$ be the other two directions. For $(p,q)\in\Xj{k}\times\Xj{l}$ let $\ell_{p,q}$ be the fibre curve in direction $j$ over $(p,q)$. The pairs with $\ell_{p,q}\subseteq\Delta$ form a proper closed subset of $\Xj{k}\times\Xj{l}$: a surface component of $\Delta$ contains the fibres over a curve of $\Xj{k}\times\Xj{l}$ if it is a pullback from the pair surface, and over a proper closed subset otherwise. Hence we may choose $s$ points $p_1,\dots,p_s$ of $\Xj{k}$ such that evaluation at them is an isomorphism on $H^0(\AA_k)$ (an information set, which exists because $h^0(\AA_k)=s$ and evaluation at any set of more than $a$ points is injective, and whose complement in $X^s$ is a proper closed subset) and such that for each $p_\alpha$ only finitely many $q$ give a fibre $\ell_{\alpha\beta}$ over $(p_\alpha,q)$ contained in $\Delta$. Then choose $s$ points $q_1,\dots,q_s$ of $\Xj{l}$ avoiding these finitely many exceptions and forming an information set for $H^0(\AA_l)$. Let $\ell_{\alpha\beta}$ be the fibre in direction $j$ over $(p_\alpha,q_\beta)$. Writing a section $\sigma\in H^0(Y,\AA)=H^0(\AA_j)\otimes H^0(\AA_k)\otimes H^0(\AA_l)$ in the bases of $H^0(\AA_k)$ and $H^0(\AA_l)$ dual to the two information sets, its restriction to $\ell_{\alpha\beta}$ is the coefficient of the basis tensor indexed by $(\alpha,\beta)$, a section of $\AA_j\cong\AA|_{\ell_{\alpha\beta}}$. Hence restriction $H^0(Y,\AA)\to\bigoplus_{\alpha,\beta}H^0(\ell_{\alpha\beta},\AA_j)$ is injective. Every section of $W$ vanishes along $\Delta$, so its restriction to $\ell_{\alpha\beta}$ vanishes on the effective divisor $\Delta|_{\ell_{\alpha\beta}}$, which has degree $\delta_j$ because $\ell_{\alpha\beta}\not\subseteq\Delta$. By Lemma~\ref{lem:short_jets}, the space of such sections has dimension at most $s-\min(\delta_j,t)$. Therefore $\dim W=s^3-r\le s^2\bigl(s-\min(\delta_j,t)\bigr)$, which is the assertion.
\end{proof}

\subsection{Interpolation on the residual curve}\label{subsec:residual_curve}

In this subsection work over $\Omega$, suppressing base-change subscripts, and use the scalar extension of $W$. The constructions here serve only to prove $T\subseteq\Delta$. They do not enlarge the arithmetic working field $K$. Put $\MM:=\AA(-\Delta)$. The sections of $W$ divided by the canonical section of $\OO_Y(\Delta)$ form a subspace $W'\subseteq H^0(Y,\MM)$ of dimension $\dim W$, whose members have no common divisorial component. Suppose $\dim W\ge2$. Choose two general sections $\sigma_1,\sigma_2\in W'$. Their zero loci have no common divisorial component (the sections of $W'$ vanishing on a given component of $Z(\sigma_1)$ form a proper subspace), so
\begin{equation}\label{eq:Cy}
C_y:=Z(\sigma_1)\cap Z(\sigma_2)
\end{equation}
is either empty or a complete-intersection curve in $Y$, Gorenstein in the latter case by Lemma~\ref{lem:koszul}. In either case it contains the reduced set
\begin{equation}\label{eq:U_def}
U:=T\setminus\Delta
\end{equation}
of points of $T$ not on $\Delta$, since every section of $W$ vanishes on $T$ and division by the equation of $\Delta$ does not affect vanishing at points off $\Delta$. If $C_y=\varnothing$, then $U=\varnothing$ immediately. For the subcurve arguments below we may therefore assume $C_y\ne\varnothing$.

\begin{lemma}[Genus bound on the residual curve]\label{lem:genus_Cy}
For every nonzero CM subcurve $V\subseteq C_y$, with $D_V:=\sum_jd_j(V)$,
\begin{equation}\label{eq:genus_Cy}
2p_a(V)-2\le(2g-2)D_V+4aD_V .
\end{equation}
\end{lemma}

\begin{proof}
Choose line bundles $\Theta_1,\Theta_2,\Theta_3$ of degree $g-1$ on $X$ with $H^0=H^1=0$ (a general bundle of degree $g-1$ lies outside $W_{g-1}$), and put $\cF:=\boxprod(\AA_j^{\otimes2}\otimes\Theta_j)$, a box product with factor degrees $2a+g-1$. By Lemma~\ref{lem:koszul} with $\MM_1=\MM_2=\MM$, $H^1(C_y,\cF|_{C_y})=0$ provided $H^1(Y,\cF)=0$, $H^2(Y,\cF\otimes\MM^{-1})=0$ and $H^3(Y,\cF\otimes\MM^{-2})=0$. The first holds by Lemma~\ref{lem:kunneth_vanishing}(a). For the second, $\cF\otimes\MM^{-1}=\cF\otimes\AA^{-1}\otimes\OO_Y(\Delta)$ and $\cF\otimes\AA^{-1}=\boxprod(\AA_j\otimes\Theta_j)=\omega_Y\otimes\boxprod\bigl(\AA_j\otimes\Theta_j\otimes\omega_X^{-1}\bigr)$, a twist of $\omega_Y$ by a box product with factor degrees $a+g-1-(2g-2)=s>0$; Lemma~\ref{lem:ambient_vanishing_D} applies with $D=\Delta$. For the third, Serre duality gives $H^3(Y,\cF\otimes\MM^{-2})^\vee=H^0\bigl(Y,\omega_Y\otimes\cF^{-1}\otimes\AA^{\otimes2}(-2\Delta)\bigr)$, and $\omega_Y\otimes\cF^{-1}\otimes\AA^{\otimes2}=\boxprod(\omega_X\otimes\Theta_j^{-1})$ has no sections, since $H^0(X,\omega_X\otimes\Theta_j^{-1})=H^1(X,\Theta_j)^\vee=0$; twisting down by the effective divisor $2\Delta$ only shrinks the space of sections. Hence $H^1(C_y,\cF|_{C_y})=0$, and the vanishing passes to every subcurve $V$ as in Lemma~\ref{lem:genus_Cij}. Riemann--Roch on $V$ gives $p_a(V)-1\le\deg(\cF|_V)=(2a+g-1)D_V$ by Lemma~\ref{lem:box_degree}.
\end{proof}

\begin{lemma}[Intersection with the fixed divisor and with the support]\label{lem:DV_UV}
Let $V\subseteq C_y$ be a nonzero CM subcurve and decompose its cycle according to the number of nonconstant projections of each component: let $d_A$, $d_P$ and $d_T$ be the sums, with multiplicities, of the projection degrees of the components with exactly one, two and three nonconstant projections respectively, so that $D_V=d_A+d_P+d_T$. Then
\begin{equation}\label{eq:DV_UV}
\Delta\cdot V\le\delta\,d_A+a\,d_P+2a\,d_T,\qquad \#(U\cap V)\le\frac N2\,d_A+\frac N2\,d_P+\frac N3\,d_T .
\end{equation}
\end{lemma}

\begin{proof}
Both quantities are additive over the cycle, up to the inequality $\#(U\cap V)\le\sum_\Sigma m_\Sigma\#(U\cap\Sigma)$, so it suffices to treat an integral component $\Sigma$ with the three possible types.

\emph{Axis type.} $\Sigma$ is a fibre curve in some direction $j$, and $\Delta\cdot\Sigma=\delta_j\le\delta$ by definition of the axis degrees, whether or not $\Sigma\subseteq\Delta$. If $\Sigma$ is not an evaluation coordinate fibre of $B_1$ it contains no grid point; if it is, it contains at most $N/2$ points of $T\supseteq U$ by Lemma~\ref{lem:shifted_word}(b).

\emph{Grid points on the other types.} If $d_k(\Sigma)>0$, then for each of the at most $N$ values $c\in B^{(1)}_k$ the fibre of $\Sigma\to\Xj{k}$ over $c$ is finite of cardinality at most $d_k(\Sigma)$, so $\#(U\cap\Sigma)\le N\min\{d_k(\Sigma):d_k(\Sigma)>0\}$. For the plane type this minimum is at most $\tfrac12(d_k+d_l)=\tfrac12D_\Sigma$, and for the threefold type at most $\tfrac13D_\Sigma$.

\emph{Intersection with $\Delta$: the nef bundles.} The linear system $W'$ on $\MM$ has no fixed divisor, so its base locus has dimension at most one. A general fibre surface $S_{jk}=\{b\}\times\Xj{j}\times\Xj{k}$ (general $b$) meets this base locus in finitely many points, so the restriction of $W'$ to $S_{jk}$ has no fixed curve and $\MM|_{S_{jk}}$ is nef by Lemma~\ref{lem:nef_system}; as all fibre surfaces in a given pair of directions are algebraically equivalent, $\MM|_{S}$ is nef for \emph{every} fibre surface $S$ in those directions, intersection numbers being invariant under algebraic equivalence. Choose a general base point $(P_1,P_2,P_3)$, so that the three fibre surfaces through it are general, and let $\MM_j$ and $\MM_{jk}$ be the restrictions of $\MM$ to the fibre curves and fibre surfaces through it, as in Proposition~\ref{prop:cube}; thus $\deg\MM_j=a-\delta_j\le a$ and each $\MM_{jk}$ is nef. By~\eqref{eq:cube}, $\MM\otimes\bigotimes_jp_j^*\MM_j$ is a tensor product of pullbacks of nef bundles, hence nef, so for every subcurve $V$,
\begin{equation}\label{eq:M_lower}
\MM\cdot V\ge-\sum_j(a-\delta_j)\,d_j(V)\ge-a\,D_V .
\end{equation}

\emph{Plane type.} $\Sigma$ lies in a fibre surface $S$ in directions $k,l$, say. On $S$ the bundle $\MM|_S$ is nef, so $\Delta\cdot\Sigma=\AA\cdot\Sigma-\MM\cdot\Sigma\le\AA\cdot\Sigma=a(d_k+d_l)=aD_\Sigma$ by Lemma~\ref{lem:box_degree}.

\emph{Threefold type.} By~\eqref{eq:M_lower}, $\Delta\cdot\Sigma=\AA\cdot\Sigma-\MM\cdot\Sigma\le aD_\Sigma+aD_\Sigma=2aD_\Sigma$.
\end{proof}

\begin{lemma}[Support on the fixed divisor]\label{lem:support_on_Delta}
Suppose that $\dim W\ge2$ and that
\begin{equation}\label{eq:mu_positive}
\mu_A:=n_*-M'-\tfrac N2-4a-\delta>0,\qquad \mu_P:=n_*-M'-\tfrac N2-5a>0,\qquad \mu_T:=n_*-M'-\tfrac N3-6a>0,
\end{equation}
and that $n_*-M'-2a>0$. Then $U=\varnothing$, that is, $T=\supp(y)\subseteq\Delta$.
\end{lemma}

\begin{proof}
If $C_y=\varnothing$, then $U\subseteq C_y$ is empty and there is nothing to prove. Assume $C_y\ne\varnothing$. Put $\QQ:=\Lambda(-\Delta)=\Lambda\otimes\OO_Y(-\Delta)$. We first show $H^1(C_y,\II_{U,C_y}\QQ|_{C_y})=0$ by Lemma~\ref{lem:fs_adjunction}. Let $V\subseteq C_y$ be a nonzero CM subcurve, with $d_A,d_P,d_T$ as in Lemma~\ref{lem:DV_UV}. By~\eqref{eq:Lambda} and Lemma~\ref{lem:box_degree}, $\deg(\Lambda|_V)\ge(2g-2+n_*-M')D_V$, so by Lemmas~\ref{lem:genus_Cy} and~\ref{lem:DV_UV}
\begin{align*}
\deg(\QQ|_V)-(2p_a(V)-2)-\#(U\cap V)
&\ge(n_*-M')D_V-4aD_V-\Delta\cdot V-\#(U\cap V)\\
&\ge\mu_A\,d_A+\mu_P\,d_P+\mu_T\,d_T>0 ,
\end{align*}
since $D_V>0$. Hence~\eqref{eq:gen1} holds for the reduced finite subscheme $U\cap V$, and $H^0(C_y,\QQ|_{C_y})\to H^0(U,\QQ|_U)$ is surjective.

Next, $H^0(Y,\QQ)\to H^0(C_y,\QQ|_{C_y})$ is surjective by Lemma~\ref{lem:koszul} with $\MM_1=\MM_2=\MM$: the required groups are $H^1(Y,\QQ\otimes\MM^{-1})$ and $H^2(Y,\QQ\otimes\MM^{-2})$. Now $\QQ\otimes\MM^{-1}=\Lambda\otimes\AA^{-1}$ is a box product with factor degrees at least $2g-2+n_*-M'-a>2g-2$, so the first group vanishes by Lemma~\ref{lem:kunneth_vanishing}(a); and $\QQ\otimes\MM^{-2}=\Lambda\otimes\AA^{-2}\otimes\OO_Y(\Delta)=\omega_Y\otimes\cF\otimes\OO_Y(\Delta)$ with $\cF$ the box product with factor degrees $|B^{(1)}_j|-\deg G'_j-2a\ge n_*-M'-2a>0$, so the second vanishes by Lemma~\ref{lem:ambient_vanishing_D}.

Consequently, for every $P_0\in U$ there is a global section $q\in H^0(Y,\QQ)$ with $q(P_0)\ne0$ and $q(P)=0$ for all $P\in U\setminus\{P_0\}$. Let $s_\Delta$ be the canonical section of $\OO_Y(\Delta)$. Then $s_\Delta q\in H^0(Y,\Lambda)$, and by Lemma~\ref{lem:dual_evaluation_bundle}(2) its values form a parity check of $S(C'_1,C'_2,C'_3)$, which contains $y$. Hence
\[
0=\sum_{P\in T}y_P\,s_\Delta(P)\,q(P)=y_{P_0}\,s_\Delta(P_0)\,q(P_0),
\]
because $s_\Delta$ vanishes at the points of $T\cap\Delta$ and $q$ at the points of $U\setminus\{P_0\}$. All three factors on the right are nonzero: $y_{P_0}\ne0$ as $P_0\in T$, $s_\Delta(P_0)\ne0$ as $P_0\notin\Delta$, and $q(P_0)\ne0$ by construction. This contradiction shows $U=\varnothing$. Since $T$ consists of $K$-rational grid points and $\Delta$ is defined over $K$, the inclusion $T\subseteq\Delta_\Omega$ proved here implies $T\subseteq\Delta$ over $K$.
\end{proof}

\subsection{Covering the grid points of a divisor of small axis degrees}\label{subsec:grid_cover}

We now work over the arithmetic field $K$. In the application, the divisor is the $K$-defined fixed divisor $\Delta$ above, and the gonality in the following lemma is $\gamma_K(X)$, bounded below by~\eqref{eq:gonality_K}.

\begin{lemma}[Grid cover]\label{lem:grid_cover}
Let $K$ be a field over which $X$ has gonality $\gamma_K(X)$, let $B'=E'_1\times E'_2\times E'_3$ be a box of evaluation points with side lengths $n_j\ge3N/4$, and let $D$ be an effective divisor on $Y_K$ with axis degrees $\delta_1,\delta_2,\delta_3$ satisfying $\max_j\delta_j<N/4$ and $2\delta_3<\gamma_K(X)$. Then the grid points of $D$ in $B'$ are covered by at most
\begin{equation}\label{eq:grid_cover}
N\bigl(\delta_1+\delta_2+2\delta_3+\min(\delta_1,\delta_2)\bigr)\le5N\max_j\delta_j
\end{equation}
grid lines of $B'$.
\end{lemma}

\begin{proof}
\emph{Evaluation planes.} For $P\in E'_j$ let $c_{j,P}$ be the multiplicity of the surface $p_j^{-1}(P)$ in $D$, and let $D'$ be $D$ with these components removed. Each such surface has intersection number $1$ with the fibre curves in direction $j$ and $0$ with the others, so $D'$ has axis degrees $\delta'_j=\delta_j-\sum_{P\in E'_j}c_{j,P}\le\delta_j$, and the grid points of $p_j^{-1}(P)$ are covered by $n_k\le N$ grid lines. The bound~\eqref{eq:grid_cover} is monotone in each $\delta_j$ with slope at least $N$, so it suffices to prove it for $D'$, and we assume from now on that $D$ contains no surface $p_j^{-1}(P)$ with $P\in E'_j$.

\emph{Bad pairs.} For $(P,Q)\in E'_1\times E'_2$ let $\ell_{P,Q}:=\{P\}\times\{Q\}\times\Xj3$, and call $(P,Q)$ \emph{bad} if $\ell_{P,Q}\subseteq D$. For fixed $P$ the surface $p_1^{-1}(P)\cong\Xj2\times\Xj3$ is not a component of $D$, so $D|_{p_1^{-1}(P)}$ is an effective divisor on it whose degree on the axes $\Xj2\times\{e\}$ is $\delta_2$; each bad $(P,Q)$ contributes the axis $\{Q\}\times\Xj3$, of intersection number $1$ with those axes, as a component, so there are at most $\delta_2$ bad pairs with first coordinate $P$, and at most $N\delta_2$ in total; symmetrically at most $N\delta_1$. Cover the bad pairs by their lines $\ell_{P,Q}\cap B'$: at most $N\min(\delta_1,\delta_2)$ lines. Since $N\min(\delta_1,\delta_2)<N^2/4<n_1n_2$, some pair $(P_0,Q_0)$ is good.

\emph{Bad rows and columns.} The pairs $(P,Q_0)$ with $P\in E'_1$ that are bad correspond to axis components $\{P\}\times\Xj3$ of $D|_{p_2^{-1}(Q_0)}$, which has degree $\delta_1$ on the axes $\Xj1\times\{e\}$; so there are at most $\delta_1$ such $P$, and symmetrically at most $\delta_2$ values $Q$ with $(P_0,Q)$ bad. For each such $P$, cover all grid points of the surface $p_1^{-1}(P)$ by $n_2\le N$ lines, and likewise for each such $Q$: at most $N(\delta_1+\delta_2)$ lines.

\emph{The remaining pairs.} Let $(P,Q)$ be a pair not yet covered, so that $(P,Q)$, $(P,Q_0)$, $(P_0,Q)$ and $(P_0,Q_0)$ are all good. For a good pair the restriction $F_{P,Q}:=D|_{\ell_{P,Q}}$ is an effective $K$-divisor of degree $D\cdot\ell_{P,Q}=\delta_3$ on $\ell_{P,Q}\cong X_K$. By Proposition~\ref{prop:cube}, applied over $K$ with a rational base point $(P_0,Q_0,R_0)$,
\[
F_{P,Q}+F_{P_0,Q_0}\ \sim\ F_{P,Q_0}+F_{P_0,Q}
\]
as divisors on $X_K$. The isomorphism~\eqref{eq:cube_fibres} holds over $K$, as noted in the proof of that proposition, so this is a linear equivalence over $K$. Both sides are effective of degree $2\delta_3<\gamma_K(X)$. If they were different, their difference would be the divisor of a nonconstant $f\in K(X_K)$ with poles bounded by $F_{P,Q_0}+F_{P_0,Q}$, that is, a nonconstant $K$-morphism $X_K\to\PP^1_K$ of degree at most $2\delta_3$, contradicting the definition of $\gamma_K(X)$. Hence the two sides are equal, and in particular $\Supp(F_{P,Q})\subseteq\Supp(F_{P,Q_0})\cup\Supp(F_{P_0,Q})$. A grid point $(P,Q,R)$ of $D$ with $R\in E'_3$ therefore has $R\in\Supp(F_{P,Q_0})$ or $R\in\Supp(F_{P_0,Q})$. For each $P\in E'_1$, the divisor $F_{P,Q_0}$ of degree $\delta_3$ contains at most $\delta_3$ rational points $R$, and the grid lines $\{P\}\times E'_2\times\{R\}$ for these $R$ cover all grid points of the first kind: at most $N\delta_3$ lines. Symmetrically, the lines $E'_1\times\{Q\}\times\{R\}$ with $R\in\Supp(F_{P_0,Q})$, at most $N\delta_3$ in number, cover the grid points of the second kind. Adding the four counts gives~\eqref{eq:grid_cover}.
\end{proof}

\subsection{The second carrier and the localisation statement}\label{subsec:second_carrier}

\begin{proposition}[Support localisation]\label{prop:localization}
Let $i\ge4$, let $0<\lambda\le7\cdot10^{-14}$, and let $x\in S_3(C_i)$ be nonzero with $w=\wt(x)\le\lambda N^3$. Then there are integers $u_1,u_2\ge1$ with
\begin{equation}\label{eq:u_bounds}
u_1<\sqrt6\,\sqrt\lambda\,N+1,\qquad u_2\le1000\sqrt5\,\sqrt\lambda\,N+1,\qquad U_0:=u_1+u_2<2239\sqrt\lambda\,N+2<\frac{N}{1600},
\end{equation}
retained sets $B_j\subseteq E$ obtained by deleting in total fewer than $2U_0$ indices, with box $B=B_1\times B_2\times B_3$ of minimal side $n_{\min}>N-2U_0$, and, over an extension field, effective divisors $D_{jk}$ on the pair surfaces with no coordinate-axis components and external class with factor degrees at most
\begin{equation}\label{eq:b_def}
b:=b_1+b_2=2g+U_0-2 ,
\end{equation}
such that:
\begin{enumerate}[label=\textup{(\alph*)}]
\item $\supp(x_B)\subseteq\Phi$, the footprint on $B$ of the carrier $J:=\sum_{j<k}p_{jk}^*D_{jk}$;
\item every row and column of every retained pair grid $B_j\times B_k$ contains at most $U_0-2$ points of $D_{jk}$, so that every fibre of $R$ meets the fibres of the other two directions in at most $\nu:=2(U_0-2)$ points of $B$;
\item $M>2g-2$ and $n_{\min}>3M+4b$, so that Theorem~\textup{\ref{thm:exact_generation}} applies to $J$ on $B$;
\item all three retained sets are safe puncturings, and the pair $(C|_{B_j},C|_{B_k})$ and $(C|_{B_j},C)$ have expansion at least $\eta$ for all $j\ne k$.
\end{enumerate}
\end{proposition}

\begin{proof}
Construct $u_1$, $B_1$, $J_1$, $y$ and $T$ as in Section~\ref{subsec:heavy_lines}; the bound on $u_1$ is~\eqref{eq:u1}. If $y=0$, put $u_2:=1$, $J_2:=0$, $B:=B_1$ and $D_{jk}:=D^{(1)}_{jk}$. Then (a) holds by Lemma~\ref{lem:shifted_word}(c), (b) by Proposition~\ref{prop:carrier}(iii) with $s=u_1$, and the remaining assertions follow from the numerical verifications below, which only use $U_0<N/1600$. Assume $y\ne0$.

\emph{Numerical verification of the hypotheses of Section~\ref{subsec:fixed_divisor}--\ref{subsec:residual_curve}.} Recall $n_*>N-2u_1$, $M'=M+2g+2u_1-2$, $a\le g+N/1000$, $u_1<N/10^5$, $g<N/26$, $M\le N/5$. By~\eqref{eq:trivial_r}, $r\le|T|\le\lambda N^3$. Next, $t\ge s/2-\tfrac12\ge N/2000-\tfrac12$ and $s^2\ge N^2/10^6$, so $ts^2\ge\tfrac{N^3}{2\cdot10^9}\bigl(1-\tfrac{1000}{N}\bigr)>\tfrac{N^3}{3\cdot10^9}>\lambda N^3\ge r$; Lemma~\ref{lem:axis_degrees} gives
\begin{equation}\label{eq:delta_bound}
\delta\le\frac{r}{s^2}\le\frac{10^6|T|}{N^2}=10^6\,\frac{\wt(y)}{N^2}\le10^6\,\lambda N\le7\cdot10^{-8}N .
\end{equation}
Also $\dim W=s^3-r\ge N^3/10^9-\lambda N^3\ge2$. For~\eqref{eq:mu_positive}: with the same bounds,
\begin{align*}
\mu_A&>\bigl(1-\tfrac15-\tfrac12\bigr)N-6g-4s-4u_1+6-\delta\\
&>\bigl(\tfrac3{10}-\tfrac6{26}-\tfrac4{1000}\bigr)N-4u_1+2-\delta=\tfrac{106}{1625}N-4u_1+2-\delta,\\
\mu_P&>\tfrac3{10}N-7g-5s-4u_1+7>\bigl(\tfrac3{10}-\tfrac7{26}-\tfrac5{1000}\bigr)N-4u_1+2=\tfrac{67}{2600}N-4u_1+2,\\
\mu_T&>\bigl(1-\tfrac15-\tfrac13\bigr)N-8g-6s-4u_1+8\\
&>\bigl(\tfrac7{15}-\tfrac8{26}-\tfrac6{1000}\bigr)N-4u_1+2=\tfrac{2983}{19500}N-4u_1+2 ,
\end{align*}
and since $4u_1+\delta<N/20000$, each of $\mu_A,\mu_P,\mu_T$ exceeds $N/40$; the smallest coefficient is $\tfrac{67}{2600}>\tfrac1{40}+\tfrac1{20000}$. Finally $n_*-M'-2a>\tfrac45N-4g-4u_1-\tfrac{2N}{1000}+2>0$. Lemma~\ref{lem:support_on_Delta} therefore gives
\begin{equation}\label{eq:T_in_Delta}
\supp(y)=T\subseteq\Delta .
\end{equation}

\emph{The second carrier.} By~\eqref{eq:delta_bound} and~\eqref{eq:gonality_K}, $\delta<N/4$ and $2\delta<N/730\le\gamma_K(X)$ for the working field $K$, over which $\Delta$ is defined; and $n_*>N-2u_1\ge3N/4$. Lemma~\ref{lem:grid_cover} applied to $\Delta$ on the box $B_1$ gives a family of
\begin{equation}\label{eq:m2}
m_2\le5N\delta\le5\cdot10^6\,\frac{\wt(y)}{N}\le5\cdot10^6\,\lambda N^2
\end{equation}
grid lines of $B_1$ covering $T$. Put $u_2:=\lfloor\sqrt{m_2}\rfloor+1$ and $b_2:=g+u_2-1$; then $u_2^2>m_2$ and $u_2\le1000\sqrt5\,\sqrt\lambda N+1$, since $\sqrt{5\cdot10^6}=1000\sqrt5$. Apply Proposition~\ref{prop:carrier} with $E^0_j=B^{(1)}_j$, the covering lines as prescribed family and $s=u_2$ (note $u_2<N/1600\le|B^{(1)}_j|$). It produces retained sets $B_j\subseteq B^{(1)}_j$, with fewer than $2u_2$ further deletions, and a carrier $J_2$ with pair divisors $D^{(2)}_{jk}$ of external class with factor degrees at most $b_2$ and no axis components, such that every row and column of $B_j\times B_k$ contains at most $u_2-1$ points of $D^{(2)}_{jk}$. Put $D_{jk}:=D^{(1)}_{jk}+D^{(2)}_{jk}$ and $J:=J_1+J_2$. The bounds~\eqref{eq:u_bounds} follow from~\eqref{eq:u1}, the bound on $u_2$, $\sqrt6+1000\sqrt5<2239$, and $2239\sqrt\lambda+2/N<2239\cdot2.646\cdot10^{-7}+2\cdot10^{-6}<5.95\cdot10^{-4}<1/1600$.

(a) Let $P\in B$ with $x_P\ne0$. By Lemma~\ref{lem:shifted_word}(c), either $P$ lies on $J_1$, or $P\in T$. In the second case $P$ lies on one of the covering lines; its two fixed coordinates are coordinates of $P$, hence retained in $B$, so the line survives in $B$ and its fibre lies in $J_2$ by Proposition~\ref{prop:carrier}(ii). In both cases $P\in\Phi$.

(b) The points of $D_{jk}$ in a row or column of $B_j\times B_k$ are those of $D^{(1)}_{jk}$, at most $u_1-1$, and those of $D^{(2)}_{jk}$, at most $u_2-1$, hence at most $U_0-2$; the crossing bound follows as in the last step of the proof of Proposition~\ref{prop:carrier}.

(c) $M>2g-2$ by Proposition~\ref{prop:Ci_parameters}. The total number of deleted indices is less than $2u_1+2u_2=2U_0$, so $n_{\min}>N-2U_0$, and
\[
n_{\min}-3M-4b>N-2U_0-\tfrac{3N}5-8g-4U_0+8>\bigl(\tfrac25-\tfrac8{26}\bigr)N-6U_0+8=\tfrac6{65}N-6U_0+8>0 ,
\]
since $6U_0<\tfrac{6N}{1600}<\tfrac6{65}N$. The pair divisors have external class with factor degrees at most $b_1+b_2=b$ (the class of a sum is the product of the classes) and no axis components.

(d) Each $B_j$ has more than $N-2U_0>\tfrac{3N}4$ points, so Proposition~\ref{prop:pair} applies.
\end{proof}

\begin{remark}[Where the constants come from]\label{rem:constants}
The parameter $s\approx N/1000$ must be large enough that $ts^2\approx s^3/2$ exceeds $r\le|T|\le\lambda N^3$, which forces $\lambda\ll s^3/N^3$, and small enough that $4a,5a,6a$ fit in the margins $\mu_A,\mu_P,\mu_T$; the density threshold $\lambda\le7\cdot10^{-14}$ is then dictated by the requirement $U_0<N/1600$, through $u_2\approx\sqrt{5N\delta}$ and $\delta\le\lambda N^3/s^2$, and the floor condition $N>10^6$ by the additive terms. The number $1600$ is the cap on $\vartheta=U_0/N$ under which the inequalities of Theorem~\ref{thm:sharp_filling} are verified. None of these constants is optimised.
\end{remark}

\section{Proof of the main theorems}\label{sec:proof}

\subsection{Weighted filling on sparse fibres}\label{subsec:weighted_filling}

\begin{lemma}[Weighted sparse-fibre filling]\label{lem:weighted_filling}
Let $\Cj1,\Cj2,\Cj3$ be codes on a box with sides $n_1,n_2,n_3$, and let $x$ be a sum of coded lines placed on a family of grid lines (``selected fibres''). Assume that every nonzero codeword of every $\Cj{j}$ has weight at least $d$, and that every selected fibre meets the selected fibres of the other two directions in at most $\nu<d$ grid points. Then
\begin{equation}\label{eq:weighted_filling}
P(x)\le\Bigl(1+\frac{2\nu}{3(d-\nu)}\Bigr)\wt(x).
\end{equation}
\end{lemma}

\begin{proof}
Combine the coded lines lying on the same fibre into one and discard the fibres on which the sum is zero. Let $\mathcal L$ be the number of remaining fibres, $S$ the sum of the weights of the codewords on them, and $w:=\wt(x)$. For a grid point $P$ let $r_P\le3$ be the number of these codewords that are nonzero at $P$, and let $I:=\sum_{P:\,r_P\ge2}r_P$ be the number of active crossing incidences. A point with $r_P\le1$ contributes the same amount, $r_P$, to $S$ and to $w$; a point with $r_P\ge2$ contributes $r_P$ to $S$ and at most $1$ to $w$. Hence $0\le S-w\le I$. A point with $r_P\ge2$ is a crossing of two selected fibres with nonzero codewords, and each fibre has at most $\nu$ crossing points, so $I\le\nu\mathcal L$. Since $S\ge d\mathcal L$,
\begin{equation}\label{eq:L_bound}
d\mathcal L-w\le S-w\le\nu\mathcal L,\qquad\text{hence}\qquad \mathcal L\le\frac{w}{d-\nu}\quad\text{and}\quad I\le\frac{\nu w}{d-\nu} .
\end{equation}
Group the fibres by direction and write $x=u_1+u_2+u_3$ with $u_j\in U_j$. Each of
\[
G_1:=\wt(u_2+u_3)+\wt(u_1),\qquad G_2:=\wt(u_1+u_3)+\wt(u_2),\qquad G_3:=\wt(u_1+u_2)+\wt(u_3)
\]
is the weight of an admissible pair filling ($y_{23}=u_2+u_3$, $y_{12}=u_1$, $y_{13}=0$, and so on), so $P(x)\le\min_jG_j\le\tfrac13(G_1+G_2+G_3)$. Compare the contributions of a single point $P$ to the three $G_j$ with its contribution to $w$. If $r_P\le1$, all four contributions agree. If $r_P=2$, say $u_1,u_2$ nonzero and $u_3=0$ at $P$: when $u_1+u_2=0$ at $P$ the contributions to $G_1,G_2,G_3$ are $2,2,0$ and to $w$ it is $0$, an average excess of $\tfrac43$; otherwise they are $2,2,1$ against $1$, an average excess of $\tfrac23$. If $r_P=3$: when $u_1+u_2+u_3=0$ at $P$ the contributions are $2,2,2$ against $0$, an excess of $2$; otherwise each $G_j$ receives at most $2$ against $1$, an excess of at most $1$. In every case the average excess is at most $\tfrac23r_P$ for $r_P\ge2$ and $0$ otherwise. Summing over the points,
\[
P(x)\le\tfrac13(G_1+G_2+G_3)\le w+\tfrac23I\le\Bigl(1+\frac{2\nu}{3(d-\nu)}\Bigr)w .
\]
The averaging is over integer weights, so it is valid in every characteristic; if $w=0$ then~\eqref{eq:L_bound} gives $\mathcal L=0$ and $x=0$.
\end{proof}

\subsection{The sharp filling theorem with explicit constants}\label{subsec:explicit}

Put
\begin{equation}\label{eq:F_and_C}
F(\vartheta):=1+\frac{4\vartheta}{3\bigl(\tfrac45-4\vartheta\bigr)}\quad(0\le\vartheta<\tfrac15),\qquad
C_\eta:=\frac{1600}{957}+\frac{3200}{799\,\eta},
\end{equation}
where $\eta$ is the expansion constant~\eqref{eq:eta}.

\begin{theorem}[Asymptotically sharp filling, explicit form]\label{thm:sharp_filling}
Let $\mu>0$, $h:=\log(1+\mu)$,
\begin{equation}\label{eq:lambda0_i0}
\lambda_0(\mu):=\min\Bigl\{7\cdot10^{-14},\ \Bigl(\frac{h}{4478\,C_\eta}\Bigr)^2\Bigr\},\qquad i_0(\mu):=\min\Bigl\{i\ge4:\ N_i\ge\frac{4C_\eta}{h}\Bigr\}.
\end{equation}
Then for every $i\ge i_0(\mu)$ and every nonzero $x\in S_3(C_i)$ with $\wt(x)\le\lambda_0(\mu)N_i^3$,
\begin{equation}\label{eq:sharp_filling}
\wt(x)\le P_i(x)\le(1+\mu)\wt(x).
\end{equation}
Moreover, for every $i\ge4$ and every nonzero $x\in S_3(C_i)$ with $\wt(x)\le7\cdot10^{-14}N_i^3$,
\begin{equation}\label{eq:crude_filling}
P_i(x)\le F\bigl(\tfrac1{1600}\bigr)\exp\Bigl(\frac{4}{1598\,\eta}\Bigr)\wt(x)=\frac{958}{957}\exp\Bigl(\frac{4}{1598\,\eta}\Bigr)\wt(x).
\end{equation}
\end{theorem}

\begin{proof}
Let $i\ge4$, $0<\lambda\le7\cdot10^{-14}$, and $x\ne0$ with $w=\wt(x)\le\lambda N^3$, $N=N_i$. Proposition~\ref{prop:localization} provides $U_0$, the box $B$ with minimal side $n_{\min}>N-2U_0$, the pair divisors $D_{jk}$ and the carrier $J$. Put $\vartheta:=U_0/N<\tfrac1{1600}$.

\emph{Decomposition on the fibres.} By Proposition~\ref{prop:localization}(a) and (c) and Theorem~\ref{thm:exact_generation}, applied to the triple $(C|_{B_j})_j$ with $G=MP_\infty$ and $a=b$, the restriction $x_B$ is a $k_0$-linear combination of coded lines of the punctured codes placed on the fibres of $R$. Every nonzero codeword of $C|_{B_j}$ has weight at least $d:=n_{\min}-M>(\tfrac45-2\vartheta)N$ (Section~\ref{subsec:one_constant}), and by Proposition~\ref{prop:localization}(b) every fibre of $R$ meets the fibres of the other directions in at most $\nu\le2\vartheta N<d$ points of $B$. Lemma~\ref{lem:weighted_filling} gives
\[
P_B(x_B)\le\Bigl(1+\frac{2\nu}{3(d-\nu)}\Bigr)\wt(x_B)\le\Bigl(1+\frac{4\vartheta}{3(\tfrac45-4\vartheta)}\Bigr)\wt(x_B)=F(\vartheta)\wt(x_B),
\]
where $P_B$ is the pair filling mass for the punctured triple.

\emph{Restoration.} By Proposition~\ref{prop:localization}(d) the hypotheses of the second part of Lemma~\ref{lem:transfer} hold, with $q<2U_0=2\vartheta N$ deleted indices in total, so
\[
P(x)\le\Gamma\bigl(w-\wt(x_B)+P_B(x_B)\bigr)\le\Gamma F(\vartheta)\,w,\qquad \Gamma\le\exp\Bigl(\frac{2q}{\eta(N-q)}\Bigr)<\exp\Bigl(\frac{4\vartheta}{\eta(1-2\vartheta)}\Bigr),
\]
using $F(\vartheta)\ge1$. For $0\le\vartheta\le\tfrac1{1600}$ the inequality $\log(1+z)\le z$ gives
\begin{equation}\label{eq:log_bound}
\log F(\vartheta)+\frac{4\vartheta}{\eta(1-2\vartheta)}\le\frac{4\vartheta}{3\bigl(\tfrac45-\tfrac4{1600}\bigr)}+\frac{4\vartheta}{\eta\bigl(1-\tfrac2{1600}\bigr)}=\Bigl(\frac{1600}{957}+\frac{3200}{799\,\eta}\Bigr)\vartheta=C_\eta\vartheta .
\end{equation}
Hence
\begin{equation}\label{eq:P_over_w}
\frac{P(x)}{w}\le\exp(C_\eta\vartheta),\qquad\vartheta<2239\sqrt\lambda+\frac2N .
\end{equation}
The bound~\eqref{eq:crude_filling} is~\eqref{eq:P_over_w} with $\vartheta<\tfrac1{1600}$ evaluated before taking logarithms: $F(\tfrac1{1600})=1+\tfrac{1}{957}$ and $\tfrac{4/1600}{\eta(1-2/1600)}=\tfrac{4}{1598\eta}$.

\emph{The sharp bound.} Now let $\lambda=\lambda_0(\mu)$ and $i\ge i_0(\mu)$. Then $2239\sqrt{\lambda_0(\mu)}\le\tfrac{2239h}{4478C_\eta}=\tfrac{h}{2C_\eta}$ and $\tfrac2N\le\tfrac{h}{2C_\eta}$, so $\vartheta<h/C_\eta$ and $C_\eta\vartheta<h$; by~\eqref{eq:P_over_w}, $P(x)\le e^hw=(1+\mu)w$. The lower bound $w\le P(x)$ is Proposition~\ref{prop:sandwich}.
\end{proof}

\begin{remark}[The first three floors]\label{rem:first_floors}
On floors $i\le3$, $N_i\le728\cdot27^2=530712$ and $7\cdot10^{-14}N_i^3<\tfrac45N_i\le N_i-M_i$, because $7\cdot10^{-14}N_i^2<1$,, while every nonzero word of $S_3(C_i)$ has weight at least $d(C_i)\ge N_i-M_i$ (the sum-code distance is at least the factor distance, Remark~\ref{rem:few_conditions}). So there are no nonzero words of weight at most $7\cdot10^{-14}N_i^3$ on these floors, and Theorem~\ref{thm:sharp_filling} is vacuous there; this is why $i_0(\mu)\ge4$ costs nothing.
\end{remark}

\subsection{Proofs of Theorems~\ref{thm:A} and~\ref{thm:B}}\label{subsec:main_proofs}

\begin{proof}[Proof of Theorem~\ref{thm:B}]
This is Theorem~\ref{thm:sharp_filling} with $\lambda_0=\lambda_0(\mu)$ and $i_0=i_0(\mu)$.
\end{proof}

\begin{corollary}[Explicit certificate]\label{cor:explicit_certificate}
Put $\lambda_1:=\lambda_0(1)=\min\{7\cdot10^{-14},(\log2/(4478C_\eta))^2\}$ and $i_1:=i_0(1)$. Then the family $\{C_i\}$ has a uniform $D=3$ certificate with constant
\begin{equation}\label{eq:rho_explicit}
\rho^{-1}=\max\Bigl\{\frac2\eta,\ \frac3{\lambda_1},\ \max_{4\le i<i_1}\max_{0\ne x\in S_3(C_i)}\varrho(x)\Bigr\}.
\end{equation}
If $C_\eta\le N_4\log2/4$, which holds whenever $\eta\ge2\cdot10^{-6}$, then $i_1=4$, the last term is absent, and $\rho^{-1}=\max\{2/\eta,\,3/\lambda_1\}$.
\end{corollary}

\begin{proof}
Theorem~\ref{thm:sharp_filling} with $\mu=1$ gives Theorem~\ref{thm:main_uniform_certificate} with $c^{-1}=2$, $\lambda_0=\lambda_1$ and $i_0=i_1$, and Remark~\ref{rem:light_words} turns it into a certificate with $\rho^{-1}=\max\{2/\eta,\,3/\lambda_1,\,\max_{i<i_1}\max_x\varrho(x)\}$. On floors $i\le3$ the words of weight at most $\lambda_1N_i^3$ do not exist (Remark~\ref{rem:first_floors}), so $\varrho\le3/\lambda_1$ there by~\eqref{eq:trivial_bounds}, and only the floors $4\le i<i_1$ remain in the last term. If $N_4=14329224\ge4C_\eta/\log2$, then $i_1=4$; the condition $C_\eta\le N_4\log2/4\approx2.48\cdot10^6$ holds as soon as $3200/(799\eta)\le2.48\cdot10^6-1.68$, that is, for $\eta\ge1.62\cdot10^{-6}$.
\end{proof}

\begin{proof}[Proof of Theorem~\ref{thm:A}]
Immediate from Corollary~\ref{cor:explicit_certificate}.
\end{proof}

\begin{proof}[Proof of Corollary~\ref{cor:B}]
Take $C_n:=C_i^\perp$ for $n=N_i$ (Remark~\ref{rem:notation_match}). On every floor its relative distance exceeds $\tfrac{8}{65}$ and its rate lies in $[\tfrac45-\tfrac1n,\tfrac{109}{130})$, and its rate tends to $\tfrac{109}{130}$, by Proposition~\ref{prop:Ci_parameters}; and its dual $C_n^\perp=C_i$ has relative distance at least $\tfrac45$. The collection … (unchanged)
\end{proof}

\begin{remark}[Orientation on the size of the constants]\label{rem:size_constants}
With $\rhoGS(\epsilon)=\epsilon^2/12$ \cite[Corollary~6.3]{GargSengupta25}, $\eta=22/6431$, $C_\eta\approx1172.4<1200$, the condition of Corollary~\ref{cor:explicit_certificate} holds, $\lambda_1\approx1.74\cdot10^{-14}$, and $\rho^{-1}=3/\lambda_1\approx1.7\cdot10^{14}$. The bound~\eqref{eq:crude_filling} then reads $P_i(x)<2.081\,\wt(x)$ for all words of weight at most $7\cdot10^{-14}N_i^3$ on floors $i\ge4$. The constant is far from optimal; its size is driven by the density threshold $\lambda$ rather than by $\eta$, and the proof gives no indication that the true constant is small. The general criterion of Section~\ref{sec:general}, evaluated at the same data, gives the same order, $\rho^{-1}\approx1.7\cdot10^{14}$. With the sharper bound on the number of conditions of Remark~\ref{rem:few_conditions} one would obtain $\lambda_1\approx3.1\cdot10^{-13}$ and $\rho^{-1}\approx10^{13}$.
\end{remark}

\begin{remark}[Four or more directions]\label{rem:four_directions}
The combinatorial parts of the argument (Sections~\ref{sec:sum_codes}, \ref{sec:transfer} and~\ref{subsec:weighted_filling}) extend to any number of directions, and so does the carrier construction. Two geometric inputs are specific to three directions. The vanishing of Lemma~\ref{lem:ambient_vanishing_D} concerns $H^{\dim Y-1}$, which for a product of $D$ curves controls interpolation on a complete intersection of $D-1$ divisors only when $D-1=2$; and the exact generation theorem passes through the curve $K$ of pairwise intersections of the cylinders, whose analogue for $D\ge4$ is a union of higher-dimensional intersections. We do not know whether the conclusion of Theorem~\ref{thm:A} holds for four copies of $C_i$.
\end{remark}

\section{A general sufficient criterion}\label{sec:general}

The proof of Theorem~\ref{thm:A} uses the tower only through numbers: the genus ratio $\beta=\tfrac1{26}$, the pole fraction $\theta=\tfrac15$, the gonality bound $N/730$, and the lengths $N_i$. This section separates these numbers by direction and records the sufficient criterion that the argument of Sections~\ref{sec:transfer}--\ref{sec:proof} actually proves. Three evaluation codes on three possibly different curves over a finite field, with arbitrary divisors, lengths differing by a bounded factor, and no restriction on the characteristic, are uniformly product-expanding as soon as five explicit inequalities in their degree and genus ratios hold (Theorem~\ref{thm:general}). The criterion is sufficient, not necessary. The argument is the one already given: the proof, in Appendix~\ref{sec:appendix_general}, consists of the replacements to be made in Sections~\ref{sec:transfer}--\ref{sec:proof} and nothing else. Two consequences are worth stating. The admissible region is asymmetric and strictly larger than the symmetric box (Example~\ref{ex:asymmetric}). And the symmetric criterion is met by an explicit tower over every non-prime finite field of order at least $512$ (Corollary~\ref{cor:field_coverage}), so that Question~\ref{q:main} has a positive answer over all these fields. The directional bookkeeping of this section was first drafted with the help of an AI system and then checked by hand; the margin functions and the examples have been verified with exact rational arithmetic by the script accompanying the paper.

\subsection{Setting and margin functions}\label{subsec:general_setting}

Let $k$ be a finite field. A \emph{triple} consists of smooth projective geometrically integral curves $X_1,X_2,X_3$ over $k$, sets $E_j\subseteq X_j(k)$ of $n_j$ distinct rational points, and divisors $G_j$ on $X_j$ with $\Supp G_j\cap E_j=\varnothing$, $\deg G_j=m_j$. Its codes are $C_j:=C_L(E_j,G_j)\subseteq k^{n_j}$ and its sum code is $S:=S(C_1,C_2,C_3)$ (Section~\ref{subsec:sum_code}). We write $g_j$ for the genus of $X_j$, $n_{\min}:=\min_jn_j$, $\mathsf V:=n_1n_2n_3$, and
\begin{equation}\label{eq:general_ratios}
\theta_j:=\frac{m_j}{n_j},\qquad \beta_j:=\frac{g_j}{n_j}\qquad(j=1,2,3),
\end{equation}
the \emph{degree ratios} and \emph{genus ratios}, in the notation of Section~\ref{sec:family} ($\theta_j$ is not the rate of $C_j$). Since the lengths may differ, the line count is weighted by length. The \emph{weighted line count} of $x\in S$ is
\begin{equation}\label{eq:weighted_lines}
L_{\mathrm w}(x):=\min\Bigl\{\sum_{j=1}^3n_j\operatorname{lines}_j(u_j)\;:\;x=u_1+u_2+u_3,\ u_j\in U_j\Bigr\},
\end{equation}
so that $L_{\mathrm w}(x)=N\,L(x)$ when all lengths equal $N$, and the triple is \emph{$\rho$-product-expanding} if $\rho\,L_{\mathrm w}(x)\le\wt(x)$ for all $x\in S$. The pair filling mass $P(x)$ is defined as in Definition~\ref{def:pair_mass}. All other notions of Sections~\ref{sec:sum_codes}--\ref{sec:transfer} apply verbatim, since they were stated for three arbitrary codes, and the trivial bounds~\eqref{eq:trivial_bounds} become $\wt(x)\le L_{\mathrm w}(x)\le3\mathsf V$.

For $w=(w_1,w_2,w_3)$ in the simplex $\mathcal D:=\{w\in\mathbb R^3_{\ge0}:w_1+w_2+w_3=1\}$ define the \emph{margin functions}
\begin{align}
\mathsf A_{ij}(w)&:=\sum_{h\in\{i,j\}}(1-\theta_h-7\beta_h)\,w_h-\min(w_i,w_j)\qquad\text{on the edge }w_i+w_j=1,\label{eq:A_margin}\\
\mathsf T(w)&:=\sum_{h=1}^3(1-\theta_h-8\beta_h)\,w_h-\min_hw_h,\label{eq:T_margin}\\
\mathsf Q_0(w)&:=\max_hw_h-\sum_{h=1}^3(\theta_h+4\beta_h)\,w_h+4\max_h\beta_hw_h,\label{eq:Q0_margin}\\
\mathsf Q_j(w)&:=1-\sum_{h=1}^3\theta_hw_h-6\sum_{h=1}^3\beta_hw_h+2\beta_jw_j+4\max_h\beta_hw_h\notag\\
&\qquad-\min(w_j,1-w_j)\qquad(j=1,2,3).\label{eq:Qa_margin}
\end{align}
They are the normalised forms of the interpolation margins of Sections~\ref{sec:generation} and~\ref{sec:localization}. $\mathsf A_{ij}$ and $\mathsf T$ are the plane-type and threefold-type margins of Lemma~\ref{lem:support_on_Delta}; $\mathsf Q_0$ is the third-layer margin and $\mathsf Q_j$ the first- and second-layer margins of Lemma~\ref{lem:step2}. Each is divided by the weighted grid degree $\sum_jn_jd_j(V)$ of the subcurve $V$ in question, with $w_j=n_jd_j(V)/\sum_hn_hd_h(V)$ (Lemma~\ref{lem:appC_localization} and Proposition~\ref{prop:appC_generation}). Each margin function is piecewise affine on $\mathcal D$, with breaks only along the lines $w_i=w_j$, $\beta_iw_i=\beta_jw_j$ and $w_j=\tfrac12$. Its minimum over $\mathcal D$, or over an edge, is therefore attained at a vertex of the subdivision of $\mathcal D$ by these lines and the boundary, and this finite test is exact. Every margin function is nonincreasing in each $\theta_j$ and in each $\beta_j$, because $\sum_h\beta_hw_h-\max_h\beta_hw_h=\min_r\sum_{h\ne r}\beta_hw_h$ and $\sum_{h\ne j}\beta_hw_h$ are nondecreasing in each $\beta_h$.

\begin{definition}[The class $\mathcal T(\mathsf L,c_{\gon},\tau,\tau_{\mathrm P})$]\label{def:general_class}
Let $\mathsf L\ge1$, $c_{\gon}>0$, $0<\tau<1$ and $0<\tau_{\mathrm P}<1$. A triple belongs to $\mathcal T(\mathsf L,c_{\gon},\tau,\tau_{\mathrm P})$ if:
\begin{enumerate}[label=\textup{(G\arabic*)},leftmargin=3em,itemsep=1pt,topsep=2pt]
\item \label{G1} $n_j\le\mathsf Ln_{\min}$ and $\gamma_k(X_j)\ge c_{\gon}n_j$ for $j=1,2,3$ \textup{(}Definition~\textup{\ref{def:gonality}}\textup{)};
\item \label{G2} $\theta_j-2\beta_j\ge\tau$ for $j=1,2,3$;
\item \label{G3} $m_i+g_i+m_j+g_j\le(1-\tau_{\mathrm P})\min(n_i,n_j)$ for all $i<j$;
\item \label{G4} $\mathsf A_{ij}\ge\tau$ on the edge $w_i+w_j=1$ for all $i<j$, and $\mathsf T\ge\tau$ on $\mathcal D$;
\item \label{G5} $\mathsf Q_0\ge\tau$ on $\mathcal D$, and $\mathsf Q_j\ge\tau$, $\mathsf Q_{j'}\ge\tau$ on $\mathcal D$ for two distinct directions $j\ne j'$.
\end{enumerate}
\end{definition}

Over a fixed field $k=\F_q$, \ref{G1} holds with $c_{\gon}=1/(q+1)$ for every triple: a $k$-morphism $X_j\to\PP^1$ of degree $d$ maps the $n_j$ points of $E_j$ to the $q+1$ rational points of $\PP^1_k$ with fibres of at most $d$ points, so $\gamma_k(X_j)\ge n_j/(q+1)$, as in Lemma~\ref{lem:gonality}(a). Condition~\ref{G2} gives $m_j\ge2g_j+\tau n_j>2g_j-2$, hence $\dim C_j=m_j+1-g_j$ and $d(C_j^\perp)\ge m_j-2g_j+2\ge\tau n_j$ by Propositions~\ref{prop:ag_parameters} and~\ref{prop:app_cauchy_duality}; the expansion argument uses only $m_j>2g_j-2$. Condition~\ref{G3} is the margin of External Input~\ref{ext:two_code_robustness} for the pair $(C_i,C_j)$ before any puncturing. Evaluating $\mathsf A_{ij}$ at a vertex of its edge gives $1-\theta_j-7\beta_j\ge\tau$ for every $j$, so \ref{G4} implies
\begin{equation}\label{eq:deltaD}
\delta_{\mathrm D}:=\min_j(1-\theta_j)\ \ge\ \tau ,\qquad d(C_j)\ \ge\ n_j-m_j\ \ge\ \delta_{\mathrm D}n_j ,
\end{equation}
the designed relative distance of the triple. In~\ref{G5} the two directions $j,j'$ are the remaining directions of the two cylinders that are peeled off first in the filtration of Lemma~\ref{lem:filtration}; the third-layer margin $\mathsf Q_0$ is the same for every labelling.

\subsection{Statements}\label{subsec:general_statements}

\begin{theorem}[General criterion]\label{thm:general}
Let $\mathsf L\ge1$, $c_{\gon}>0$ and $\tau,\tau_{\mathrm P}\in(0,1)$. There is a constant $\rho=\rho(\mathsf L,c_{\gon},\tau,\tau_{\mathrm P})>0$ such that every triple in $\mathcal T(\mathsf L,c_{\gon},\tau,\tau_{\mathrm P})$, of any length, is $\rho$-product-expanding, and its codes satisfy $\dim C_j/n_j\ge\tau$, $d(C_j)/n_j\ge\delta_{\mathrm D}\ge\tau$ and $d(C_j^\perp)/n_j\ge\tau$. Moreover, for every $\mu>0$ there are $\lambda_0(\mu)>0$ and $n_0(\mu)$, depending only on $\mathsf L,c_{\gon},\tau,\tau_{\mathrm P},\mu$, such that every triple in the class with $n_{\min}\ge n_0(\mu)$ satisfies
\begin{equation}\label{eq:general_sharp}
\wt(x)\ \le\ P(x)\ \le\ (1+\mu)\wt(x)\qquad\text{for every nonzero }x\in S\text{ with }\wt(x)\le\lambda_0(\mu)\mathsf V .
\end{equation}
The constants are explicit; they are given in Proposition~\ref{prop:general_constants}.
\end{theorem}

\begin{proposition}[Explicit constants]\label{prop:general_constants}
With $\mathsf L,c_{\gon},\tau,\tau_{\mathrm P}$ as in Theorem~\textup{\ref{thm:general}} and $\delta_{\mathrm D}$ as in~\eqref{eq:deltaD}, put
\begin{equation}\label{eq:general_constants}
\begin{aligned}
\alpha_{\mathrm{gen}}&:=\rhoGS(\tau_{\mathrm P}/2), &
\vartheta_0&:=\min\{\tau,\tau_{\mathrm P}\}/20, &
\zeta&:=\tau/40,\\
\eta'_{\mathrm{gen}}&:=\frac{\alpha_{\mathrm{gen}}(\delta_{\mathrm D}-4\vartheta_0)}{1+\alpha_{\mathrm{gen}}}, &
\mathsf L'&:=\frac{\mathsf L}{1-2\vartheta_0}, &
\eta_{\mathrm{gen}}&:=\frac{\eta'_{\mathrm{gen}}}{\mathsf L'+(\mathsf L'-1)/(\delta_{\mathrm D}-4\vartheta_0)},\\
\Upsilon&:=\sqrt{\frac{12\mathsf L^3}{\tau}}+\frac{40\sqrt5\,\mathsf L^2}{\tau}, &
n_0&:=\Bigl\lceil\frac{80}{\min\{\tau,\tau_{\mathrm P}\}}\Bigr\rceil, &
C_{\mathrm{gen}}&:=\frac{4}{3(\delta_{\mathrm D}-4\vartheta_0)}+\frac{4}{\eta_{\mathrm{gen}}(1-2\vartheta_0)},\\
\lambda_{\max}&:=\min\Bigl\{\frac{\zeta^3}{5\mathsf L^3},\ \frac{\zeta^2\tau}{10\mathsf L^3},\ \frac{\zeta^2c_{\gon}}{4\mathsf L^3},\ \Bigl(\frac{\vartheta_0}{2\Upsilon}\Bigr)^2\Bigr\}. \span\span\span\span
\end{aligned}
\end{equation}
Let the triple lie in $\mathcal T(\mathsf L,c_{\gon},\tau,\tau_{\mathrm P})$ with $n_{\min}\ge n_0$.
\begin{enumerate}[label=\textup{(\alph*)},leftmargin=2.4em]
\item \textup{(}explicit filling\textup{)} For $0<\lambda\le\lambda_{\max}$ and every nonzero $x\in S$ with $\wt(x)\le\lambda\mathsf V$,
\begin{equation}\label{eq:general_filling}
\wt(x)\ \le\ P(x)\ \le\ \exp\Bigl(C_{\mathrm{gen}}\bigl(\Upsilon\sqrt\lambda+2/n_{\min}\bigr)\Bigr)\,\wt(x).
\end{equation}
\item \textup{(}sharp filling\textup{)} Theorem~\textup{\ref{thm:general}} holds with $h:=\log(1+\mu)$, $\lambda_0(\mu):=\min\{\lambda_{\max},(h/(2\Upsilon C_{\mathrm{gen}}))^2\}$ and $n_0(\mu):=\max\{n_0,\lceil4C_{\mathrm{gen}}/h\rceil\}$.
\item \textup{(}uniform expansion\textup{)} Theorem~\textup{\ref{thm:general}} holds with
\begin{equation}\label{eq:general_rho}
\rho^{-1}:=\max\Bigl\{\frac2{\eta_{\mathrm{gen}}},\ \frac3{\lambda_0(1)},\ 3\mathsf L^3n_0(1)^3\Bigr\}.
\end{equation}
\end{enumerate}
The constants depend on the triple only through $\delta_{\mathrm D}$, which may be replaced by $\tau$ throughout.
\end{proposition}

The constants are of the same form as those of Theorem~\ref{thm:sharp_filling}, with $\vartheta_0$, $\zeta$ and $\Upsilon$ in place of $\tfrac1{1600}$, $\tfrac1{1000}$ and $2239$. They are chosen for ease of verification and are not optimised. For the limiting data of the tower, $\theta_j=\tfrac15$, $\beta_j=\tfrac1{26}$, $\mathsf L=1$, $c_{\gon}=\tfrac1{730}$, one finds $\tau=\tfrac2{65}$, $\tau_{\mathrm P}=\tfrac{34}{65}$, $\delta_{\mathrm D}=\tfrac45$. With $\rhoGS(\epsilon)=\epsilon^2/12$ the formulas give $\eta_{\mathrm{gen}}\approx4.5\cdot10^{-3}$, $C_{\mathrm{gen}}\approx900$, $\lambda_0(1)\approx1.7\cdot10^{-14}$ and $\rho^{-1}\approx1.7\cdot10^{14}$, the same order as in Remark~\ref{rem:size_constants}: the general constants reproduce those of the tower proof because the two proofs are the same.

\begin{corollary}[Symmetric region]\label{cor:general_symmetric}
Let $k=\F_q$ be a fixed finite field, and let $\theta,\beta\ge0$ satisfy
\begin{equation}\label{eq:symmetric_region}
2\beta<\theta,\qquad 3\theta+8\beta<1,\qquad \theta+7\beta<\tfrac12 .
\end{equation}
Fix $\tau_0>0$. All triples over $k$ with a common length $n$, genus ratios $\beta_j\le\beta$ and degree ratios $2\beta_j+\tau_0\le\theta_j\le\theta$ are uniformly $\rho$-product-expanding, with $\rho$ depending only on $q,\theta,\beta,\tau_0$. They have rate at least $\theta_j-\beta_j$, relative distance at least $1-\theta$ and dual relative distance at least $\tau_0$. The region~\eqref{eq:symmetric_region} is nonempty exactly when $\beta<\tfrac1{18}$, and then it is the interval $2\beta<\theta<\min\{(1-8\beta)/3,\ \tfrac12-7\beta\}$.
\end{corollary}

The proof is in Appendix~\ref{subsec:appC_corollaries}.

\begin{corollary}[Field coverage]\label{cor:field_coverage}
Let $q=p^m$ with $p$ prime and $m\ge2$. If $q\ge512$, there is an explicit recursive tower of curves over $\F_q$ whose limiting genus ratio is below $\tfrac1{18}$, and consequently an explicit family of one-point codes over $\F_q$ whose duals satisfy \textup{(i)}--\textup{(iii)} of Question~\textup{\ref{q:main}} \textup{(}Remark~\textup{\ref{rem:notation_match})}. For the nineteen non-prime field orders $q<512$, the towers of~\cite{GS95,BBGS15} do not reach the region~\eqref{eq:symmetric_region}.
\end{corollary}

The proof, also in Appendix~\ref{subsec:appC_corollaries}, is a table of the limits of the towers of~\cite{GS95,BBGS15}.

Thus the smallest covered orders are $512$ in characteristic~$2$ (via the tower over $\F_{2^9}$ with the partition $9=5+4$), $729$ in characteristic~$3$, $625$ in characteristic~$5$, $2401$ in characteristic~$7$, $1331$ in characteristic~$11$, $2197$ in characteristic~$13$, $4913$ and $6859$ in characteristics $17$ and $19$, and $p^2$ for every prime $p\ge23$. At $q=2197$ the tower is the cubic tower of~\cite{BGS08}, with limiting genus ratio $\tfrac5{112}$, and at $\theta=\tfrac3{20}$ the margins are $\tfrac3{80}$, $\tfrac{67}{420}$ and $\tfrac9{140}$. Prime fields are not reached: the towers of~\cite{GS95,BBGS15} are defined over non-prime fields, and we know of no explicit tower over a prime field whose limit exceeds $18$. The threshold $\tfrac1{18}$ describes the region~\eqref{eq:symmetric_region} only; it is not a nonexistence statement for smaller fields.

\begin{example}[The tower at other pole fractions]\label{ex:tower_window}
For the family of Definition~\ref{def:ag_code_family} with $M_i=\lfloor\theta N_i\rfloor$ in place of $\lfloor N_i/5\rfloor$, Corollary~\ref{cor:general_symmetric} with $\beta=\tfrac1{26}$ gives a uniform certificate for every pole fraction
\[
\tfrac1{13}<\theta<\tfrac3{13},
\]
the two upper constraints of~\eqref{eq:symmetric_region} coinciding at $\beta=\tfrac1{26}$. This contains the value $\theta=\tfrac15$ of Theorem~\ref{thm:A}, where the three margins are $\tfrac12-\theta-7\beta=\tfrac13-\theta-\tfrac83\beta=\tfrac2{65}$ and $\tfrac23-\theta-8\beta=\tfrac{31}{195}$. It extends past the bound $\theta<0.2115$ of Remark~\ref{rem:admissible_theta}, which came from the fixed puncturing fraction $\tfrac34$ and the fixed $\epsilon=\tfrac13$ of Proposition~\ref{prop:pair}. The same corollary applies to the Garcia--Stichtenoth tower over $\F_{\ell^2}$ for every prime power $\ell\ge23$, with $\beta=1/(\ell-1)$; in particular $\ell=25$ is admissible, so the choice $\ell=27$ in Section~\ref{sec:family} is a matter of the verified constants, not of the method. The constant $\rho$ obtained in this way is the one of Proposition~\ref{prop:general_constants}(c), which is worse than that of Corollary~\ref{cor:explicit_certificate} at $\theta=\tfrac15$.
\end{example}

\begin{example}[An asymmetric triple]\label{ex:asymmetric}
Take three one-point codes of a common length on the tower, with degree ratios $(\theta_1,\theta_2,\theta_3)=(\tfrac7{25},\tfrac3{25},\tfrac3{25})$ and genus ratio $\tfrac1{26}$. The exact minima are $\min\mathsf A_{12}=\min\mathsf A_{13}=\tfrac2{65}$, $\min\mathsf A_{23}=\tfrac{36}{325}$, $\min\mathsf T=\tfrac{181}{975}$, $\min\mathsf Q_0=\tfrac{56}{975}$, attained at the barycentre, $\min\mathsf Q_j=\tfrac{12}{65}$ for every $j$, and $\min_j(\theta_j-2\beta_j)=\tfrac{14}{325}$. So the triple lies in $\mathcal T(1,\tfrac1{730},\tfrac2{65},\tfrac{34}{65})$ and is uniformly expanding, with the first code of rate close to $\tfrac{157}{650}$ and dual relative distance about $\tfrac{66}{325}$. The symmetric test fails at $\theta=\max_j\theta_j=\tfrac7{25}$: there $3\theta+8\beta=\tfrac{373}{325}>1$ and $\theta+7\beta=\tfrac{357}{650}>\tfrac12$. Unequal lengths are admitted as well. One curve of the tower with three evaluation subsets of relative sizes $1,\tfrac{21}{20},\tfrac{11}{10}$ and the same degree ratios has margins $\tfrac{17}{1560}$ (plane), $\tfrac{1126}{20475}$ (third layer) and $\tfrac{269}{6825}$ (dual distance), so it lies in $\mathcal T(\tfrac{11}{10},\tfrac1{730},\tfrac{17}{1560},\tfrac{818}{1625})$.
\end{example}

\subsection{Boundaries}\label{subsec:general_boundaries}

\begin{remark}[Where the coefficients come from]\label{rem:general_coefficients}
The two symmetric upper bounds in~\eqref{eq:symmetric_region} have the following origin. In $\theta+7\beta<\tfrac12$: $2\beta$ is the pole shift $m'_j-m_j\approx2g_j$ of the first carrier; $4\beta$ is the genus bound~\eqref{eq:genus_Cy} on all CM subcurves of the residual curve, which costs $4a_j\approx4g_j$ per direction because the Koszul argument needs the auxiliary bundle $\cF$ to have factor degree $2a_j+g_j-1$; and $\beta$ is the intersection $\Delta\cdot\Sigma\le a_kd_k+a_ld_l$ with the fixed divisor on a fibre surface. The $\tfrac12$ is the cluster bound $\min(n_kd_k,n_ld_l)\le\tfrac12(n_kd_k+n_ld_l)$. In $3\theta+8\beta<1$: the third-layer cluster may use $\tfrac23$ of the weighted grid degree, and the genus bound~\eqref{eq:gen_genus_Cij} costs $2(H_d-H_{\max})\le\tfrac43H_d$ with $b_j\approx2g_j$. Adjunction on the whole residual curve would cost only $2a_j$ per direction; the loss of the factor $2$ occurs in passing to arbitrary CM subcurves, where no Cartier adjunction is available. Whether $7$ and $8$ can be lowered is open, and so is any form of necessity: failure of an interpolation inequality is not a counterexample to expansion.
\end{remark}

\begin{remark}[One-sidedness in general]\label{rem:general_onesided}
The one-sidedness of Remark~\ref{rem:schur_family} needs only two directions. If $2m_j+1<n_j$ for some $j$, then $C_j*C_j\subseteq C_L(E_j,2G_j)$ is proper. A vector $v\ne0$ orthogonal to it gives the matrix $\operatorname{diag}(v)\in C_j^\perp\otimes k^{n_j}+k^{n_j}\otimes C_j^\perp$, and placed on a single slice of the cube it is a nonzero word of $S(C_j^\perp,C_j^\perp,C_j^\perp)$ meeting every line at most once. So $(C_j^\perp,C_j^\perp,C_j^\perp)$ has expansion at most $1/n_j$ whenever $\theta_j<\tfrac12-1/(2n_j)$. In the region of Theorem~\ref{thm:general} this holds for at least two of the three directions: evaluating $\mathsf A_{ij}$ at the midpoint of its edge gives $\theta_i+\theta_j\le1-2\tau-7(\beta_i+\beta_j)<1-2\tau$, so at most one direction has $\theta_j\ge\tfrac12-\tau$, and $\tau n_j\ge80>\tfrac12$. A single direction may have $\theta_j$ above $\tfrac12$; for instance $(\theta_1,\theta_2,\theta_3)=(\tfrac{11}{20},\tfrac3{20},\tfrac3{20})$ with $\beta_j=\tfrac1{500}$ satisfies~\ref{G2}--\ref{G5} with $\tau=\tfrac{67}{1500}$. The expansion asserted in the theorem is that of the evaluation codes. The Schur fullness $(C_j^\perp)^{*3}=k^{n_j}$ holds by Proposition~\ref{prop:schur_fullness}, applied to the ambient divisor of degree $n_j-m_j+2g_j-2$ of Remark~\ref{rem:schur_family}, as soon as $n_j-m_j\ge3$ and $3m_j<2n_j+4g_j-4$, in particular whenever $3m_j<2n_j-4$.
\end{remark}

\begin{remark}[What each hypothesis does]\label{rem:general_hypotheses}
The gonality hypothesis in~\ref{G1} is used once, in Lemma~\ref{lem:grid_cover}, to turn the linear equivalence~\eqref{eq:cube_fibres} of two effective divisors of degree $2\delta_3$ into an equality. It is automatic over a fixed field by point counting, and it is exactly what the Reed--Solomon families of Section~\ref{subsec:rs_witness} lack, since $\PP^1$ has gonality $1$ (Corollary~\ref{cor:H_insufficient}). Geometric gonality over $\Omega$ is not a substitute: the argument is made over the arithmetic field $K$ of Lemma~\ref{lem:gonality}(b). The condition $m_j>2g_j-2$ supplies the unique residue representatives of Lemma~\ref{lem:dual_evaluation_bundle}(1) on the selected fibres; \ref{G2} strengthens it to a designed dual distance. Nothing about the curves beyond genus, gonality and rational points enters: the recursive equations of a tower, its ramification and its splitting behaviour are used only to compute $\beta$ and $N_i$ in Section~\ref{sec:family}. The characteristic is unrestricted; characteristic $3$ plays no role in the proof of Theorem~\ref{thm:A} either. Four or more directions remain outside the method for the reasons of Remark~\ref{rem:four_directions}, and the mixed collections of~\cite{GG25b} remain open: for a pair $(C_j,C_j^\perp)$ the margin~\eqref{eq:gs_margin} reads $m_j+(n_j-m_j+2g_j-2)+2g_j+4<n_j$, which fails by $4g_j+2$.
\end{remark}

\section*{Acknowledgements}
C.~Ding thanks Ted Chinburg for useful conversations and for carefully proofreading the manuscript. The authors thank Louis Golowich and Venkatesan Guruswami for helpful conversations.

\paragraph{Use of AI systems.}AI language models were used in the preparation of this manuscript in the following ways: to produce first drafts of the exact generation theorem of Sections~\ref{sec:generation} under the discussions with the authors, and of the directional bookkeeping of Section~\ref{sec:general} and Appendix~\ref{sec:appendix_general}, all of which were then checked line by line by the authors; to write the scripts that verify the margin functions, the explicit constants and the field-coverage table with exact rational arithmetic. All mathematical statements, proofs and decisions about what to claim are the authors' own, and the authors take full responsibility for the content.

\paragraph{Code availability and formal verification.}
The margin functions, the explicit constants and the field-coverage table of Section~\ref{sec:general} and Appendix~\ref{sec:appendix_general} were checked with exact rational arithmetic. A Lean~4 development that formally verifies the principal conclusions of this paper is available at~\cite{LeanD3}.
The Lean development has as its endpoints the uniform expansion of Theorem~\ref{thm:A} on every floor $i\ge1$ (with a different, cruder explicit constant), the sharp filling bound of Theorem~\ref{thm:B} for all words below explicit thresholds in $\mu$, $i$ and $N_i$, and the rate, distance and dual-distance assertions of Corollary~\ref{cor:B}; the Schur fullness assertion of Corollary~\ref{cor:B} and the closed genus formula of Section~\ref{sec:family} are not among the formal endpoints, the latter being replaced by the formally proved genus-ratio limit $g_i/N_i\to\tfrac1{26}$. The development is pinned to Lean v4.31.0, Mathlib and a Riemann--Roch library for function fields, and its endpoint checks admit no axioms beyond \texttt{propext}, \texttt{Classical.choice} and \texttt{Quot.sound}. The correspondence between the formal definitions and the ones of this paper, and the reproduction procedure, are described in the repository.

\appendix
\section{Evaluation codes: proofs and trivialisations} \label{sec:appendix_ag_parameters}

This appendix gives the proofs of Propositions~\ref{prop:ag_parameters} and~\ref{prop:app_cauchy_duality}, following~\cite[Section~2.2]{Stichtenoth09}, and a fuller account of the choice of trivialisation discussed in Remark~\ref{rem:app_generator_matrices}. As in Definition~\ref{def:app_line_bundle_ev}, $P_1,\dots,P_n$ are distinct rational points of $X$, $E=P_1+\dots+P_n$, and $G$ is a divisor whose support is disjoint from $E$.

\subsection{The parameters of evaluation codes} \label{subsec:appendix_parameters_proof}

\begin{proof}[Proof of Proposition~\ref{prop:ag_parameters}]
The whole argument rests on the following equivalence. Let $E'$ be the sum of some of the points $P_m$. Then a function $f\in L(G)$ vanishes at every point of $E'$ if and only if $f\in L(G-E')$. Indeed, since $\Supp(G)$ is disjoint from $E$, the coefficient of $G$ at each $P_m$ is zero, so the condition $(f)+G\ge 0$ gives $v_{P_m}(f)\ge 0$. Thus $f$ is regular at $P_m$, and its value $f(P_m)\in k(P_m)=\F_q$ is its class modulo $\mathfrak m_{P_m}$, so that $f(P_m)=0$ if and only if $v_{P_m}(f)\ge 1$. Since the points $P_m$ are distinct, $E'$ has coefficient $1$ at each of its points, and for a point $P_m$ of $E'$ the inequality $v_{P_m}(f)\ge 1$ says precisely that the coefficient of $(f)+G-E'$ at $P_m$ is non-negative. At every other place the divisors $(f)+G-E'$ and $(f)+G$ have the same coefficient. Hence $(f)+G-E'\ge 0$ exactly when $f$ vanishes at all points of $E'$.

Taking $E'=E$ shows that the kernel of $\ev_E$ is $L(G-E)$. Each $P_m$ is rational, so $\deg(E)=n$. If $\deg(G)<n$, then $\deg(G-E)<0$, and $L(G-E)=0$ because a nonzero function has as many zeros as poles. So $\ev_E$ is injective. For the distance, let $c=\ev_E(f)$ be a nonzero codeword of weight $w$. Then $f\ne 0$, and $f$ vanishes at the $n-w$ points $P_m$ at which $c$ has a zero coordinate. If $E'$ is the sum of these points, then $\deg(E')=n-w$ and $f\in L(G-E')$ by the equivalence. Since $f\ne0$, the same fact about zeros and poles gives $\deg(G-E')\ge 0$, that is, $w\ge n-\deg(G)$. In other words, a nonzero function in $L(G)$ vanishes at no more than $\deg(G)$ of the points $P_m$; this part of the argument does not use the injectivity of $\ev_E$. When $\ev_E$ is injective, $\dim C_L(E,G)=\dim L(G)$, and Theorem~\ref{thm:app_riemann_roch} gives $\dim L(G)\ge\deg(G)+1-g$, with equality when $\deg(G)>2g-2$ by~\eqref{eq:rr_exact}.
\end{proof}

\subsection{The dual of an evaluation code} \label{subsec:appendix_dual_proof}

\begin{proof}[Proof of Proposition~\ref{prop:app_cauchy_duality}]
Let $f\in L(G)$ and $\omega\in H^0(X,\omega_X(E-G))$. Then $(f\omega)=(f)+(\omega)\ge -G+(G-E)=-E$, so the differential $f\omega$ is regular outside $E$ and has at most simple poles at the points of $E$. Since $f$ is regular at these points, the residue theorem and~\eqref{eq:residue_product} give
$0=\sum_{P}\Res_P(f\omega)=\sum_{m=1}^n f(P_m)\,\Res_{P_m}(\omega)$,
which says that $\res_E(\omega)$ is orthogonal to $\ev_E(f)$. Hence $C_\Omega(E,G)\subseteq C_L(E,G)^\perp$. A differential in the kernel of $\res_E$ has no pole on $E$, so the kernel is $H^0(X,\omega_X(-G))$, which is isomorphic to $L(K_X-G)$ and vanishes when $\deg(G)>2g-2$. For the dimensions, Theorem~\ref{thm:app_riemann_roch} applied to $K_X+E-G$ and to $K_X-G$ gives
\begin{align*}
\dim C_\Omega(E,G) &= \dim L(K_X+E-G)-\dim L(K_X-G)\\ &= n+\dim L(G-E)-\dim L(G) = n-\dim C_L(E,G),
\end{align*}
where the last equality uses that $L(G-E)$ is the kernel of $\ev_E$. The two codes therefore coincide.

For the parameters of the dual, suppose $2g-2<\deg G<n$. There is a differential $\eta$ with $v_{P_m}(\eta)=-1$ and $\Res_{P_m}(\eta)=1$ for all $m$ (by the strong approximation theorem; see~\cite[Section~2.2]{Stichtenoth09}, where this differential is used to prove the same statement). Mapping $\omega\in H^0(X,\omega_X(E-G))$ to $\omega/\eta\in L(E-G+(\eta))$ identifies $C_\Omega(E,G)$ with $C_L(E,G')$, $G':=E-G+(\eta)$, a divisor disjoint from $E$ since $(\eta)$ has coefficient $-1$ at each $P_m$, of degree $n-\deg G+2g-2\in(2g-2,n)$. Proposition~\ref{prop:ag_parameters} gives $\dim C_L(E,G')=n-\deg G+g-1$ and $d(C_L(E,G'))\ge n-\deg G'=\deg G-(2g-2)$.
\end{proof}

\subsection{Values of sections and the choice of trivialisation} \label{subsec:appendix_trivialisation}

If $f_1,\dots,f_\kappa$ is a basis of $L(G)$, the matrix $\bigl(f_a(P_m)\bigr)_{a,m}$ is a generator matrix of $C_L(E,G)$ when $\ev_E$ is injective, with entries in $\F_q$ because the $f_a$ are functions. For sections of a general line bundle $\cF$ in place of $\OO_X(G)$ this is no longer automatic, and a choice has to be made at every evaluation point. Near a point $P$ the bundle is trivial: there are an open set $U\ni P$ and a local generator $e$ with $\cF|_U=\OO_U\cdot e$, so that every section can be written $s=f\,e$ with $f$ regular at $P$. The value of $s$ at $P$ is its class $s(P)=f(P)\,e_P$ in the fibre $\cF_P\otimes k(P)$, a one-dimensional vector space over $k(P)$ with no preferred basis. The scalar $f(P)$ is the coordinate of $s(P)$ with respect to $e_P$. If $e_P$ is replaced by $\lambda e_P$ with $\lambda\in\F_q^*$, the coordinate of every section at $P$ is divided by $\lambda$, and so is the column of the generator matrix that belongs to $P$. Choices $\lambda_1,\dots,\lambda_n$ at the points $P_1,\dots,P_n$ replace the code by its image under the diagonal matrix $\operatorname{diag}(\lambda_1^{-1},\dots,\lambda_n^{-1})$, and its dual by the image under $\operatorname{diag}(\lambda_1,\dots,\lambda_n)$. A rescaling of coordinates preserves length, dimension and the Hamming weight of every word, and rescales Schur products coordinatewise, so none of the coding-theoretic quantities of this paper is affected. The generalised Reed--Solomon codes~\cite[Chapter~10]{MS77} are an instance of this freedom, and it is the sense in which Lemma~\ref{lem:MDSbundle} speaks of the evaluation code of an auxiliary line bundle $\BB$.

In the two situations in which codes are defined in this paper, the fibre has a canonical basis vector and no choice is involved. First, let $\cF=\OO_X(G)$ and $P\notin\Supp(G)$. Then $G$ vanishes near $P$, the constant function $1$ is a local generator, and the coordinate of a section is the value $f(P)$ of the function itself. Second, consider the bundles of differentials at a point $P$ where a pole of order at most one is allowed. If $t$ is a local parameter at $P$, then $dt/t$ is a local generator, and the coordinate of $\omega=f\,dt/t$ is $f(P)=\Res_P(\omega)$. The class of $dt/t$ in the fibre does not depend on $t$: if $t'=ut$ with $u$ a unit at $P$, then $dt'/t'=dt/t+du/u$, and $du/u$ is regular at $P$, so it vanishes in the fibre. The residue therefore identifies the fibre canonically with $k(P)$; this uses that the pole is at most simple. The two canonical choices are compatible with each other by~\eqref{eq:residue_product}: the coordinate of the product of a function regular at $P$ and a differential with at most a simple pole at $P$ is the product of the coordinates. This is why the residue code of Proposition~\ref{prop:app_cauchy_duality} is orthogonal to the evaluation code exactly, and why the values~\eqref{eq:residue_value} of sections of the bundle $\LME$ on $Y$ involve no choice. In Section~\ref{sec:localization} a third situation occurs, codes of general line bundles $\cG_j$; there a divisor representative disjoint from the evaluation points is fixed (Remark~\ref{rem:shifted_codes}), which reduces to the first situation.

\section{Interpolation on Gorenstein curves: proof of Lemma~\ref{lem:fs_adjunction}} \label{sec:appendix_curves}

Throughout, $\kappa$ is an algebraically closed field and $Z$ is a projective Gorenstein curve over $\kappa$: a projective scheme of pure dimension one, without embedded points, whose dualizing sheaf $\omega_Z$ is invertible. The curves to which Lemma~\ref{lem:fs_adjunction} is applied in the text are complete intersections in a smooth threefold (Lemma~\ref{lem:koszul}) and effective Cartier divisors on smooth surfaces~\cite[Proposition~III.7.11]{Hartshorne77}; both are Gorenstein. The lemma is a form of the automatic adjunction of Catanese, Franciosi, Hulek and Reid~\cite[Lemma~2.4 and Theorem~1.1]{CFHR99}; we include a complete proof in the form used here, since the statement must be applied to nonreduced subcurves and to clusters of positive length.

\subsection{Preliminaries} \label{subsec:app_curve_prelim}

\paragraph{(A) Purity.} A coherent sheaf $\cG$ on a curve is \emph{pure} if it has no nonzero subsheaf with finite support. A subsheaf of a pure sheaf is pure. A closed subscheme $V\subseteq Z$ of pure dimension one is Cohen--Macaulay, that is, without embedded points, if and only if $\OO_V$ is pure; and then every line bundle on $V$ and every subsheaf of one is pure. Two consequences are used repeatedly: a map from a pure sheaf that is injective at every generic point of its support is injective, because its kernel has finite support; and a map into a pure sheaf that vanishes at every generic point is zero, because its image has finite support.

\paragraph{(B) Dualizing sheaves and Serre duality.} For a projective scheme $V$ of dimension one there is a dualizing sheaf $\omega_V$ with a functorial isomorphism $\Hom_{\OO_V}(\cG,\omega_V)\cong H^1(V,\cG)^\vee$ for coherent $\cG$~\cite[Proposition~III.7.5]{Hartshorne77}; it is unique up to isomorphism. If $V$ is moreover Cohen--Macaulay of pure dimension one, Serre duality holds in the form $\operatorname{Ext}^i(\cG,\omega_V)\cong H^{1-i}(V,\cG)^\vee$~\cite[Theorem~III.7.6]{Hartshorne77}, and in particular $h^0(\omega_V)=h^1(\OO_V)$ and $h^1(\omega_V)=h^0(\OO_V)$, so that
\begin{equation}\label{eq:app_chi_omega}
\chi(\omega_V)=-\chi(\OO_V).
\end{equation}
For a closed subscheme $V\subseteq Z$ of pure dimension one, the sheaf
\begin{equation}\label{eq:app_omegaV}
\omega_V:=\cHom_{\OO_Z}(\OO_V,\omega_Z)=\{\text{local sections of }\omega_Z\text{ annihilated by }\II_{V,Z}\}\;\subseteq\;\omega_Z
\end{equation}
is a dualizing sheaf of $V$: for an $\OO_V$-module $\cG$, adjunction gives $\Hom_{\OO_V}(\cG,\cHom_{\OO_Z}(\OO_V,\omega_Z))=\Hom_{\OO_Z}(\cG,\omega_Z)\cong H^1(Z,\cG)^\vee=H^1(V,\cG)^\vee$. Since $\omega_Z$ is invertible, hence pure, so is $\omega_V$.

\paragraph{(C) Artinian stalks.} At a generic point $\eta$ of a component of $Z$ the local ring $A:=\OO_{Z,\eta}$ is Artinian and Gorenstein, so $A$ is its own canonical module, and for an ideal $I\subseteq A$ with quotient $B:=A/I$ the module $(0:_AI)=\Hom_A(B,A)$ is the canonical module of $B$, of length $\length(B)$ by Matlis duality~\cite[Chapter~21]{Eisenbud95}. A quotient $B$ of $A$ is Gorenstein if and only if its canonical module is cyclic.

\paragraph{(D) Riemann--Roch on $V$.} For a Cohen--Macaulay subcurve $V$ and a line bundle $\cF$ on $V$, the degree is defined by $\deg(\cF)=\chi(\cF)-\chi(\OO_V)$, as in~\eqref{eq:curve_degree}, so that Riemann--Roch $\chi(\cF)=\deg(\cF)+1-p_a(V)$ holds by definition. For a finite subscheme $T'\subset V$, the sequence $0\to\II_{T',V}\cF\to\cF\to\cF|_{T'}\to0$ is exact, and $\cF|_{T'}$, a line bundle on a finite scheme, has length $\length(T')$; hence
\begin{equation}\label{eq:app_chi_ideal}
\chi(\II_{T',V}\cF)=\deg(\cF)+\chi(\OO_V)-\length(T').
\end{equation}

\subsection{Proof of Lemma~\ref{lem:fs_adjunction}} \label{subsec:app_fs_proof}

Let $Z$, $T$ and $\cF$ be as in Lemma~\ref{lem:fs_adjunction}, and write $\II_T$ for $\II_{T,Z}$. Suppose, for a contradiction, that $H^1(Z,\II_T\cF)\ne0$.

\paragraph{Step 1: a nonzero map to $\omega_Z$.} By (B) there is a nonzero homomorphism $\alpha:\II_T\cF\to\omega_Z$.

\paragraph{Step 2: the annihilator subcurve.} Let $\mathfrak a:=\Ann_{\OO_Z}(\im\alpha)\subseteq\OO_Z$ and let $V\subseteq Z$ be the closed subscheme with ideal $\mathfrak a$. Since $\im\alpha\ne0$, $\mathfrak a$ is a proper ideal and $V\ne\varnothing$. We claim that $V$ is a Cohen--Macaulay subcurve, that is, that $\OO_V=\OO_Z/\mathfrak a$ is pure. Indeed, $\OO_V$ acts faithfully on $\im\alpha$: a local section $t$ of $\OO_Z$ acts as zero on $\im\alpha$ exactly when $t\in\mathfrak a$. If $\mathcal T\subseteq\OO_V$ were a nonzero subsheaf with finite support, then $\mathcal T\cdot\im\alpha\subseteq\im\alpha\subseteq\omega_Z$ would be a subsheaf with finite support of the pure sheaf $\omega_Z$, hence zero, so $\mathcal T$ would act as zero on $\im\alpha$ and would vanish in $\OO_V$. Thus $\OO_V$ is pure; in particular $V$ has no isolated points, so it has pure dimension one, and no embedded points. By~\eqref{eq:app_omegaV} and the definition of $\mathfrak a$, $\im\alpha\subseteq\omega_V\subseteq\omega_Z$.

\paragraph{Step 3: factoring through $V$.} Since $\omega_V$ is an $\OO_V$-module, $\alpha$ kills $\mathfrak a\cdot\II_T\cF$ and factors through $\II_T\cF\otimes\OO_V=(\II_T/\mathfrak a\II_T)\otimes\cF$, where we used that $\cF$ is locally free. The natural map $\II_T/\mathfrak a\II_T\to\OO_V$ has image $(\II_T+\mathfrak a)/\mathfrak a=\II_{T\cap V,V}$, the ideal in $V$ of the scheme-theoretic intersection $T\cap V$, and kernel $(\II_T\cap\mathfrak a)/\mathfrak a\II_T$, which vanishes away from the finite set $T$ (where $\II_T=\OO_Z$) and hence has finite support. Its image under $\alpha$ in the pure sheaf $\omega_V$ is therefore zero, and $\alpha$ induces a homomorphism of $\OO_V$-modules
\[
\bar\alpha:\II_{T\cap V,V}\,\cF|_V\longrightarrow\omega_V ,\qquad \im\bar\alpha=\im\alpha .
\]

\paragraph{Step 4: $\bar\alpha$ is an isomorphism at the generic points of $V$.} Let $\eta$ be the generic point of a component of $V$, and put $A:=\OO_{Z,\eta}$, $I:=\mathfrak a_\eta$, $B:=A/I=\OO_{V,\eta}$. Since $T$ is finite, $\eta\notin T$, so $(\II_{T\cap V,V})_\eta=B$ and the source of $\bar\alpha_\eta$ is $\cF_\eta\otimes B\cong B$, a free $B$-module of rank one generated by the class $e$ of a local generator of $\cF$. The image $\im\alpha_\eta$ is generated by $m:=\bar\alpha_\eta(e)$, and its annihilator in $A$ is $I$ by the definition of $\mathfrak a$ (annihilators commute with localisation for finitely generated modules), so $\im\alpha_\eta\cong A/I=B$ as $A$-modules; in particular $\length(\im\alpha_\eta)=\length(B)$. On the other hand $\im\alpha_\eta\subseteq(\omega_V)_\eta=\Hom_A(B,A)=(0:_AI)$, which has length $\length(B)$ by (C). A submodule of the same finite length is the whole module, so $\im\alpha_\eta=(\omega_V)_\eta$, and $\bar\alpha_\eta:B\to(\omega_V)_\eta$ is a surjection between $A$-modules of the same finite length, hence an isomorphism. (Incidentally, $(\omega_V)_\eta$ is cyclic, so $B$ is Gorenstein: the annihilator subcurve is generically Gorenstein, as in~\cite{CFHR99}.)

\paragraph{Step 5: $\bar\alpha$ is injective with finite cokernel.} The source of $\bar\alpha$ is a subsheaf of the line bundle $\cF|_V$ on the Cohen--Macaulay curve $V$, hence pure by (A). Its kernel vanishes at every generic point of $V$ by Step~4, so it has finite support and is zero. The cokernel $Q$ of $\bar\alpha$ also vanishes at every generic point, so it has finite support and $\chi(Q)=h^0(Q)=\length(Q)\ge0$.

\paragraph{Step 6: the Euler characteristic count.} From the exact sequence $0\to\II_{T\cap V,V}\cF|_V\xrightarrow{\bar\alpha}\omega_V\to Q\to0$ and~\eqref{eq:app_chi_omega}, which applies because $V$ is Cohen--Macaulay,
\[
\chi\bigl(\II_{T\cap V,V}\cF|_V\bigr)=\chi(\omega_V)-\length(Q)\le\chi(\omega_V)=-\chi(\OO_V).
\]
By~\eqref{eq:app_chi_ideal} applied to $T':=T\cap V$,
\[
\chi\bigl(\II_{T\cap V,V}\cF|_V\bigr)=\deg(\cF|_V)+\chi(\OO_V)-\length(T\cap V).
\]
Combining the two, $\deg(\cF|_V)-\length(T\cap V)\le-2\chi(\OO_V)=2p_a(V)-2$. By Step~2, $V$ is a nonzero Cohen--Macaulay subcurve of $Z$, so this contradicts the hypothesis~\eqref{eq:gen1}. Hence $H^1(Z,\II_{T,Z}\cF)=0$. The equivalent formulation follows from the exact sequence $H^0(Z,\cF)\to H^0(T,\cF|_T)\to H^1(Z,\II_{T,Z}\cF)\to H^1(Z,\cF)\to0$.\hfill$\square$

\begin{remark}[What the proof uses] \label{rem:app_fs_uses}
Only standard inputs are used: Serre duality on a projective Cohen--Macaulay curve, the description~\eqref{eq:app_omegaV} of the dualizing sheaf of a subcurve, Matlis duality for Artinian Gorenstein rings, and purity. No external input of the paper is involved. The choice of $V$ in Step~2 is forced: it is the smallest subcurve through which $\alpha$ factors, and it is exactly what makes $\bar\alpha$ generically bijective in Step~4. The hypothesis must be checked on every Cohen--Macaulay subcurve because the proof cannot predict which $V$ the hypothetical class $\alpha$ selects; when $Z$ is a Cartier divisor on a smooth surface, the subcurves are the effective subdivisors $V\le Z$ and $V$ is cut out by the orders of vanishing of $\alpha$ along the components, which recovers the form of the criterion in~\cite{CFHR99}.
\end{remark}

\section{Proof of the general criterion}\label{sec:appendix_general}

This appendix proves Theorem~\ref{thm:general} and Proposition~\ref{prop:general_constants} by going through Sections~\ref{sec:transfer}--\ref{sec:proof} and recording what changes. Throughout, a triple in $\mathcal T(\mathsf L,c_{\gon},\tau,\tau_{\mathrm P})$ with $n_{\min}\ge n_0$ is fixed, the constants are those of~\eqref{eq:general_constants}, and the notation of Section~\ref{subsec:general_setting} is in force.

\subsection{What is reused}\label{subsec:appC_reused}

Sections~\ref{sec:ag_background} and~\ref{sec:product_geometry} are stated for arbitrary curves: the threefold is $Y=X_1\times X_2\times X_3$, with $\omega_Y=\boxprod\omega_{X_j}$ and the K\"unneth formula unchanged. The following results are proved for arbitrary codes, curves or line bundles and are used as they stand.
\begin{center}
\renewcommand{\arraystretch}{1.15}
\begin{tabular}{@{}>{\raggedright\arraybackslash}p{0.40\textwidth}>{\raggedright\arraybackslash}p{0.56\textwidth}@{}}
\toprule
result & how it is used below\\
\midrule
Lemmas~\ref{lem:rectangle}, \ref{lem:transfer}, \ref{lem:scalars}; Prop.~\ref{prop:sandwich} & transfer of expansion between lengths and codes (Lemma~\ref{lem:appC_two_directions})\\
Lemmas~\ref{lem:MDSbundle}, \ref{lem:short_jets}, \ref{lem:grid_ideal}, \ref{lem:gluing}, \ref{lem:K_CM} & the interpolation apparatus, unchanged\\
Lemma~\ref{lem:fs_adjunction}; Prop.~\ref{prop:cube} & adjunction on CM subcurves (Prop.~\ref{prop:appC_generation}, Lemma~\ref{lem:appC_localization}); theorem of the cube (Section~\ref{subsec:appC_localization})\\
Lemma~\ref{lem:gonality}(b) & $\gamma_K(X_j)=\gamma_k(X_j)\ge c_{\gon}n_j$ over the working field $K$\\
Lemmas~\ref{lem:filtration}, \ref{lem:layer_lengths}, \ref{lem:axis_degrees} & layer lengths (Prop.~\ref{prop:appC_generation}) and axis degrees (Section~\ref{subsec:appC_localization}), with weights $n_j$\\
Prop.~\ref{prop:carrier} & carriers, with a line bundle $\BB_j$ of degree $g_j+s-1$ on each $X_j$\\
Lemma~\ref{lem:weighted_filling} & weighted filling (Section~\ref{subsec:appC_final})\\
\bottomrule
\end{tabular}
\end{center}
In Proposition~\ref{prop:carrier} the pair divisors have factor degree at most $g_j+s-1$ in direction $j$. The working field is $K=k(V)$ with $V$ a product of Picard varieties of the three curves and affine spaces, as in Section~\ref{subsec:fields}.

\subsection{Two directions}\label{subsec:appC_two}

Call a puncturing of the triple \emph{admissible} if it deletes in total at most $2\vartheta_0n_{\min}$ indices. Each retained set then has at least $(1-2\vartheta_0)n_j$ points, which exceeds $m_j$ by~\eqref{eq:deltaD}, so the restrictions are injective and the punctured codes have relative distance at least $1-\theta_j/(1-2\vartheta_0)\ge\delta_{\mathrm D}-4\vartheta_0$.

\begin{lemma}[Admissible puncturings have expansion $\eta_{\mathrm{gen}}$]\label{lem:appC_two_directions}
Every pair of codes of an admissible puncturing has expansion at least $\eta_{\mathrm{gen}}$. Consequently $\eta_{\mathrm{gen}}L_{\mathrm w}(x)\le P(x)$ for all $x\in S$, and the second part of Lemma~\textup{\ref{lem:transfer}} holds for every admissible puncturing with the constant $\eta_{\mathrm{gen}}$.
\end{lemma}

\begin{proof}
Let two codes of an admissible puncturing be restricted to a common length $n'\ge(1-2\vartheta_0)\min(n_i,n_j)$. Condition~\ref{G3} and $\vartheta_0\le\tau_{\mathrm P}/8$ give
\[
4+m_i+g_i+m_j+g_j\le4+(1-\tau_{\mathrm P})\min(n_i,n_j)<\bigl(1-\tfrac12\tau_{\mathrm P}\bigr)n' ,
\]
because $(1-\tfrac12\tau_{\mathrm P})(1-2\vartheta_0)-(1-\tau_{\mathrm P})\ge\tfrac14\tau_{\mathrm P}$ and $\tfrac14\tau_{\mathrm P}\min(n_i,n_j)\ge20>4$ for $n_{\min}\ge n_0$. External Input~\ref{ext:two_code_robustness}, which allows different curves, gives $\alpha_{\mathrm{gen}}$-robustness, and Lemma~\ref{lem:robust_to_expansion} gives expansion at least $\eta'_{\mathrm{gen}}$. For unequal retained lengths $n\ge n'$ one has $n/n'\le\mathsf L'$ and $(n-n')/d(C')\le(\mathsf L'-1)/(\delta_{\mathrm D}-4\vartheta_0)$, so Lemma~\ref{lem:rectangle} gives expansion at least $\eta_{\mathrm{gen}}$ as defined in~\eqref{eq:general_constants}. This is the general form of Proposition~\ref{prop:pair}. Applying it slice by slice as in the proof of Proposition~\ref{prop:sandwich} gives $\eta_{\mathrm{gen}}L_{\mathrm w}(x)\le P(x)$. The second part of Lemma~\ref{lem:transfer} was stated for three copies of one code, but its proof, which punctures the three coordinates one after the other, applies verbatim to three different codes as long as every intermediate pair has expansion at least $\eta_{\mathrm{gen}}$; for an admissible puncturing all intermediate pairs are pairs of an admissible puncturing.
\end{proof}

\subsection{Directional exact generation}\label{subsec:appC_generation}

\begin{proposition}[Directional exact generation]\label{prop:appC_generation}
Let $E'_j\subseteq E_j$ have sizes $n'_j$, let $C_j=C_L(E'_j,G_j)$ with $m_j>2g_j-2$, and let the pair divisors $D_{jk}$ have no axis components and external class with factor degree at most $b_j$ in direction $j$. For $d=(d_1,d_2,d_3)\ne0$ with $d_j\ge0$ put $S_d:=\sum_jn'_jd_j$, $H_d:=\sum_jb_jd_j$ and $H_{\max}:=\max_jb_jd_j$. Suppose that, for two distinct directions $j_0\ne j_1$,
\begin{align}
&n'_j-m_j-2b_j>0\quad(j=1,2,3),\label{eq:gen_free}\\
&(n'_i-m_i-2b_i)d_i+(n'_j-m_j-2b_j)d_j>\min(n'_id_i,n'_jd_j)\notag\\
&\hspace{12em}(i<j,\ d_i+d_j>0,\ d_i\le b_j,\ d_j\le b_i),\label{eq:gen_surface}\\
&\max_jn'_jd_j-\textstyle\sum_jm_jd_j-2(H_d-H_{\max})>0,\label{eq:gen_third}\\
&S_d-\textstyle\sum_jm_jd_j-(H_d-b_{j_0}d_{j_0})-2(H_d-H_{\max})>\min(n'_{j_0}d_{j_0},\,S_d-n'_{j_0}d_{j_0}),\ \text{ and the same with $j_1$ for $j_0$}.\label{eq:gen_first}
\end{align}
Then $S_B\cap k^\Phi=\sum_{\ell\subset R}V_\ell$, as in Theorem~\textup{\ref{thm:exact_generation}}.
\end{proposition}

\begin{proof}
Label the cylinders so that $\Pi_1$ and $\Pi_2$ are the two with remaining directions $j_0$ and $j_1$. The proof of Theorem~\ref{thm:exact_generation} goes through with the following replacements.
\begin{itemize}[leftmargin=2em,itemsep=2pt,topsep=2pt]
\item \emph{Lemma~\ref{lem:config}(c) and the assembly step.} $\LL(-J)$ has factor degrees at least $2g_j-2+n'_j-m_j-2b_j>2g_j-2$ by~\eqref{eq:gen_free}.
\item \emph{Lemma~\ref{lem:step1}.} For $\Pi_1=D_{12}\times\Xj3$ the $\Xj3$-factor has degree at least $2g_3-2+n'_3-m_3-2b_3>2g_3-2$. For a subdivisor $V\le D_{12}$ with coordinate degrees $u=d_1(V)\le b_2$ and $v=d_2(V)\le b_1$ one has $V^2\le2uv\le b_1u+b_2v$ by Hodge index, so adjunction gives $2p_a(V)-2\le(2g_1-2+b_1)u+(2g_2-2+b_2)v$, while $\deg(\cF_{12}|_V)\ge(2g_1-2+n'_1-m_1-b_1)u+(2g_2-2+n'_2-m_2-b_2)v$ and $\length(Z\cap V)\le\min(n'_1u,n'_2v)$. The margin of Lemma~\ref{lem:fs_adjunction} is the left side of~\eqref{eq:gen_surface} minus its right side.
\item \emph{Lemma~\ref{lem:genus_Cij}.} The auxiliary box products $\AA^{(r)}$ have factor degrees $g_r-1$ in direction $r$ and $g_j-1+b_j$ for $j\ne r$, and the conclusion becomes
\begin{equation}\label{eq:gen_genus_Cij}
2p_a(V)-2\le\sum_j(2g_j-2)d_j+2(H_d-H_{\max}) .
\end{equation}
\item \emph{Lemma~\ref{lem:step2}.} $\deg(\LL|_V)=\sum_j(2g_j-2+n'_j-m_j)d_j$; the cylinder $\Pi_1$ has $\Pi_1\cdot V\le H_d-b_{j_0}d_{j_0}$, its factor degree in the remaining direction $j_0$ being $0$; and the cluster of the first layer has length at most $\min(n'_{j_0}d_{j_0},S_d-n'_{j_0}d_{j_0})$ by Lemma~\ref{lem:layer_lengths}. So the first-layer margin is the left side of~\eqref{eq:gen_first} minus its right side, the second layer is the same with $j_1$ in place of $j_0$, and the third-layer margin, with cluster at most $S_d-\max_jn'_jd_j$, is the left side of~\eqref{eq:gen_third}.
\end{itemize}
Everything else in Section~\ref{sec:generation}, including the descent to $k$, is unchanged.
\end{proof}

Normalising by $S_d$ with $w_j:=n'_jd_j/S_d$, $\theta'_j:=m_j/n'_j$ and $\beta'_j:=b_j/(2n'_j)$ turns~\eqref{eq:gen_third} into $\mathsf Q_0(w)>0$ and~\eqref{eq:gen_first} into $\mathsf Q_{j_0}(w)>0$ and $\mathsf Q_{j_1}(w)>0$, with $(\theta'_j,\beta'_j)$ in place of $(\theta_j,\beta_j)$. This is the origin of the margin functions~\eqref{eq:Q0_margin} and~\eqref{eq:Qa_margin}.

\subsection{Support localisation}\label{subsec:appC_localization}

Let $0<\lambda\le\lambda_{\max}$ and let $x\in S$ be nonzero with $\wt(x)\le\lambda\mathsf V$. Call a line in direction $j$ \emph{heavy} if it carries more than $\tau n_j/4$ points of $\supp(x)$. Counting incidences direction by direction, the number of heavy lines is $\mathsf m_1<\sum_j4\wt(x)/(\tau n_j)\le12\lambda\mathsf L^3n_{\min}^2/\tau$, and $u_1:=\lfloor\sqrt{\mathsf m_1}\rfloor+1<\sqrt{12\mathsf L^3\lambda/\tau}\,n_{\min}+1$. Proposition~\ref{prop:carrier} with $s=u_1$ gives the first carrier $J_1$, with factor degree at most $b_{1,j}:=g_j+u_1-1$ in direction $j$, retained sets of sizes $n'_j\ge n_j-2u_1$, and the shifted word $y$ of Lemma~\ref{lem:shifted_word}, which lies in the sum code of the codes $C'_j=C_L(B^{(1)}_j,G'_j)$ with $\deg G'_j\le m'_j:=m_j+2g_j+2u_1-2$, and whose lines in direction $j$ carry at most $\tau n_j/4$ points of $T=\supp(y)$. If $y=0$, put $u_2:=1$ and $J_2:=0$ and pass directly to~\eqref{eq:appC_theta} and Lemma~\ref{lem:appC_box}, as in the proof of Proposition~\ref{prop:localization}; assume $y\ne0$. Put $s:=\lceil\zeta n_{\min}\rceil$, $t:=\lfloor s/2\rfloor$, $a_j:=g_j+s-1$, choose $\AA_j$ of degree $a_j$ on $X_j$ by Lemma~\ref{lem:short_jets}, and define $W$, $r$ and $\Delta$ as in Section~\ref{subsec:fixed_divisor}. As in~\eqref{eq:trivial_r}, the trivial bound
\begin{equation}\label{eq:trivial_codim}
r=\codim W\le|T|\le\wt(x)\le\lambda\mathsf V\le\lambda\mathsf L^3n_{\min}^3
\end{equation}
suffices. Since $\zeta n_{\min}\ge2$, $ts^2\ge\tfrac14\zeta^3n_{\min}^3>\lambda\mathsf L^3n_{\min}^3\ge r$ by the first term of $\lambda_{\max}$, so Lemma~\ref{lem:axis_degrees} gives $\delta_j\le r/s^2\le\delta_*:=\lambda\mathsf L^3n_{\min}/\zeta^2$ for each $j$, and $\dim W\ge s^3-r\ge\tfrac45\zeta^3n_{\min}^3\ge2$. The second and third terms of $\lambda_{\max}$ give $\delta_*\le\tau n_{\min}/10$ and $2\delta_*\le c_{\gon}n_{\min}/2<\gamma_K(X_j)$.

\begin{lemma}[Localisation margins]\label{lem:appC_localization}
Assume $u_1\le\vartheta_0n_{\min}$. Then the three margins of Lemma~\textup{\ref{lem:support_on_Delta}}, for components of axis type in direction $j$, plane type in directions $k,l$, and threefold type, satisfy
\begin{align}
\mu_A^{(j)}&:=n'_j-m'_j-4a_j-\delta_j-\tfrac14\tau n_j\ \ge\ n_{\min}\bigl(\tfrac34-\tfrac15-\tfrac1{10}-\tfrac1{20}-\tfrac1{10}\bigr)\tau\ >\ 0,\label{eq:appC_muA}\\
\mu_P^{(kl)}(d)&:=\sum_{h\in\{k,l\}}(n'_h-m'_h-5a_h)d_h-\min(n_kd_k,n_ld_l)\ \ge\ S^0_{kl}\bigl(1-\tfrac15-\tfrac18-\tfrac1{16}\bigr)\tau\ >\ 0,\label{eq:appC_muP}\\
\mu_T(d)&:=\sum_j(n'_j-m'_j-6a_j)d_j-\min_jn_jd_j\ \ge\ S^0_d\bigl(1-\tfrac15-\tfrac3{20}-\tfrac3{40}\bigr)\tau\ >\ 0,\label{eq:appC_muT}
\end{align}
where $S^0_{kl}:=n_kd_k+n_ld_l$ and $S^0_d:=\sum_jn_jd_j$. Hence $T\subseteq\Delta$.
\end{lemma}

\begin{proof}
Lemma~\ref{lem:genus_Cy} becomes $2p_a(V)-2\le\sum_j(2g_j-2+4a_j)d_j(V)$, with $\cF=\boxprod(\AA_j^{\otimes2}\otimes\Theta_j)$ of factor degrees $2a_j+g_j-1$ and the same three vanishing statements (the twist $\cF\otimes\AA^{-1}\otimes\omega_Y^{-1}$ has factor degrees $s>0$). In Lemma~\ref{lem:DV_UV}, with $\deg\MM_j=a_j-\delta_j$, the bounds read $\Delta\cdot\Sigma\le\delta_j$, $\le a_kd_k+a_ld_l$ and $\le2\sum_ja_jd_j$ for the three types, and $\#(U\cap\Sigma)\le\tau n_j/4$, $\le\min(n_kd_k,n_ld_l)$ and $\le\min_jn_jd_j$ respectively. With $\deg(\Lambda|_V)=\sum_j(2g_j-2+n'_j-m'_j)d_j$, the margins of Lemma~\ref{lem:support_on_Delta} are the three quantities displayed, and
\begin{align*}
\mu_A^{(j)}&\ge n_j\bigl(1-\theta_j-6\beta_j-\tfrac14\tau\bigr)-4u_1-4s-\delta_*+6,\\
\mu_P^{(kl)}(d)&\ge S^0_{kl}\Bigl(\mathsf A_{kl}(w)-\frac{4u_1+5s}{n_{\min}}\Bigr),\qquad
\mu_T(d)\ \ge\ S^0_d\Bigl(\mathsf T(w)-\frac{4u_1+6s}{n_{\min}}\Bigr),
\end{align*}
where $w_h=n_hd_h/S^0_{kl}$, respectively $w_j=n_jd_j/S^0_d$, using $n'_j-m'_j\ge n_j(1-\theta_j-2\beta_j)-4u_1+2$, $a_j=g_j+s-1$ and $S^0_{kl}\ge n_{\min}(d_k+d_l)$. Now insert $u_1\le\vartheta_0n_{\min}\le\tau n_{\min}/20$, $s\le\zeta n_{\min}+1$ with $\zeta=\tau/40$, $n_{\min}\ge80/\tau$, $\delta_*\le\tau n_{\min}/10$, and $1-\theta_j-7\beta_j\ge\tau$, $\mathsf A_{kl}\ge\tau$, $\mathsf T\ge\tau$ from~\ref{G4}: the displayed lower bounds follow term by term. The remaining hypothesis of Lemma~\ref{lem:support_on_Delta}, $n'_j-m'_j-2a_j>0$, follows from $1-\theta_j-4\beta_j\ge\tau$ in the same way. Hence $T\subseteq\Delta$.
\end{proof}

Lemma~\ref{lem:grid_cover} holds on $Y=X_1\times X_2\times X_3$ with the same proof, the cube identity~\eqref{eq:cube_fibres} being an identity of line bundles on $X_3$ and the gonality hypothesis being $2\delta_3<\gamma_K(X_3)$; with sides $n'_j\le\mathsf Ln_{\min}$ the count is at most $5\mathsf Ln_{\min}\max_j\delta_j\le5\mathsf Ln_{\min}\delta_*$, and the existence of a good pair needs only $\delta_*<\min_jn'_j$. So $\mathsf m_2\le5\mathsf L^4\lambda n_{\min}^2/\zeta^2$ and $u_2:=\lfloor\sqrt{\mathsf m_2}\rfloor+1<(\sqrt5\mathsf L^2/\zeta)\sqrt\lambda\,n_{\min}+1$. The second carrier $J_2$ is constructed by Proposition~\ref{prop:carrier} with $s=u_2$, and the pair divisors $D_{jk}:=D^{(1)}_{jk}+D^{(2)}_{jk}$ of $J:=J_1+J_2$ have factor degree at most $b_j:=2g_j+U_0-2$ in direction $j$, where $U_0:=u_1+u_2$. By the choice of $\Upsilon$ and the fourth term of $\lambda_{\max}$,
\begin{equation}\label{eq:appC_theta}
\vartheta:=\frac{U_0}{n_{\min}}<\Upsilon\sqrt\lambda+\frac2{n_{\min}}\le\vartheta_0 ,
\end{equation}
using $n_{\min}\ge n_0\ge4/\vartheta_0$. Since $u_1\le U_0$, this justifies the assumption of Lemma~\ref{lem:appC_localization} and the applicability of Proposition~\ref{prop:carrier} ($u_1,u_2\le n_{\min}/8$). As in Proposition~\ref{prop:localization}(a),(b), $\supp(x_B)$ lies on the footprint of $J$ in the final box $B$, every fibre of $R$ meets the fibres of the other directions in at most $2(U_0-2)$ points of $B$, fewer than $2U_0\le2\vartheta_0n_{\min}$ indices have been deleted in total, and the puncturing is admissible.

\subsection{The final box}\label{subsec:appC_box}

\begin{lemma}[The final box satisfies the generation hypotheses]\label{lem:appC_box}
The divisor $J$ on the final box $B$ satisfies~\eqref{eq:gen_free}--\eqref{eq:gen_first} with $j_0,j_1$ the two directions of~\ref{G5}. Consequently $x_B$ is a $k$-linear combination of coded lines on the fibres of $R$.
\end{lemma}

\begin{proof}
On $B$ we have $n'_j\ge n_j-2\vartheta_0n_{\min}\ge(1-2\vartheta_0)n_j$ and $b_j\le2g_j+\vartheta_0n_{\min}$. Write $S^0_d:=\sum_jn_jd_j$ and $S^0_{ij}:=n_id_i+n_jd_j$ for the undeleted weighted degrees, $w_j:=n_jd_j/S^0_d$ (respectively $w_h:=n_hd_h/S^0_{ij}$ on an edge), and $H^0_d:=\sum_j2g_jd_j$, $H^0_{\max}:=\max_j2g_jd_j$. Then
\[
S_d\ge(1-2\vartheta_0)S^0_d,\quad
H_d-b_{j_0}d_{j_0}\le(H^0_d-2g_{j_0}d_{j_0})+\vartheta_0S^0_d,\quad
H_d-H_{\max}=\min_r\sum_{j\ne r}b_jd_j\le(H^0_d-H^0_{\max})+\vartheta_0S^0_d,
\]
and $\min(n'_{j_0}d_{j_0},S_d-n'_{j_0}d_{j_0})\le\min(n_{j_0}d_{j_0},S^0_d-n_{j_0}d_{j_0})$, $\max_jn'_jd_j\ge(1-2\vartheta_0)\max_jn_jd_j$. Substituting, the left side minus the right side of each hypothesis is bounded below as follows:
\begin{align*}
\eqref{eq:gen_first}:&\quad S^0_d\,\bigl(\mathsf Q_{j_0}(w)-5\vartheta_0\bigr),\ \text{and likewise with }j_1,\\
\eqref{eq:gen_third}:&\quad S^0_d\,\bigl(\mathsf Q_0(w)-4\vartheta_0\bigr),\\
\eqref{eq:gen_surface}:&\quad S^0_{ij}\Bigl(\sum_{h\in\{i,j\}}(1-\theta_h-4\beta_h)w_h-\min(w_i,w_j)-4\vartheta_0\Bigr)\ \ge\ S^0_{ij}\bigl(\mathsf A_{ij}(w)-4\vartheta_0\bigr),\\
\eqref{eq:gen_free}:&\quad n_j\bigl(1-\theta_j-4\beta_j-4\vartheta_0\bigr)\ \ge\ \tfrac45\tau n_j .
\end{align*}
All four are positive by~\ref{G4} and~\ref{G5}, since $5\vartheta_0\le\tfrac14\tau$. Proposition~\ref{prop:appC_generation} gives the conclusion.
\end{proof}

\subsection{Weighted filling, restoration, and the constants}\label{subsec:appC_final}

\begin{proof}[Proof of Proposition~\textup{\ref{prop:general_constants}}]
(a) Every nonzero codeword of $C_j|_{B_j}$ has weight at least $d_{\min}:=\min_j(n'_j-m_j)\ge(\delta_{\mathrm D}-2\vartheta)n_{\min}$, and the crossing bound is $\nu\le2\vartheta n_{\min}<d_{\min}$, so Lemma~\ref{lem:weighted_filling} gives $P_B(x_B)\le F_{\mathrm{gen}}(\vartheta)\wt(x_B)$ with $F_{\mathrm{gen}}(\vartheta):=1+4\vartheta/(3(\delta_{\mathrm D}-4\vartheta))$. Lemma~\ref{lem:transfer}, second part, in the form of Lemma~\ref{lem:appC_two_directions} and with fewer than $2\vartheta n_{\min}$ deleted indices in total, gives $P(x)\le\Gamma F_{\mathrm{gen}}(\vartheta)\wt(x)$ with $\Gamma\le\exp(4\vartheta/(\eta_{\mathrm{gen}}(1-2\vartheta)))$, and $\log F_{\mathrm{gen}}(\vartheta)+4\vartheta/(\eta_{\mathrm{gen}}(1-2\vartheta))\le C_{\mathrm{gen}}\vartheta$ for $\vartheta\le\vartheta_0$ by $\log(1+z)\le z$. With~\eqref{eq:appC_theta} this is~\eqref{eq:general_filling}.

(b) As in the last step of the proof of Theorem~\ref{thm:sharp_filling}: $C_{\mathrm{gen}}\Upsilon\sqrt{\lambda_0(\mu)}\le h/2$ and $2C_{\mathrm{gen}}/n_{\min}\le h/2$, so the exponent in~\eqref{eq:general_filling} is at most $h=\log(1+\mu)$.

(c) Take $\mu=1$. A nonzero $x$ with $\wt(x)\le\lambda_0(1)\mathsf V$ on a triple with $n_{\min}\ge n_0(1)$ has $L_{\mathrm w}(x)\le P(x)/\eta_{\mathrm{gen}}\le2\wt(x)/\eta_{\mathrm{gen}}$ by Lemma~\ref{lem:appC_two_directions}; if $\wt(x)>\lambda_0(1)\mathsf V$ then $L_{\mathrm w}(x)\le3\mathsf V<3\wt(x)/\lambda_0(1)$; and if $n_{\min}<n_0(1)$ then $L_{\mathrm w}(x)\le3\mathsf V\le3\mathsf L^3n_0(1)^3\le3\mathsf L^3n_0(1)^3\wt(x)$ for every nonzero $x$.

Finally, replacing $\delta_{\mathrm D}$ by the smaller number $\tau$ in~\eqref{eq:general_constants} only decreases $\eta_{\mathrm{gen}}$ and increases $C_{\mathrm{gen}}$, so the statements remain true.
\end{proof}

\begin{proof}[Proof of Theorem~\textup{\ref{thm:general}}]
The parameter bounds $\dim C_j/n_j\ge\tau$, $d(C_j)/n_j\ge\tau$, $d(C_j^\perp)/n_j\ge\tau$ were recorded after Definition~\ref{def:general_class}. Uniform expansion with the constant~\eqref{eq:general_rho} is Proposition~\ref{prop:general_constants}(c), and~\eqref{eq:general_sharp} is Proposition~\ref{prop:general_constants}(b). By the last sentence of the proposition, $\delta_{\mathrm D}$ may be replaced by $\tau$ in all constants, after which $\rho$, $\lambda_0(\mu)$ and $n_0(\mu)$ depend only on $\mathsf L,c_{\gon},\tau,\tau_{\mathrm P}$ and $\mu$, as asserted.
\end{proof}

\subsection{Proofs of the corollaries}\label{subsec:appC_corollaries}

\begin{proof}[Proof of Corollary~\textup{\ref{cor:general_symmetric}}]
By the monotonicity of the margin functions it suffices to evaluate them at $\theta_j=\theta$, $\beta_j=\beta$. There
\[
\mathsf A_{ij}\ge\tfrac12-\theta-7\beta,\qquad
\mathsf T\ge\tfrac23-\theta-8\beta,\qquad
\mathsf Q_0=(1+4\beta)\max_hw_h-\theta-4\beta\ge\tfrac13-\theta-\tfrac83\beta,\qquad
\mathsf Q_j\ge\tfrac12-\theta-3\beta .
\]
The first two bounds are immediate from $\min(w_i,w_j)\le\tfrac12$ and $\min_hw_h\le\tfrac13$. For the last, consider $w_j$. If $w_j\le\tfrac13$, then $\max_hw_h\ge(1-w_j)/2$ and $\mathsf Q_j\ge1-\theta-4\beta-w_j\ge\tfrac23-\theta-4\beta$. If $\tfrac13\le w_j\le\tfrac12$, then $\max_hw_h\ge w_j$ and $\mathsf Q_j\ge1-\theta-6\beta+6\beta w_j-w_j\ge\tfrac12-\theta-3\beta$. If $w_j\ge\tfrac12$, then $\mathsf Q_j\ge w_j+6\beta w_j-\theta-6\beta\ge\tfrac12-\theta-3\beta$. All four lower bounds are positive under~\eqref{eq:symmetric_region}: the first is its third inequality, the third is its second inequality, the second exceeds $\tfrac16-\beta>0$, and the fourth exceeds the first. The pair condition~\ref{G3} holds with $\tau_{\mathrm P}:=1-2\theta-2\beta>0$, and~\ref{G1} with $\mathsf L=1$, $c_{\gon}=1/(q+1)$. Take $\tau:=\min\{\tfrac12-\theta-7\beta,\ \tfrac23-\theta-8\beta,\ \tfrac13-\theta-\tfrac83\beta,\ \tau_0\}$ and apply Theorem~\ref{thm:general}. For the last assertion, the interval $(2\beta,\min\{(1-8\beta)/3,\tfrac12-7\beta\})$ is nonempty if and only if $2\beta<\tfrac12-7\beta$ and $2\beta<(1-8\beta)/3$, that is, $\beta<\tfrac1{18}$ and $\beta<\tfrac1{14}$.
\end{proof}

\begin{proof}[Proof of Corollary~\textup{\ref{cor:field_coverage}}]
For even $m$ the Garcia--Stichtenoth tower over $\F_{\ell^2}$ with $\ell=p^{m/2}$ has limit $\lim N(F_i)/g(F_i)=\ell-1$~\cite{GS95}. For odd $m=2e+1\ge3$, Theorem~1.1 of Bassa, Beelen, Garcia and Stichtenoth~\cite{BBGS15} gives an explicit recursive tower over $\F_{p^m}$ with
\[
\liminf_i\frac{N(F_i)}{g(F_i)}\ \ge\ \Lambda(p,e):=\frac{2(p^{e+1}-1)}{p+1+(p-1)/(p^e-1)}=\Bigl(\frac{1}{2(p^{e+1}-1)}+\frac1{2(p^e-1)}\Bigr)^{-1},
\]
the harmonic mean of $p^{e+1}-1$ and $p^e-1$. The criterion $\beta<\tfrac1{18}$ asks for a limit above $18$. For even $m$ this is $\ell\ge23$, and the twelve orders $q=\ell^2\le361$ with $\ell\le19$ are excluded. For odd $m$, $\Lambda(p,e)$ is at least its smaller argument $p^e-1$ and increases with $p$ and with $e$, so the following values decide every case:
\begin{center}
\renewcommand{\arraystretch}{1.15}
\begin{tabular}{@{}cccl@{}}
\toprule
$m$ & excluded $(p,m)$: $\Lambda<18$ & first covered $(p,m)$: $\Lambda>18$ & consequence\\
\midrule
$3$ & $(2,3),(3,3),(5,3),(7,3)$ & $(11,3)$, $\Lambda=\tfrac{240}{13}$ & all $p\ge11$ covered\\
$5$ & $(2,5),(3,5)$ & $(5,5)$, $\Lambda=\tfrac{1488}{37}$ & all $p\ge5$ covered\\
$7$ & $(2,7)$ & $(3,7)$, $\Lambda=\tfrac{2080}{53}$ & all $p\ge3$ covered\\
$9$ & --- & $(2,9)$, $\Lambda=\tfrac{465}{23}$ & all $p$ covered\\
$\ge11$ & --- & $p^e-1\ge31$ & all $p$ covered\\
\bottomrule
\end{tabular}
\end{center}
The seven excluded odd-exponent orders are $8,27,125,343,32,243,128$. Reading $q$ as $\ell^n$ for any divisor $n\ge2$ of $m$ and applying~\cite[Theorem~1.2]{BBGS15} to a coprime partition $n=j+k$ with $j>k$ other than the balanced one over $p$ gives nothing further: the resulting harmonic mean is at most $2(\ell^k-1)$, where $\ell^k\le p^{m/2-1}$ when $m$ is even and $\ell^k\le p^{(m-3)/2}$ when $m$ is odd, while $2(p^{c-1}-1)<p^c-1$ for all $c\ge1$. The two lists together are exactly the non-prime orders $q\ge512$ and $q<512$. On each floor of such a tower take $E_i:=X_i(\F_q)\setminus\{P_i\}$ for a rational point $P_i$ and $G_i:=\lfloor\theta n_i\rfloor P_i$, with $\theta$ in the interval of Corollary~\ref{cor:general_symmetric} for $\beta:=\tfrac12\bigl(\tfrac1{18}+\limsup_ig_i/n_i\bigr)$. The genus ratio is below $\beta$ on all sufficiently large floors, and the corollary applies with $\tau_0:=\tfrac12(\theta-2\beta)$. Explicitness is as in Section~\ref{subsec:codes_Ci}: the towers are given by recursive equations and the codes by Hess' algorithm.
\end{proof}


\end{document}